\documentclass[twocolumn]{autart}    % Enable this line and disable the 
\DeclareSymbolFont{newfont}{OML}{cmm}{m}{it}% Computer Modern math font
\DeclareMathSymbol{\Epsilon}{3}{newfont}{15}% Symbol 15
\DeclareMathSymbol{\Varrho}{3}{newfont}{37}% Symbol 37
\usepackage{amsfonts}
\usepackage{amsmath}
\usepackage{cancel}
\usepackage{enumerate}
\usepackage{subcaption}
\usepackage{appendix}
\usepackage{steinmetz}
\usepackage{esvect}
\usepackage{amssymb}
\usepackage{stfloats}
\newtheorem{definition}{Definition}
\newtheorem{corollary}{Corollary}
\newtheorem{lemma}{Lemma}
\newtheorem{assumption}{Assumption}
\newtheorem{remark}{Remark}
\usepackage{mathptmx}
\usepackage{graphicx}          % Include this line if your 
\usepackage{tikz} 
\usepackage{multirow}
\usetikzlibrary{calc,patterns,arrows,shapes.arrows,intersections}
\usepackage{mathtools}
\DeclarePairedDelimiter\abs{\lvert}{\rvert}%
\DeclarePairedDelimiter\norm{\lVert}{\rVert}%
\makeatletter
\let\oldabs\abs
\def\abs{\@ifstar{\oldabs}{\oldabs*}}
\let\oldnorm\norm
\def\norm{\@ifstar{\oldnorm}{\oldnorm*}}
\makeatother
\DeclareMathAlphabet{\mathcal}{OMS}{cmsy}{m}{n}

\begin{document}

\begin{frontmatter}

\title{Frequency-Domain Stability Method for Reset Control Systems \thanksref{footnoteinfo}} % Title, preferably not more 

\thanks[footnoteinfo]{This paper was not presented at any IFAC 
meeting. Corresponding author S.~H.~HosseinNia. Tel. +31152784248.}

\author[Ali]{Ali Ahmadi Dastjerdi}\ead{A.AhmadiDastjerdi@tudelft.nl},    % Add the 
\author[Astolfi]{Alessandro Astolfi}\ead{a.astolfi@imperial.ac.uk},               % e-mail address 
\author[Ali]{S. Hassan HosseinNia}\ead{S.H.HosseinNiaKani@tudelft.nl}  % (ead) as shown

\address[Ali]{Department of Precision and Microsystem  Engineering, Delft University of Technology, 2826 CD Delft, The Netherlands}  % Please supply                                              
\address[Astolfi]{Department of Electrical and Electronic Engineering, Imperial College London, London, SW7 2AZ, UK}             % full addresses
       % here.

                             % keyword list or with the 
                                          % help of the Automatica 
                                          % keyword wizard

\begin{abstract}                          % Abstract of not more than 200 words.
Reset controllers have the potential to enhance the performance of high-precision industrial motion systems. However, similar to other non-linear controllers, the stability analysis for these controllers is complex and often requires parametric model of the system, which may hinder their applicability. In this paper a frequency-domain approach for assessing stability properties of control systems with first and second order reset elements is developed. The proposed approach is also able to determine uniformly bounded-input bounded-state (UBIBS) property for reset control systems in the case of resetting to non-zero values. An illustrative example to demonstrate the effectiveness of the proposed approach in using frequency response measurements to assess stability properties of reset control systems is presented.
\end{abstract}

\end{frontmatter}

\section{Introduction}\label{sec_4:1}
High-tech precision industrial applications have control requirements which are hard to fulfil by means of linear controllers. One way to increase the performance of these systems is to replace linear controllers with non-linear ones, for instance reset controllers. Owing to their simple structure, these controllers have attracted significant attention from academia and industry~\cite{clegg1958nonlinear,beker2004fundamental,aangenent2010performance,forni2011reset,villaverde2011reset,banos2011reset,van2017frequency,hosseinnia2013fractional,guo2015analysis,muri2}. In particular, reset controllers have been utilized to improve the performance of several mechatronic systems (see, e.g.~\cite{horowitz1975non,hazeleger2016second,guo2009frequency,van2018hybrid,chen2019development,valerio2019reset,saikumar2019constant,muri3,muri4,muris}). 

The first reset element was introduced by Clegg~\cite{clegg1958nonlinear} in 1958. The Clegg Integrator (CI) is an integrator which resets its state to zero whenever its input signal is zero. To provide additional design freedom and flexibility, extensions of the CI including First Order Reset Elements (FORE)~\cite{zaccarian2005first,horowitz1975non}, Generalized First Order Reset Elements (GFORE)~\cite{saikumar2019constant}, Second Order Reset Elements (SORE)~\cite {hazeleger2016second}, Generalized Second Order Reset Elements (GSORE)~\cite{saikumar2019constant}, and Second Order Single State Reset Elements (SOSRE)~\cite{Nima} have been developed. Moreover, to improve the performances of these controllers several methods such as reset bands~\cite{barreiro2014reset,banos2014tuning}, fixed reset instants, partial reset techniques (resetting to a non-zero value or resetting a selection of the controller states)~\cite{zheng2007improved}, use of shaping filters in the reset instants line~\cite{Caipaper}, and the PI+CI approach~\cite{zheng2007improved} have also been investigated. 

Similar to every control system, stability is one of the most essential requirements of reset control systems~\cite{khalil2002nonlinear,beker2004fundamental,van2017frequency,guo2015analysis,banos2011reset,onevsic2008stability,banos2010reset,rifai2006compositional}. Stability properties for reset control systems have been studied using quadratic Lyapunov functions~\cite{banos2011reset,guo2015analysis,polenkova2012stability,vettori2014geometric}, reset instants dependent methods~\cite{banos2010reset,banos2007reset,paesa2011design}, passivity, small gain, and IQC approaches~\cite{khalil2002nonlinear,griggs2007stability,carrasco2010passivity,hollot1997stability}. However, most of these methods are complex, require parametric models of the system and the solution of LMI's, and are only applicable to specific types of systems. Thus, since industry often favors the use of frequency-domain methods, these methods are not well matched with the current control design requirements in industry. To overcome this challenge, some frequency-domain approaches for assessing stability properties of reset control systems have been proposed~\cite{beker1999stability,beker2004fundamental,van2017frequency}. A method for determining stability properties of a FORE in closed-loop with a mass-spring-damper system has been developed in~\cite{beker1999stability}. However, this method is only applicable to a specific type of systems. Under the specific reset condition $e(t)u(t)<\dfrac{u^2(t)}{\varepsilon}$, for some $\varepsilon>0$, in which $e$ and $u$ are the input and the output of the reset element, respectively, the approach in~\cite{van2017frequency} is applicable to reset control systems. However, this method is not applicable to traditional reset control systems in which the reset condition is $e(t)=0$.    

The $H_\beta$ condition is one of the most widely-used methods for assessing stability properties of reset control systems~\cite{beker2004fundamental,banos2010reset,guo2015analysis}. In particular, when the base linear system of the reset element has a first order transfer function, it gives sufficient frequency-domain conditions for uniform bounded-input bounded-state (UBIBS) stability. However, assessing the $H_\beta$ condition in the frequency-domain is not intuitive, especially for high order transfer function plants. In addition, the effect of a shaping filter in the reset line on the $H_\beta$ condition has not been studied yet. Furthermore, there is a lack of methods to assess the $H_\beta$ condition for GSORE using Frequency Response Function (FRF) measurements. Finally, the $H_\beta$ condition is not applicable to assess UBIBS stability of reset control systems in the case of partial reset techniques. Hence, obtaining a general easy-to-use frequency-domain method for assessing UBIBS stability of reset control systems is an important open question. 

In this paper, on the basis of the $H_\beta$ condition, novel frequency-domain stability conditions for control systems with first and second order reset elements with a shaping filter in the reset line are proposed. This approach allows for assessing UBIBS stability of reset control systems in the frequency-domain. In this approach, the $H_\beta$ condition does not have to be explicitly tested and stability properties are directly determined on the basis of the FRF measurements of the base linear open-loop system. In addition, the approach can be used in the case of partial reset techniques. 

The remainder of the paper is organized as follows. In Section~\ref{sec_4:2} preliminaries about reset elements are presented and the problem is formulated. The frequency-domain approaches for assessing stability properties of control systems with first and second order reset elements are presented in Section~\ref{sec_4:3} and Section~\ref{sec_4:4}, respectively. In Section~\ref{sec_4:5} the effectiveness of these approaches is demonstrated via a practical example. Finally, conclusions and suggestions for future studies are given in Section~\ref{sec_4:6}.
%%%%%%%%%%%%%%%%Technical part
%%%%%%%%%%%%%%%%%%%%Prelimenries
\section{Preliminaries}\label{sec_4:2}
In this section the description of reset elements and the $H_\beta$ condition are breifly recalled and some preliminaries are given. The focus of the paper is on the single-input single-output (SISO) control architecture illustrated in Fig.~\ref{F_4-01}. The closed-loop system consists of a linear plant with transfer function $G(s)$ (which we assume strictly proper), linear controllers with proper transfer functions $C_{L_1}(s)$ and $C_{L_2}(s)$, a reset element with base transfer function $C_R(s)$, and a shaping filter with a proper stable transfer function $\mathcal{C}_s(s)$. 

The state-space representation of the reset element is
\begin{figure}[!t]
	\centering
	\resizebox{\columnwidth}{!}{%
\tikzset{every picture/.style={line width=0.75pt}} %set default line width to 0.75pt        
\begin{tikzpicture}[x=0.75pt,y=0.75pt,yscale=-1,xscale=1]
%uncomment if require: \path (0,253); %set diagram left start at 0, and has height of 253
%Shape: Path Data [id:dp5038892938875503] 
\draw  [fill={rgb, 255:red, 172; green, 172; blue, 172 }  ,fill opacity=0.2 ][dash pattern={on 1.69pt off 2.76pt}][line width=1.5]  (68.36,49.72) -- (162.89,49.72) -- (162.89,127.46) -- (364.83,127.46) -- (364.83,49.72) -- (592.79,49.72) .. controls (615.83,49.72) and (634.5,66.58) .. (634.5,87.37) -- (634.5,200.34) .. controls (634.5,221.14) and (615.83,238) .. (592.79,238) -- (68.36,238) .. controls (45.33,238) and (26.65,221.14) .. (26.65,200.34) -- (26.65,87.37) .. controls (26.65,66.58) and (45.33,49.72) .. (68.36,49.72) -- cycle ;
%Shape: Rectangle [id:dp8872257254473093] 
\draw  [line width=1.5]  (384.5,56) -- (463,56) -- (463,122) -- (384.5,122) -- cycle ;
%Shape: Ellipse [id:dp19980167715656805] 
\draw  [line width=1.5]  (45.63,86.45) .. controls (45.63,78.84) and (52.3,72.68) .. (60.53,72.68) .. controls (68.76,72.68) and (75.43,78.84) .. (75.43,86.45) .. controls (75.43,94.06) and (68.76,100.22) .. (60.53,100.22) .. controls (52.3,100.22) and (45.63,94.06) .. (45.63,86.45) -- cycle ;
%Straight Lines [id:da0385434449901394] 
\draw [line width=1.5]    (622.77,91.37) -- (623,214) -- (61,214) -- (62.48,104.22) ;
\draw [shift={(62.53,100.22)}, rotate = 450.77] [fill={rgb, 255:red, 0; green, 0; blue, 0 }  ][line width=0.08]  [draw opacity=0] (11.61,-5.58) -- (0,0) -- (11.61,5.58) -- cycle    ;
%Straight Lines [id:da05885102875442705] 
\draw [line width=1.5]    (2.5,88) -- (41.63,88.41) ;
\draw [shift={(45.63,88.45)}, rotate = 180.6] [fill={rgb, 255:red, 0; green, 0; blue, 0 }  ][line width=0.08]  [draw opacity=0] (11.61,-5.58) -- (0,0) -- (11.61,5.58) -- cycle    ;
%Straight Lines [id:da39309315496373975] 
\draw [line width=1.5]    (598.65,91.05) -- (650.5,91) ;
\draw [shift={(654.5,91)}, rotate = 539.95] [fill={rgb, 255:red, 0; green, 0; blue, 0 }  ][line width=0.08]  [draw opacity=0] (11.61,-5.58) -- (0,0) -- (11.61,5.58) -- cycle    ;
%Straight Lines [id:da6065506865121402] 
\draw [line width=1.5]    (464.49,90.5) -- (489.5,90.93) ;
\draw [shift={(493.5,91)}, rotate = 180.99] [fill={rgb, 255:red, 0; green, 0; blue, 0 }  ][line width=0.08]  [draw opacity=0] (11.61,-5.58) -- (0,0) -- (11.61,5.58) -- cycle    ;
%Shape: Rectangle [id:dp5467349831333912] 
\draw  [line width=1.5]  (550.17,61.09) -- (598.5,61.09) -- (598.5,119) -- (550.17,119) -- cycle ;
%Straight Lines [id:da663322415899998] 
\draw [line width=1.5]    (153,89) -- (238.5,89) ;
\draw [shift={(242.5,89)}, rotate = 180] [fill={rgb, 255:red, 0; green, 0; blue, 0 }  ][line width=0.08]  [draw opacity=0] (11.61,-5.58) -- (0,0) -- (11.61,5.58) -- cycle    ;
%Straight Lines [id:da00759412524357761] 
\draw [line width=1.5]    (324.49,89.5) -- (378,89.97) ;
\draw [shift={(382,90)}, rotate = 180.5] [fill={rgb, 255:red, 0; green, 0; blue, 0 }  ][line width=0.08]  [draw opacity=0] (11.61,-5.58) -- (0,0) -- (11.61,5.58) -- cycle    ;
%Shape: Ellipse [id:dp029726587066210453] 
\draw  [line width=1.5]  (490.63,90.45) .. controls (490.63,82.84) and (497.3,76.68) .. (505.53,76.68) .. controls (513.76,76.68) and (520.43,82.84) .. (520.43,90.45) .. controls (520.43,98.06) and (513.76,104.22) .. (505.53,104.22) .. controls (497.3,104.22) and (490.63,98.06) .. (490.63,90.45) -- cycle ;
%Straight Lines [id:da20690026255567973] 
\draw [line width=1.5]    (506,32) -- (505.57,72.68) ;
\draw [shift={(505.53,76.68)}, rotate = 270.6] [fill={rgb, 255:red, 0; green, 0; blue, 0 }  ][line width=0.08]  [draw opacity=0] (11.61,-5.58) -- (0,0) -- (11.61,5.58) -- cycle    ;
%Shape: Path Data [id:dp5923398308383396] 
\draw  [line width=1.5]  (283,59.09) -- (283,63.09) -- (323,63.09) -- (323,120) -- (282.17,120) -- (282.17,116) -- (242.17,116) -- (242.17,59.09) -- (283,59.09) -- cycle ;
%Straight Lines [id:da023018556967414172] 
\draw [line width=1.5]    (520.43,90.45) -- (545.44,90.88) ;
\draw [shift={(549.44,90.95)}, rotate = 180.99] [fill={rgb, 255:red, 0; green, 0; blue, 0 }  ][line width=0.08]  [draw opacity=0] (11.61,-5.58) -- (0,0) -- (11.61,5.58) -- cycle    ;
%Straight Lines [id:da43982489498145316] 
\draw  [dash pattern={on 4.5pt off 4.5pt}]  (216,160) -- (242,159) -- (242.17,116) ;
%Shape: Rectangle [id:dp4251942074669439] 
\draw  [line width=1.5]  (182.17,146.09) -- (214,146.09) -- (214,176) -- (182.17,176) -- cycle ;
%Shape: Rectangle [id:dp8630536426570494] 
\draw  [line width=1.5]  (108.17,60.09) -- (153,60.09) -- (153,118) -- (108.17,118) -- cycle ;
%Straight Lines [id:da39503099255064567] 
\draw  [dash pattern={on 4.5pt off 4.5pt}]  (168,91) -- (167.5,162) -- (180.5,162) ;
%Straight Lines [id:da0661931984677957] 
\draw [line width=1.5]    (75.43,88.45) -- (106,88.94) ;
\draw [shift={(110,89)}, rotate = 180.91] [fill={rgb, 255:red, 0; green, 0; blue, 0 }  ][line width=0.08]  [draw opacity=0] (11.61,-5.58) -- (0,0) -- (11.61,5.58) -- cycle    ;

% Text Node
\draw (60.53,87.61) node  [font=\large,xscale=1.4,yscale=1.4]  {$-$};
% Text Node
\draw (15.84,72.76) node  [font=\large,xscale=1.4,yscale=1.4]  {$r$};
% Text Node
\draw (642.62,68.76) node  [font=\large,xscale=1.4,yscale=1.4]  {$y$};
% Text Node
\draw (229,146) node  [font=\large,xscale=1.4,yscale=1.4]  {$e_{r}$};
% Text Node
\draw (431,16) node  [font=\large,xscale=1.4,yscale=1.4] [align=left] {{\fontfamily{ptm}\selectfont {\large \textbf{ \ \ \ Liner }}}\\{\fontfamily{ptm}\selectfont {\large \textbf{Controller}}}};
% Text Node
\draw (574,30) node  [font=\large,xscale=1.4,yscale=1.4] [align=left] {{\fontfamily{ptm}\selectfont \textbf{{\large Plant}}}};
% Text Node
\draw (504.53,91.61) node  [font=\large,xscale=1.4,yscale=1.4]  {$+$};
% Text Node
\draw (506.84,15.76) node  [font=\large,xscale=1.4,yscale=1.4]  {$d$};
% Text Node
\draw (285,88) node  [font=\large,xscale=1.4,yscale=1.4]  {$\mathbf{C_{\mathbf{R}}}$};
% Text Node
\draw (286,23) node  [font=\large,xscale=1.4,yscale=1.4] [align=left] {{\fontfamily{ptm}\selectfont {\large \textbf{ \ \ \ Reset }}}\\{\fontfamily{ptm}\selectfont {\large \textbf{Controller}}}};
% Text Node
\draw (423.75,89) node  [font=\large,xscale=1.4,yscale=1.4]  {$\mathbf{C_{\mathbf{L}_{2}}}$};
% Text Node
\draw (344.35,74.76) node  [font=\large,xscale=1.4,yscale=1.4]  {$u_{r}$};
% Text Node
\draw (574.34,90.04) node  [font=\large,xscale=1.4,yscale=1.4]  {$\mathbf{G}$};
% Text Node
\draw (42,208) node  [font=\large,xscale=1.4,yscale=1.4]  {$\mathcal{L}$};
% Text Node
\draw (198.09,161.04) node  [font=\large,xscale=1.4,yscale=1.4]  {$\mathcal{C}_s$};
% Text Node
\draw (130.59,89.04) node  [font=\large,xscale=1.4,yscale=1.4]  {$\mathbf{C_{\mathbf{L}_{1}}}$};
% Text Node
\draw (88.84,72.76) node  [font=\large,xscale=1.4,yscale=1.4]  {$e$};
% Text Node
\draw (197.84,70.76) node  [font=\large,xscale=1.4,yscale=1.4]  {$u_{1}$};
% Text Node
\draw (131,16) node  [font=\large,xscale=1.4,yscale=1.4] [align=left] {{\fontfamily{ptm}\selectfont {\large \textbf{ \ \ \ Liner }}}\\{\fontfamily{ptm}\selectfont {\large \textbf{Controller}}}};
\end{tikzpicture}}
	\caption{The closed-loop architecture of a reset control system}
	\label{F_4-01}
\end{figure}
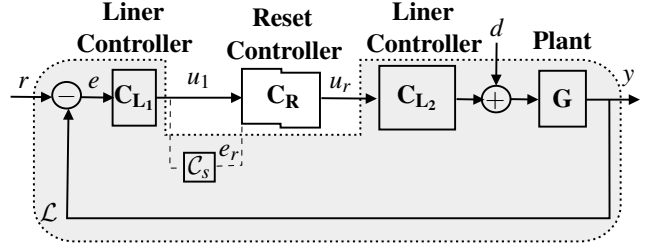
\begin{equation}\label{E_4-21}
\resizebox{\columnwidth}{!}{$
\left\{
\begin{aligned}
\dot{x}_r(t)&=A_rx_r(t)+B_ru_1(t), &e_r(t)&\neq0,  \\
x_r(t^+)&=A_\rho x_r(t), &e_r(t)&=0\land (I-A_\rho)x_r(t)\neq0, \\
u_r(t)&=C_rx_r(t)+D_ru_1(t),
\end{aligned}
\right.
$}
\end{equation}
in which $x_r(t)\in\mathbb{R}^{n_r}$ is the vector containing the reset state, $A_r$, $B_r$, $C_r$, and $D_r$ are the dynamic matrices of the reset element, $A_\rho$ is the reset matrix, which determines the values of the reset state after the reset action, and $u_1(t)\in\mathbb{R}$ and $u_r(t)\in\mathbb{R}$ are the input and output of the reset element, respectively. The transfer function $C_r(sI-A_r)^{-1}B_r+D_r$ is called the base transfer function of the reset element. The base transfer function in case of GFORE is (in all cases $\omega_r>0$)
\begin{equation}\label{E_4-22}
C_R(s)=\dfrac{1}{\displaystyle\frac{s}{\omega_r}+1},
\end{equation}	
for CI and Proportional Clegg Integrator (PCI) one has
\setlength{\arraycolsep}{0.0em}
\begin{align}
C_R(s)&=\dfrac{1}{s},\label{E_4-2-23}\\
C_R(s)&=1+\dfrac{\omega_r}{s},\label{E_4-23}
\end{align}
\setlength{\arraycolsep}{5pt}and for GSORE one has
\begin{equation}\label{E_4-24}
C_R(s)=\dfrac{1}{s^2+2\xi\omega_rs+\omega_r^2},\ \xi>0.
\end{equation}
Thus, for GFORE, $A_r=-C_r=-\omega_r$ ($\omega_r$ is the so-called corner frequency), $D_r=0$, and $B_r=1$, whereas for the PCI, $A_r=0$, $C_r=\omega_r$, and $B_r=D_r=1$. In the case of CI, $A_r=D_r=0$, $B_r=C_r=1$, and if we consider the controllable canonical form realization for GSORE, we obtain
\begin{equation}\label{E_4-25}
A_r=\begin{bmatrix}-2\xi\omega_r & -\omega_r^2\\ 1 & 0
\end{bmatrix},\ B_r=\begin{bmatrix} 1 \\ 0
\end{bmatrix},\ C_r=\begin{bmatrix} 0 & 1
\end{bmatrix},\text{ and } D_r=0.
\end{equation} 
\newline Let $\mathcal{L}$ be the linear time-invariant (LTI) part of the system, see Fig.~\ref{F_4-01}, with input $u_r(t)\in\mathbb{R}$, external disturbance $w(t)=\begin{bmatrix} r(t)& d(t)\end{bmatrix}^T\in\mathbb{R}^2$, and outputs $y(t)\in\mathbb{R}$, $e_r(t)\in\mathbb{R}$, and $u_1(t)\in\mathbb{R}$. The state-space realization of $\mathcal{L}$ is given by equations
\begin{equation}\label{E_4-26}
\mathcal{L}:\left\{
\begin{aligned}
\dot{\zeta}(t)&=A\zeta(t)+B_uu_r(t)+Bw(t),\\
y(t)&=C\zeta(t),\\
e_r(t)&=C_e\zeta(t)+D_er(t),\\
u_1(t)&=C_u\zeta(t)+D_1r(t),
\end{aligned}
\right.
\end{equation}
where $\zeta(t)\in\mathbb{R}^{n_p}$ describes the states of the plant and of the linear controllers ($n_p$ is the number of states of the whole linear part), and $A$, $B$, $B_u$, and $C$ are the corresponding dynamic matrices. The closed-loop state-space representation of the overall system can, therefore, be written as
\begin{equation}\label{E_4-27}
\left\{
\begin{aligned}
\dot{x}(t)&=\bar{A}x(t)+\bar{B}w(t), & e_r(t)&\neq 0,\\
x(t^+)&=\bar{A}_\rho x(t), & e_r(t)&=0\land (I-\bar{A}_\rho)x(t)\neq0,  \\
y(t)&=\bar{C}x(t),\\
e_r(t)&=\bar{C}_ex(t)+D_er(t),
\end{aligned}
\right.
\end{equation}
where $x(t)=\begin{bmatrix} x_r(t)^T& \zeta(t)^T\end{bmatrix}^T\in\mathbb{R}^{n_r+n_p}$, 
$\bar{C}=\begin{bmatrix} 0_{1\times n_r} & C \end{bmatrix}$,  
$\bar{B}=\begin{bmatrix}0_{n_r\times 2}\\B\end{bmatrix}+\begin{bmatrix} B_rD_{1} & 0_{n_r\times 1} \\ B_uD_rD_{1} & 0_{n_p\times 1} \end{bmatrix}$, 
$\bar{C}_{e}=\begin{bmatrix} 0_{1\times n_r} & C_{e} \end{bmatrix}$, 
$\bar{A}=\begin{bmatrix} A_r & B_rC_u \\ B_uC_r & A+B_uD_rC_u\end{bmatrix}$, 
and $\bar{A}_\rho=\begin{bmatrix}A_\rho & 0_{n_r\times n_p} \\ 0_{n_p\times n_r} & I_{n_p\times n_p} \end{bmatrix}$. %$\bar{C}_{u}=\begin{bmatrix} 0_{1\times n_r} & C_{u} \end{bmatrix}$
\begin{definition}\label{D_4D0}
A time $\bar{T}>0$ is called a reset instant for the reset control system~(\ref{E_4-27}) if $e_R(\bar{T})=0\land (I-\bar{A}_\rho)x(T)\neq0$. For any given initial condition and input $w$ the resulting set of all reset instants defines the reset sequence $\{t_k\}$, with $t_k\leq t_{k+1}$, for all $k\in\mathbb{N}$. The reset instants $t_k$ have the well-posedness property if for any initial condition $x_0$ and any input $w$, all the reset instants are distinct, and there exists $\lambda>0$ such that, for all $k\in\mathbb{N}$, $\lambda\leq t_{k+1}-t_k$ \cite{banos2016impulsive,banos2011reset}.
\end{definition}
One of the methods for determining stability properties of reset control systems is the $H_{\beta}$ condition~\cite{beker2004fundamental,banos2011reset,banos2010reset,guo2015analysis,hollot2001establishing}, which is briefly recalled. Let  
\begin{equation}\label{E_4x-28}
C_0=[\Varrho\quad\beta C],\quad B_0=\begin{bmatrix} I_{n_r\times n_r} \\ 0_{n_p\times n_r} \end{bmatrix},\quad \Varrho=\Varrho^T>0,\quad\Varrho\in\mathbb{R}^{n_r\times n_r},
\end{equation}
and $\beta\in\mathbb{R}^{n_r\times 1}.$ The $H_\beta$ condition~\cite{beker2004fundamental,banos2011reset,banos2010reset,guo2015analysis,hollot2001establishing} states that the zero equilibrium of the reset control system~(\ref{E_4-27}) with $C_{L_1}=\mathcal{C}_s=1$ and $w=0$ is globally uniformly asymptotically stable if and only if there exist $\Varrho=\Varrho^T>0$ and $\beta$ such that the transfer function 
\begin{equation}\label{E_4-29}
H(s)=C_0(sI-\bar{A})^{-1}B_0
\end{equation} 
is Strictly Positive Real (SPR), $(\bar{A},B_0)$ and $(\bar{A},C_0)$ are controllable and observable, respectively, and
\begin{equation}\label{E_4-30}
A_\rho^T\Varrho A_\rho-\Varrho\ \leq0.
\end{equation}
Evaluating the $H_\beta$ condition requires finding the parameters $\Varrho$ and $\beta$, which may be very difficult when the system has a high order transfer function. Furthermore, in the case of GSORE there is no direct frequency-domain method to assess this condition. Besides, the UBIBS property of GSORE and of GFORE have not yet been studied, and the effects of the shaping filter on the $H_\beta$ condition have not been considered yet. In the current paper, frequency-domain methods to determine stability properties without finding $\Varrho$ and $\beta$ for GFORE and of GSORE with considering the shaping filter are proposed.
\begin{assumption}\label{AS_41}
There are infinitely many reset instants and $\displaystyle\lim_{k\to\infty} t_k=\infty$. 
\end{assumption}
Assumption~\ref{AS_41} is introduced to rule out a trivial situation. In fact, if there are finitely many reset instants, then there exists a $T_K\in[0,\infty)$ such that for all $t\geq T_K$ the reset control system~(\ref{E_4-27}) is a linear stable system provided the $H_\beta$ condition is satisfied. In addition to Assumption~\ref{AS_41}, we need the following assumption, which is instrumental to study the UBIBS property of reset control systems.
\begin{assumption}\label{AS_42}
In the case of partial reset technique, if $A_\rho$ has the structure
$$
A_\rho=\begin{bmatrix}
I_{\tilde{n}_r} & 0\\
0& A^\prime_{n^\prime_r}
\end{bmatrix},
$$
then $A_r$ has the structure
$$
A_r=\begin{bmatrix}
A_{r_1} & A_{r_2}\\
0_{\tilde{n}_r\times n^\prime_r}& A_{r_3}
\end{bmatrix}.
$$
\end{assumption}
\begin{remark}\label{R_4s00_2}
{\rm In the case of GFORE, GSORE, PCI, and CI in which all states of the reset element reset, Assumption~\ref{AS_42} holds.}
\end{remark}
Before stating the main theorem, an important technical lemma, which is instrumental for all proofs, is formulated and proved.
\begin{lemma}\label{L_41}
Consider the reset control system~(\ref{E_4-27}). Suppose that
\begin{itemize}
\item Assumption~\ref{AS_41} holds;
	\item $A_\rho^T\Varrho A_\rho-\Varrho<0$;
	\item the $H_\beta$ condition holds;
	\item at least one of the following conditions holds:
		\begin{enumerate}
		\item $\mathcal{C}_s=1$ and Assumption~\ref{AS_42} holds;
		\item the reset instants have the well-posedness property.  
		\end{enumerate}
\end{itemize}
Then the reset control system~(\ref{E_4-27}) has a well-defined unique left-continuous solution for any initial condition $x_0$ and any input $w$ which is a Bohl function\footnote{See definition Bohl function in~\cite{banos2016impulsive}}. In addition, this solution is UBIBS and the reset instants have the well-posedness property.
\end{lemma} 
\begin{pf}
	See Appendix~\ref{aap_41}.	
\end{pf} 
%%%%%%%%%%%%%%%%%%%%%%%%%%%%%%%%%%%%%GFORE
\section{Stability analysis of reset control systems with first order reset elements}\label{sec_4:3}
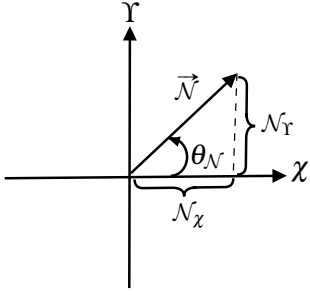
\begin{figure}[!t]
 	\centering
	\resizebox{0.5\columnwidth}{!}{%
 	\tikzset{every picture/.style={line width=0.75pt}} %set default line width to 0.75pt        
 	\begin{tikzpicture}[x=0.75pt,y=0.75pt,yscale=-1,xscale=1]
 	%uncomment if require: \path (0,246); %set diagram left start at 0, and has height of 246
 	%Straight Lines [id:da6055645258040092] 
 	\draw [line width=1.5]    (9,150) -- (231,148.04) ;
 	\draw [shift={(235,148)}, rotate = 539.49] [fill={rgb, 255:red, 0; green, 0; blue, 0 }  ][line width=0.08]  [draw opacity=0] (11.61,-5.58) -- (0,0) -- (11.61,5.58) -- cycle    ;
 	%Straight Lines [id:da49677046907762845] 
 	\draw [line width=1.5]    (108.75,238.01) -- (109.24,32.01) ;
 	\draw [shift={(109.25,28.01)}, rotate = 450.14] [fill={rgb, 255:red, 0; green, 0; blue, 0 }  ][line width=0.08]  [draw opacity=0] (11.61,-5.58) -- (0,0) -- (11.61,5.58) -- cycle    ;
 	%Straight Lines [id:da3706867081231975] 
 	\draw [line width=1.5]    (110,146) -- (192.09,68.74) ;
 	\draw [shift={(195,66)}, rotate = 496.74] [fill={rgb, 255:red, 0; green, 0; blue, 0 }  ][line width=0.08]  [draw opacity=0] (11.61,-5.58) -- (0,0) -- (11.61,5.58) -- cycle    ;
 	%Shape: Arc [id:dp3484281447845041] 
 	\draw  [draw opacity=0][line width=1.5]  (140.24,118.13) .. controls (140.81,118.04) and (141.4,117.99) .. (141.99,117.99) .. controls (149.16,117.98) and (154.99,124.69) .. (155,132.97) .. controls (155.02,141.26) and (149.22,147.99) .. (142.04,148.01) .. controls (142.03,148.01) and (142.02,148.01) .. (142,148.01) -- (142.02,133) -- cycle ; \draw  [line width=1.5]  (140.24,118.13) .. controls (140.81,118.04) and (141.4,117.99) .. (141.99,117.99) .. controls (149.16,117.98) and (154.99,124.69) .. (155,132.97) .. controls (155.02,141.26) and (149.22,147.99) .. (142.04,148.01) .. controls (142.03,148.01) and (142.02,148.01) .. (142,148.01) ;
 	%Straight Lines [id:da4482726275271183] 
 	\draw    (150,121) -- (143.12,118.98) ;
 	\draw [shift={(140.24,118.13)}, rotate = 376.40999999999997] [fill={rgb, 255:red, 0; green, 0; blue, 0 }  ][line width=0.08]  [draw opacity=0] (8.93,-4.29) -- (0,0) -- (8.93,4.29) -- cycle    ;
 	%Straight Lines [id:da2830703782839088] 
 	\draw  [dash pattern={on 4.5pt off 4.5pt}]  (195,66) -- (192,147) ;
 	%Shape: Brace [id:dp06778232658161376] 
 	\draw  [line width=1.5]  (195,147) .. controls (199.67,147.06) and (202.03,144.76) .. (202.09,140.09) -- (202.37,118.09) .. controls (202.46,111.42) and (204.83,108.12) .. (209.5,108.18) .. controls (204.83,108.12) and (202.54,104.76) .. (202.62,98.09)(202.59,101.09) -- (202.9,76.09) .. controls (202.96,71.42) and (200.66,69.06) .. (195.99,69) ;
 	%Shape: Brace [id:dp36142542073028316] 
 	\draw  [line width=1.5]  (113,151) .. controls (113.06,155.67) and (115.42,157.97) .. (120.09,157.91) -- (142.59,157.62) .. controls (149.26,157.54) and (152.62,159.83) .. (152.68,164.5) .. controls (152.62,159.83) and (155.92,157.46) .. (162.59,157.37)(159.59,157.41) -- (185.09,157.08) .. controls (189.76,157.02) and (192.06,154.66) .. (192,149.99) ;
%Shape: Circle [id:dp8459499845851355] 
%\draw  [line width=1.2]  (215,26.5) .. controls (215,20.15) and (220.15,15) .. (226.5,15) .. controls (232.85,15) and (238,20.15) .. (238,26.5) .. controls (238,32.85) and (232.85,38) .. (226.5,38) .. controls (220.15,38) and (215,32.85) .. (215,26.5) -- cycle ;
%Shape: Circle [id:dp9488456942872145] 
%\draw  [line width=1.2]  (29,27.5) .. controls (29,21.15) and (34.15,16) .. (40.5,16) .. controls (46.85,16) and (52,21.15) .. (52,27.5) .. controls (52,33.85) and (46.85,39) .. (40.5,39) .. controls (34.15,39) and (29,33.85) .. (29,27.5) -- cycle ;
%Shape: Circle [id:dp05587054045189843] 
%\draw  [line width=1.2]  (25,223.5) .. controls (25,217.15) and (30.15,212) .. (36.5,212) .. controls (42.85,212) and (48,217.15) .. (48,223.5) .. controls (48,229.85) and (42.85,235) .. (36.5,235) .. controls (30.15,235) and (25,229.85) .. (25,223.5) -- cycle ;
%Shape: Circle [id:dp7774300727276198] 
%\draw  [line width=1.2]  (213,222.5) .. controls (213,216.15) and (218.15,211) .. (224.5,211) .. controls (230.85,211) and (236,216.15) .. (236,222.5) .. controls (236,228.85) and (230.85,234) .. (224.5,234) .. controls (218.15,234) and (213,228.85) .. (213,222.5) -- cycle ;
 
 	% Text Node
 	\draw (246,145) node  [font=\Large]  {$\chi $};
 	% Text Node
 	\draw (110,17) node  [font=\Large]  {$\Upsilon $};
 	% Text Node
 	\draw (171,132) node   [font=\Large] {$\theta _{\mathcal{N}}$};
 	% Text Node
 	\draw (155,77) node  [font=\large]  {$\vv{\mathcal{N}}$};
 	% Text Node
 	\draw (228,107) node  [font=\large]  {$\mathcal{N}_{\Upsilon }$};
 	% Text Node
 	\draw (157,180) node  [font=\large]  {$\mathcal{N}_{\chi }$};
% Text Node
%\draw (229,26.5) node  [font=\large]  {$\mathcal{I}_{1}$};
% Text Node
%\draw (43,27.5) node  [font=\large]  {$\mathcal{I}_{2}$};
% Text Node
%\draw (39,222) node  [font=\large]  {$\mathcal{I}_{3}$};
% Text Node
%\draw (227,222) node  [font=\large]  {$\mathcal{I}_{4}$};
 	\end{tikzpicture}}
 	\caption{Representation of the NSV in the $\chi-\Upsilon$ plane}
 	\label{F_4-02}
 \end{figure}

In this section frequency-domain methods for assessing stability properties of the reset control system~(\ref{E_4-27}) with GFORE~(\ref{E_4-22}), CI~(\ref{E_4-2-23}), and PCI~(\ref{E_4-23}) are proposed on the basis of the $H_\beta$ condition. To this end, the Nyquist Stability Vector (NSV=$\vv{\mathcal{N}}(\omega)\in\mathbb{R}^2$) in a plane with axis $\chi-\Upsilon$ (see Fig.~\ref{F_4-02}) is defined as follows. 
 \begin{definition}\label{D_41}
 	The Nyquist Stability Vector is, for all $\omega\in\mathbb{R}^+$, the vector 
	\setlength{\arraycolsep}{0.0em}
\begin{eqnarray}
\vv{\mathcal{N}}(\omega)&{=}&[\mathcal{N}_\chi \quad \mathcal{N}_\Upsilon]^T\nonumber\\
&{=}&[\Re(L(j\omega)\mathcal{C}_s(j\omega)\kappa(j\omega)) \quad \Re(\kappa(j\omega)C_R(j\omega))]^T,\nonumber
\end{eqnarray}
\setlength{\arraycolsep}{5pt}in which $L(s)=C_{L_1}(s)C_R(s)C_{L_2}(s)G(s)$, $L(j\omega)=a(\omega)+b(\omega)j$, and $\kappa(j\omega)=1+L^*(j\omega)$ ($L^*(j\omega)$ is the conjugate of $L(j\omega)$).
  \end{definition}  
 %%%%%%%%%%%%%%%%%%%%%%%%%%%%%%%%Figure
For simplicity, and without loss of generality, let $\phase{\vv{\mathcal{N}}(\omega)}=\theta_{\mathcal{N}}\in[-\frac{\pi}{2},\ \frac{3\pi}{2})$ and define the open sets
$$\mathcal{I}_1=\left\{\omega\in\mathbb{R}^+|\ 0<\phase{\vv{\mathcal{N}}(\omega)}<\frac{\pi}{2}\right\},$$ 
$$\mathcal{I}_2=\left\{\omega\in\mathbb{R}^+|\ \dfrac{\pi}{2}<\phase{\vv{\mathcal{N}}(\omega)}<\pi\right\},$$
$$\mathcal{I}_3=\left\{\omega\in\mathbb{R}^+|\ \pi<\phase{\vv{\mathcal{N}}(\omega)}<\dfrac{3\pi}{2}\right\},$$
$$\mathcal{I}_4=\left\{\omega\in\mathbb{R}^+|\ -\dfrac{\pi}{2}<\phase{\vv{\mathcal{N}}(\omega)}<0\right\}.$$
Let $L(s)\mathcal{C}_s(s)=\dfrac{K_ms^m+K_{m-1}s^{m-1}+...+K_0}{s^n+K^\prime_{n-1}s^{n-1}+...+K^\prime_0}$ and $\mathcal{C}_s(s)=\dfrac{K_{s_m}s^{m_s}+K_{s_{m-1}}s^{m_s-1}+...+K_{s_0}}{K^\prime_{s_n}s^{n_s}+K^\prime_{s_{n-1}}s^{n_s-1}+...+1}$. On the basis of the definition of the NSV, systems of Type I and of Type II, which are used to assess stability properties of the reset control system~(\ref{E_4-27}), are defined.  
\begin{definition}\label{D_42}
The reset control system~(\ref{E_4-27}) is of Type I if the following conditions hold. 
\begin{enumerate}[(1)]
\item If $C_{L_1}(s)C_{L_2}(s)G(s)$ has at least one pole at the origin, then $K_{s_0}>0$.
\item In the case of CI~(\ref{E_4-2-23}), $K_{s_0}<0$.
\item For all $\omega\in\mathcal{M}=\{\omega\in\mathbb{R}^+|\ \mathcal{N}_\chi(\omega)=0\}$ one has $\mathcal{N}_\Upsilon(\omega)>0$.
\item For all $\omega\in\mathcal{Q}=\{\omega\in\mathbb{R}^+|\ \mathcal{N}_\Upsilon(\omega)=0\}$ one has $\mathcal{N}_\chi(\omega)>0$.
\item At least one of the following statements is true:
\begin{enumerate}
	\item $\forall\ \omega\in\mathbb{R}^+:\ \mathcal{N}_\Upsilon(\omega)\geq0.$
	\item $\forall\ \omega\in\mathbb{R}^+:\ \mathcal{N}_\chi(\omega)\geq0.$
	\item Let $\delta_1=\underset{\omega\in\mathcal{I}_4}{\max}\abs{\dfrac{\mathcal{N}_\Upsilon(\omega)}{\mathcal{N}_\chi(\omega)}}$ and $\Psi_1=\underset{\omega\in\mathcal{I}_2}{\min}\abs{\dfrac{\mathcal{N}_\Upsilon(\omega)}{\mathcal{N}_\chi(\omega)}}$. Then $\delta_1<\Psi_1$ and $\mathcal{I}_3=\varnothing$.
\end{enumerate}
	\end{enumerate}	
 \end{definition}
\begin{remark}\label{R_4s1}
{\rm Let
\begin{equation}\label{E_4-333}
\theta_{1}=\underset{\omega\in\mathbb{R}^+}{\min}\phase{\vv{\mathcal{N}}(\omega)}\text{ and }\theta_{2}=\underset{\omega\in\mathbb{R}^+}{\max}\phase{\vv{\mathcal{N}}(\omega)}.
\end{equation}
Then the conditions identifying Type I systems are equivalent to the following conditions. 
		\begin{enumerate}[(1)]
		\item If $C_{L_1}(s)C_{L_2}(s)G(s)$ has at least one at the origin, then $K_{s_0}>0$.  
		\item In the case of CI~(\ref{E_4-2-23}), $K_{s_0}<0$.
\item The condition \begin{equation}\label{E_4-3333}
\left(-\dfrac{\pi}{2}<\theta_{1}<\pi\right)\ \land\ \left(-\dfrac{\pi}{2}<\theta_{2}<\pi\right)\ \land\ (\theta_{2}-\theta_{1}<\pi)
\end{equation}
holds.	
\end{enumerate}}
\end{remark}
\begin{definition}\label{D_43}
The reset control system~(\ref{E_4-27}) is of Type II if the following conditions hold. 
	\begin{enumerate}[(1)]
		\item If $C_{L_1}(s)C_{L_2}(s)G(s)$ has at least one at the origin, then $K_{s_0}<0$.
		\item In the case of CI~(\ref{E_4-2-23}), $K_{s_0}>0$.
		\item For all $\omega\in\mathcal{M}$ one has $\mathcal{N}_\Upsilon(\omega)>0$.
		\item For all $\omega\in\mathcal{Q}$ one has $\mathcal{N}_\chi(\omega)<0$. 
		\item At least, one of the following statements is true:
		\begin{enumerate}
			\item $\forall\ \omega\in\mathbb{R}^+:\ \mathcal{N}_\Upsilon(\omega)\geq0$;
			\item $\forall\ \omega\in\mathbb{R}^+:\ \mathcal{N}_\chi(\omega)\leq0$;
			\item Let $\delta_2=\underset{\omega\in\mathcal{I}_3}{\max}\abs{\dfrac{\mathcal{N}_\Upsilon(\omega)}{\mathcal{N}_\chi(\omega)}}$ and $\Psi_2=\underset{\omega\in\mathcal{I}_1}{\min}\abs{\dfrac{\mathcal{N}_\Upsilon(\omega)}{\mathcal{N}_\chi(\omega)}}$. Then, $\delta_2<\Psi_2$ and $\mathcal{I}_4=\varnothing$.  
		\end{enumerate}
	\end{enumerate}	
\end{definition}		
\begin{remark}\label{R_4s2}
	{\rm The conditions identifying Type II systems are equivalent to the following conditions. 
		\begin{enumerate}[(1)]
		\item If $C_{L_1}(s)C_{L_2}(s)G(s)$ has at least one at the origin, then $K_{s_0}<0$.
		\item In the case of CI~(\ref{E_4-2-23}), $K_{s_0}>0$.
	    \item The condition \begin{equation}\label{E_4-555}
	\left(0<\theta_{1}<\dfrac{3\pi}{2}\right)\ \land\ \left(0<\theta_{2}<\dfrac{3\pi}{2}\right)\ \land\ (\theta_{2}-\theta_{1}<\pi)
	\end{equation} holds.
	\end{enumerate}}	
\end{remark}
 \begin{thm}\label{T_41}
 	The zero equilibrium of the reset control system (\ref{E_4-27}) with GFORE~(\ref{E_4-22}), or CI~(\ref{E_4-2-23}), or PCI~(\ref{E_4-23}) is globally uniformly asymptotically stable when $w=0$, and the system has the UBIBS property for any input $w$ which is a Bohl function if all of the following conditions are satisfied.	
 	\begin{itemize}
 		\item The base linear system is stable and the open-loop transfer function does not have any pole-zero cancellation.
		\item In the case of CI~(\ref{E_4-2-23}), $C_{L_1}(s)C_{L_2}(s)G(s)$ does not have any pole at the origin and $n-m=2$. 
 		\item The reset control system (\ref{E_4-27}) is either of Type I and/or of Type II. 
		\item $A_\rho=\gamma,\ -1<\gamma<1.$
		\item $\mathcal{C}_s(s)=1$ and/or the reset instants have the well-posedness property.
 	\end{itemize}
 \end{thm}
\begin{pf}
For $w(t)=0$, for all $t\geq0$, reset happens when $x(t)\in \text{ker}(\bar{C}_e)$. Looking at the proof of the $H_\beta$ condition, which is given in~\cite{beker2004fundamental,guo2015analysis,banos2011reset}, when there is a shaping filter in the reset line, $C_0$ in the $H_\beta$ condition is changed to
\begin{equation}\label{E_4-F1}
C_0=[\Varrho\quad\beta \bar{C}_e].
\end{equation} 
 Theorem \ref{T_41} is now proved in several steps.
 	\begin{itemize}
 		\item Step 1: It is shown that there is a $\beta$ and $\Varrho >0$ such that $\Re(H(j\omega))>0$, for all $\omega\in\mathbb{R}^+$. 
 		\item Step 2: For systems with poles at the origin it is shown that $\displaystyle\lim_{\omega\to 0} \Re(H(j\omega))>0$.
 		\item Step 3: It is shown that either $\displaystyle\lim_{s\to \infty} H(s)>0$ or $\displaystyle\lim_{\omega\to \infty} \omega^2\Re(H(j\omega))>0.$
 		\item Step 4: It is shown that $(A,C_0)$ and $(A,B_0)$ are observable and controllable, respectively.
		\end{itemize}  
Step 1: For simplicity take $\beta^\prime=-\beta$ and $\Varrho^\prime=\dfrac{\Varrho}{C_r}$. The transfer function (\ref{E_4-29}) with the modified $C_0$ as in~(\ref{E_4-F1}) can be rewritten as (see also Fig.~\ref{F_4-03})
\begin{equation}\label{E_4E-32}
H(s)=\dfrac{y}{r}=\dfrac{\beta^\prime L(s)\mathcal{C}_s(s)+\Varrho^\prime C_R(s)}{1+L(s)}.
\end{equation}
 %%%%%%%%%%%%%%%%%%%%%%%%%%%%%%%%Figure
 \begin{figure}[!t]
 	\centering
	\resizebox{\columnwidth}{!}{%
 	\tikzset{every picture/.style={line width=0.75pt}} %set default line width to 0.75pt        
\begin{tikzpicture}[x=0.75pt,y=0.75pt,yscale=-1,xscale=1]
%uncomment if require: \path (0,208); %set diagram left start at 0, and has height of 208
%Shape: Rectangle [id:dp8872257254473093] 
\draw  [line width=1.5]  (229.5,64) -- (289,64) -- (289,117) -- (229.5,117) -- cycle ;
%Shape: Ellipse [id:dp19980167715656805] 
\draw  [line width=1.5]  (38.63,89.45) .. controls (38.63,81.84) and (45.3,75.68) .. (53.53,75.68) .. controls (61.76,75.68) and (68.43,81.84) .. (68.43,89.45) .. controls (68.43,97.06) and (61.76,103.22) .. (53.53,103.22) .. controls (45.3,103.22) and (38.63,97.06) .. (38.63,89.45) -- cycle ;
%Straight Lines [id:da0385434449901394] 
\draw [line width=1.5]    (479,98) -- (480.5,198) -- (54.5,198) -- (53.57,107.22) ;
\draw [shift={(53.53,103.22)}, rotate = 449.41] [fill={rgb, 255:red, 0; green, 0; blue, 0 }  ][line width=0.08]  [draw opacity=0] (11.61,-5.58) -- (0,0) -- (11.61,5.58) -- cycle    ;
%Straight Lines [id:da6065506865121402] 
\draw [line width=1.5]    (289.49,92.5) -- (314.5,92.93) ;
\draw [shift={(318.5,93)}, rotate = 180.99] [fill={rgb, 255:red, 0; green, 0; blue, 0 }  ][line width=0.08]  [draw opacity=0] (11.61,-5.58) -- (0,0) -- (11.61,5.58) -- cycle    ;
%Straight Lines [id:da00759412524357761] 
\draw [line width=1.5]    (186.49,92.5) -- (224,92.95) ;
\draw [shift={(228,93)}, rotate = 180.69] [fill={rgb, 255:red, 0; green, 0; blue, 0 }  ][line width=0.08]  [draw opacity=0] (11.61,-5.58) -- (0,0) -- (11.61,5.58) -- cycle    ;
%Shape: Path Data [id:dp5923398308383396] 
\draw  [line width=1.5]  (145,62.09) -- (145,66.09) -- (185,66.09) -- (185,123) -- (144.17,123) -- (144.17,119) -- (104.17,119) -- (104.17,62.09) -- (145,62.09) -- cycle ;
%Shape: Rectangle [id:dp4251942074669439] 
\draw  [line width=1.5]  (500.17,81.09) -- (532,81.09) -- (532,111) -- (500.17,111) -- cycle ;
%Straight Lines [id:da0661931984677957] 
\draw [line width=1.5]    (68.43,91.45) -- (99,91.94) ;
\draw [shift={(103,92)}, rotate = 180.91] [fill={rgb, 255:red, 0; green, 0; blue, 0 }  ][line width=0.08]  [draw opacity=0] (11.61,-5.58) -- (0,0) -- (11.61,5.58) -- cycle    ;
%Straight Lines [id:da34467405320406397] 
\draw [line width=1.5]    (376.49,94.5) -- (403,94.93) ;
\draw [shift={(407,95)}, rotate = 180.94] [fill={rgb, 255:red, 0; green, 0; blue, 0 }  ][line width=0.08]  [draw opacity=0] (11.61,-5.58) -- (0,0) -- (11.61,5.58) -- cycle    ;
%Straight Lines [id:da022191737056530547] 
\draw [line width=1.5]    (464.49,97.5) -- (497,97.05) ;
\draw [shift={(501,97)}, rotate = 539.22] [fill={rgb, 255:red, 0; green, 0; blue, 0 }  ][line width=0.08]  [draw opacity=0] (11.61,-5.58) -- (0,0) -- (11.61,5.58) -- cycle    ;
%Straight Lines [id:da822440506762085] 
\draw [line width=1.5]    (533.49,96.5) -- (560,96.07) ;
\draw [shift={(564,96)}, rotate = 539.06] [fill={rgb, 255:red, 0; green, 0; blue, 0 }  ][line width=0.08]  [draw opacity=0] (11.61,-5.58) -- (0,0) -- (11.61,5.58) -- cycle    ;
%Straight Lines [id:da2013061122362101] 
\draw [line width=1.5]    (1,90) -- (36,90) ;
\draw [shift={(40,90)}, rotate = 180] [fill={rgb, 255:red, 0; green, 0; blue, 0 }  ][line width=0.08]  [draw opacity=0] (11.61,-5.58) -- (0,0) -- (11.61,5.58) -- cycle    ;
%Shape: Rectangle [id:dp5064137799607067] 
\draw  [line width=1.5]  (316.5,66) -- (376,66) -- (376,119) -- (316.5,119) -- cycle ;
%Shape: Rectangle [id:dp5099807740867237] 
\draw  [line width=1.5]  (404.5,69) -- (464,69) -- (464,122) -- (404.5,122) -- cycle ;
%Shape: Triangle [id:dp6872168669491976] 
\draw  [line width=1.5]  (595.6,96.75) -- (563.4,121.87) -- (563.6,70.93) -- cycle ;
%Shape: Triangle [id:dp06219002161491982] 
\draw  [line width=1.5]  (428.6,30.75) -- (396.4,55.87) -- (396.6,4.93) -- cycle ;
%Straight Lines [id:da3832375292532042] 
\draw [line width=1.5]    (199,91) -- (200,29) -- (390,29.98) ;
\draw [shift={(394,30)}, rotate = 180.3] [fill={rgb, 255:red, 0; green, 0; blue, 0 }  ][line width=0.08]  [draw opacity=0] (11.61,-5.58) -- (0,0) -- (11.61,5.58) -- cycle    ;
%Straight Lines [id:da7623898981871877] 
\draw [line width=1.5]    (428.6,30.75) -- (631,32) -- (631.49,77.68) ;
\draw [shift={(631.53,81.68)}, rotate = 269.39] [fill={rgb, 255:red, 0; green, 0; blue, 0 }  ][line width=0.08]  [draw opacity=0] (11.61,-5.58) -- (0,0) -- (11.61,5.58) -- cycle    ;
%Shape: Ellipse [id:dp9645205509132346] 
\draw  [line width=1.5]  (616.63,95.45) .. controls (616.63,87.84) and (623.3,81.68) .. (631.53,81.68) .. controls (639.76,81.68) and (646.43,87.84) .. (646.43,95.45) .. controls (646.43,103.06) and (639.76,109.22) .. (631.53,109.22) .. controls (623.3,109.22) and (616.63,103.06) .. (616.63,95.45) -- cycle ;
%Straight Lines [id:da4432835278242917] 
\draw [line width=1.5]    (595.6,96.75) -- (612.64,95.7) ;
\draw [shift={(616.63,95.45)}, rotate = 536.46] [fill={rgb, 255:red, 0; green, 0; blue, 0 }  ][line width=0.08]  [draw opacity=0] (11.61,-5.58) -- (0,0) -- (11.61,5.58) -- cycle    ;
%Straight Lines [id:da43486905417643107] 
\draw [line width=1.5]    (646.43,95.45) -- (673,95.06) ;
\draw [shift={(677,95)}, rotate = 539.1600000000001] [fill={rgb, 255:red, 0; green, 0; blue, 0 }  ][line width=0.08]  [draw opacity=0] (11.61,-5.58) -- (0,0) -- (11.61,5.58) -- cycle    ;

% Text Node
\draw (53.53,90.61) node  [font=\large,xscale=1.4,yscale=1.4]  {$-$};
% Text Node
\draw (660.62,75.76) node  [font=\large,xscale=1.4,yscale=1.4]  {$y$};
% Text Node
\draw (147,91) node  [font=\large,xscale=1.4,yscale=1.4]  {$\mathbf{C_{\mathbf{R}}}$};
% Text Node
\draw (259.25,90.5) node  [font=\large,xscale=1.4,yscale=1.4]  {$\mathbf{C_{\mathbf{L}_{2}}}$};
% Text Node
\draw (346.25,92.5) node  [font=\large,xscale=1.4,yscale=1.4]  {$\mathbf{G}$};
% Text Node
\draw (516.09,96.04) node  [font=\large,xscale=1.4,yscale=1.4]  {$\mathcal{C}_{s}$};
% Text Node
\draw (434.25,95.5) node  [font=\large,xscale=1.4,yscale=1.4]  {$\mathbf{C_{\mathbf{L}_{1}}}$};
% Text Node
\draw (12.62,71.76) node  [font=\large,xscale=1.4,yscale=1.4]  {$r$};
% Text Node
\draw (574.55,97.47) node  [font=\large,xscale=1.4,yscale=1.4]  {$\beta ^{\prime }$};
% Text Node
\draw (407.55,31.47) node  [font=\large,xscale=1.4,yscale=1.4]  {$\Varrho ^{\prime }$};
% Text Node
\draw (631.53,96.61) node  [font=\large,xscale=1.4,yscale=1.4]  {$+$};
\end{tikzpicture}}
 	\caption{The block diagram of the $H_\beta$ condition for the closed-loop architecture Fig.~\ref{F_4-01} with GFORE or PCI}
 	\label{F_4-03}
 \end{figure}
Thus\footnote{Omitting arguments for simplicity}	
\begin{equation}\label{E_4E-33}
\Re(H(j\omega))=\dfrac{\beta^\prime\mathcal{N}_\chi +\Varrho^\prime\mathcal{N}_\Upsilon}{(a+1)^2+b^2}.
\end{equation}
Define now the vector $\vv{\xi}$ in the $\chi-\Upsilon$ plane as $\vv{\xi}=[\beta^\prime\ \ \Varrho^\prime]^T$. Using Definition \ref{D_41}, equation (\ref{E_4E-33}) can be re-written as 
\begin{equation}\label{E_4E-34}
\Re(H(j\omega))=\dfrac{\vv{\xi}\cdot\vv{\mathcal{N}}}{(a+1)^2+b^2}.
\end{equation}
Therefore 
\begin{equation}\label{E_4E-36}
\begin{array}{*{35}{c}}
\forall\omega\in\mathbb{R}^+:\ \Re(H(j\omega))>0\iff\vv{\xi}\cdot\vv{\mathcal{N}}>0\iff\\
-\frac{\pi}{2}<\phase{(\vv{\xi},\vv{\mathcal{N}})}<\frac{\pi}{2}\ \land\ \abs{\vv{\mathcal{N}}}\neq 0\ \land\ \abs{\vv{\mathcal{\xi}}}\neq 0.
\end{array}
\end{equation}
The rest of the proof of this step are the same as the proof of Step 1 provided in~\cite{AliCDC}.
\newline Step 2: When the open-loop system has poles at the origin and $C_R$ is a GFORE, equation (\ref{E_4E-32}) becomes
\begin{equation}\label{E_4E-44}
%\resizebox{\columnwidth}{!}{$
	\displaystyle\lim_{\omega\to 0}\Re(H(j\omega))=K_{s_0}\beta^\prime>0,
	%$}
\end{equation}
whereas in the case of PCI and CI when $C_{L_1}(s)C_{L_2}(s)G(s)$ does not have any pole at the origin, (\ref{E_4E-32}) becomes 
\begin{equation}\label{E_4-445}
%\resizebox{\columnwidth}{!}{$
	\displaystyle\lim_{\omega\to 0}\Re(H(j\omega))=K_{s_0}\beta^\prime+\Varrho^\prime\dfrac{\omega_r}{C_{L_1}(0)C_{L_2}(0)G(0)}>0.
	%$}
\end{equation}
Setting $\vv{\mathcal{N}^{\prime}}=[K_{s_0}\quad\dfrac{\omega_r}{C_{L_1}(0)C_{L_2}(0)G(0)}]^T$, yields
\begin{equation}\label{E_4E-447}
\displaystyle\lim_{\omega\to 0}\Re(H(j\omega))=\vv{\xi}\cdot\vv{\mathcal{N}^{\prime}}.
\end{equation}  
In addition
\begin{equation}\label{E_4E-47100}
\phase{\vv{\mathcal{N}^{\prime}}}=\displaystyle\lim_{\omega\to 0}\phase{\vv{\mathcal{N}}}\xRightarrow[]{\ (\ref{E_4-333})\ }\theta_{1}\leq\phase{\vv{\mathcal{N}^{\prime}}}\leq\theta_{2}.
\end{equation}
As a result, by Step 1, $\displaystyle\lim_{\omega\to 0} \Re(H(j\omega))=\vv{\xi}\cdot\vv{\mathcal{N}^{\prime}}>0$. For PCI, when $C_{L_1}(s)C_{L_2}(s)G(s)$ has poles at the origin,
\begin{equation}\label{E_4E-446}
\displaystyle\lim_{\omega\to 0}\Re(H(j\omega))=K_{s_0}\beta^\prime>0.
\end{equation}
Note that for CI in equations~(\ref{E_4-445})-(\ref{E_4E-47100}), $\omega_r=1$. It is therefore concluded that if $C_{L_1}(s)C_{L_2}(s)G(s)$ has poles at the origin, then $K_{s_0}\beta^\prime>0$. If $C_{L_1}(s)C_{L_2}(s)G(s)$ does not have any pole at the origin, $\beta$ can be either positive or negative.    
\newline Step 3: In the case of GFORE with $n-m=2$, setting $\vv{\mathcal{N}^{\prime\prime}}=[-K_n\quad\omega_r^2]^T$ yields
\begin{equation}\label{E_4E-52}
\displaystyle\lim_{\omega\to \infty} \omega^2\Re(H(j\omega))=-\beta^\prime K_n+\Varrho^\prime\omega_r^2=\vv{\xi}\cdot\vv{\mathcal{N}^{\prime\prime}}.
\end{equation}  
In addition, 
\begin{equation}\label{E_4E-471}
\phase{\vv{\mathcal{N}^{\prime\prime}}}=\displaystyle\lim_{\omega\to\infty}\phase{\vv{\mathcal{N}}}\xRightarrow[]{\ (\ref{E_4-333})\ }\theta_{1}\leq\phase{\vv{\mathcal{N}^{\prime\prime}}}\leq\theta_{2}.
\end{equation}
Thus, by Step 1 $\displaystyle\lim_{\omega\to \infty} \omega^2\Re(H(j\omega))=\vv{\xi}\cdot\vv{\mathcal{N}^{\prime\prime}}>0$. For GFORE with $n-m>2$, $\displaystyle\lim_{\omega\to \infty}\omega^2\Re(H(j\omega))=\Varrho^\prime\omega_r^2>0$.  
For PCI $\displaystyle\lim_{s\to \infty}H(s)=\Varrho^\prime>0$. Moreover, in the case of CI when $n-m>2$,
\begin{equation}\label{Ap_43-1}
\displaystyle\lim_{\omega\to \infty} \omega^2\Re(H(j\omega))=0,
\end{equation}  
which implies that $H(s)$ is not SPR in the case of $n-m>2$. Whereas in the case of CI with $n-m=2$,
\begin{equation}\label{Ap_43-2}
\displaystyle\lim_{\omega\to \infty} \omega^2\Re(H(j\omega))=-K_{s_0}\beta^\prime>0,
\end{equation}  
which means that in the case of CI, $C_{L_1}(s)C_{L_2}(s)G(s)$ must not have any pole at the origin.
\newline Step 4: In order to show that the pairs $(A,C_0)$ and $(A,B_0)$ are observable and controllable, respectively, it is sufficient to show that the denominator and the numerator of $H(s)$ do not have any common root. Let $a_0+jb_0$ be a root of the denominator. Then
\begin{equation}\label{E_4-50-51-1}
%\resizebox{\columnwidth}{!}{$
1+R_L(a_0,b_0)+jI_L(a_0,b_0)=0\Rightarrow
\begin{cases}
R_L(a_0,b_0)=-1,\\
I_L(a_0,b_0)=0.
\end{cases}
%$}
\end{equation}  
Now, the numerator must not have a root at $a_0+jb_0$, that is  
\begin{equation}\label{E_4-50-51-2}
\resizebox{\columnwidth}{!}{$
	\begin{array}{*{35}{c}}
	\beta^\prime\left(R_{\mathcal{C}_s}(a_0,b_0)+jI_{\mathcal{C}_s}(a_0,b_0)\right)\neq\Varrho^\prime\left(R_{C_R}(a_0,b_0)+jI_{C_R}(a_0,b_0)\right)\\ \Rightarrow
	\beta^\prime R_{\mathcal{C}_s}(a_0,b_0)\neq\Varrho^\prime R_{C_R}(a_0,b_0)\ \lor \ \beta^\prime I_{\mathcal{C}_s}(a_0,b_0)\neq\Varrho^\prime I_{C_R}(a_0,b_0).
\end{array}
$}
\end{equation}
Therefore, using Step 1 and (\ref{E_4-50-51-2}) it is possible to find a pair $(\beta^\prime,\Varrho^\prime)$ such that $H(s)$ does not have any pole-zero cancellation. According to Step 1-4, $H(s)$ is SPR~\cite{khalil2002nonlinear}, $(\bar{A},C_0)$ is observable and $(\bar{A},B_0)$ is controllable, and the base linear system is stable. Moreover, since $-1<\gamma<1$, one has that $A_\rho^T\Varrho A_\rho-\Varrho<0$. As a result, the $H_\beta$ condition is satisfied for the reset control system (\ref{E_4-27}) with GFORE~(\ref{E_4-22}), or CI~(\ref{E_4-2-23}), or PCI~(\ref{E_4-23}). Hence, the zero equilibrium of the reset control system (\ref{E_4-27}) is globally uniformly asymptotically stable when $w=0$, and according to Lemma \ref{L_41}, it has the UBIBS property for any initial condition $x_0$ and any input $w$ which is a Bohl function.      
\end{pf}
\begin{corollary}\label{CO_41} 
Let $\mathcal{C}_s(s)=1$, $\theta_L=\phase{L(j\omega)}$, and $\theta_{C_R}=\phase{C_R(j\omega)}$. Suppose that the base linear system of the reset control system (\ref{E_4-27}) is stable, $A_\rho=\gamma,\ -1<\gamma<1$, $L(s)$ and the open-loop system does not have any pole-zero cancellation. Then the zero equilibrium of the reset control system (\ref{E_4-27}) with GFORE~(\ref{E_4-22}), or CI~(\ref{E_4-2-23}), or PCI~(\ref{E_4-23}) is globally uniformly asymptotically stable when $w=0$, and the system has the UBIBS property for any input $w$ which is a Bohl function if at least one of the following conditions hold.
\begin{enumerate}
\item For all $\omega\in\mathbb{R}^+$, $\sin(\theta_L)\geq0$.
\item For all $\omega\in\mathbb{R}^+$, $\cos(\theta_L-\theta_{C_R})\geq0$ and the reset element is not CI~(\ref{E_4-2-23}).	
\end{enumerate}   	
\end{corollary}
\begin{pf}
When $\mathcal{C}_s(s)=1$, $\mathcal{N}_\chi(\omega)=a(\omega)^2+b(\omega)^2+b(\omega)$. By Hypothesis 1, $b(\omega)\geq0$, for all $\omega\in\mathbb{R}^+$, which implies that $\mathcal{N}_\chi(\omega)>0$. Thus, the reset control system (\ref{E_4-27}) is of Type I. In addition, defining $C_R(j\omega)=a_R(\omega)+jb_R(\omega)$, yields $\mathcal{N}_\Upsilon(\omega)=a(\omega)a_R(\omega)+b(\omega)b_R(\omega)+a_R(\omega)$. By Hypothesis 2,   	
\begin{equation}\label{E_4-co}
\resizebox{\columnwidth}{!}{$
\forall\ \omega\in\mathbb{R}^+:\ \cos(\theta_L-\theta_{C_R})\geq0\Rightarrow \dfrac{a(\omega)a_R(\omega)+b(\omega)b_R(\omega)}{|L(j\omega)C_R(j\omega)|}\geq0,
$}
\end{equation}
and since $a_R(\omega)>0$ in the cases of PCI and GFORE, $\mathcal{N}_\Upsilon(\omega)>0$, for all $\omega\in\mathbb{R}^+$. Therefore, the reset control system (\ref{E_4-27}) is of Type I and/or Type II, hence the claim.  
\end{pf}
In~\cite{mechNima} the GFORE, CI and PCI architectures have been modified to improve the performance of reset control systems. Using the same procedure as Theorem~\ref{T_41} a frequency-domain method to assess stability properties of these reset control systems illustrated in Fig.~\ref{F_4-04} is proposed. 

 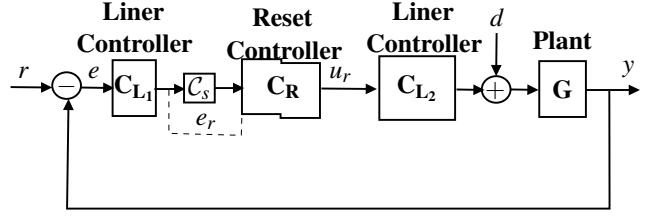
\begin{figure}[!t]
 	\centering
	\resizebox{\columnwidth}{!}{%
 	\tikzset{every picture/.style={line width=0.75pt}} %set default line width to 0.75pt        
\begin{tikzpicture}[x=0.75pt,y=0.75pt,yscale=-1,xscale=1]
%uncomment if require: \path (0,233); %set diagram left start at 0, and has height of 233

%Shape: Rectangle [id:dp8872257254473093] 
\draw  [line width=1.5]  (384.5,56) -- (463,56) -- (463,122) -- (384.5,122) -- cycle ;
%Shape: Ellipse [id:dp19980167715656805] 
\draw  [line width=1.5]  (45.63,86.45) .. controls (45.63,78.84) and (52.3,72.68) .. (60.53,72.68) .. controls (68.76,72.68) and (75.43,78.84) .. (75.43,86.45) .. controls (75.43,94.06) and (68.76,100.22) .. (60.53,100.22) .. controls (52.3,100.22) and (45.63,94.06) .. (45.63,86.45) -- cycle ;
%Straight Lines [id:da0385434449901394] 
\draw [line width=1.5]    (622.77,91.37) -- (623,214) -- (61,214) -- (62.48,104.22) ;
\draw [shift={(62.53,100.22)}, rotate = 450.77] [fill={rgb, 255:red, 0; green, 0; blue, 0 }  ][line width=0.08]  [draw opacity=0] (11.61,-5.58) -- (0,0) -- (11.61,5.58) -- cycle    ;
%Straight Lines [id:da05885102875442705] 
\draw [line width=1.5]    (2.5,88) -- (41.63,88.41) ;
\draw [shift={(45.63,88.45)}, rotate = 180.6] [fill={rgb, 255:red, 0; green, 0; blue, 0 }  ][line width=0.08]  [draw opacity=0] (11.61,-5.58) -- (0,0) -- (11.61,5.58) -- cycle    ;
%Straight Lines [id:da39309315496373975] 
\draw [line width=1.5]    (598.65,91.05) -- (650.5,91) ;
\draw [shift={(654.5,91)}, rotate = 539.95] [fill={rgb, 255:red, 0; green, 0; blue, 0 }  ][line width=0.08]  [draw opacity=0] (11.61,-5.58) -- (0,0) -- (11.61,5.58) -- cycle    ;
%Straight Lines [id:da6065506865121402] 
\draw [line width=1.5]    (464.49,90.5) -- (489.5,90.93) ;
\draw [shift={(493.5,91)}, rotate = 180.99] [fill={rgb, 255:red, 0; green, 0; blue, 0 }  ][line width=0.08]  [draw opacity=0] (11.61,-5.58) -- (0,0) -- (11.61,5.58) -- cycle    ;
%Shape: Rectangle [id:dp5467349831333912] 
\draw  [line width=1.5]  (550.17,61.09) -- (598.5,61.09) -- (598.5,119) -- (550.17,119) -- cycle ;
%Straight Lines [id:da00759412524357761] 
\draw [line width=1.5]    (324.49,89.5) -- (378,89.97) ;
\draw [shift={(382,90)}, rotate = 180.5] [fill={rgb, 255:red, 0; green, 0; blue, 0 }  ][line width=0.08]  [draw opacity=0] (11.61,-5.58) -- (0,0) -- (11.61,5.58) -- cycle    ;
%Shape: Ellipse [id:dp029726587066210453] 
\draw  [line width=1.5]  (490.63,90.45) .. controls (490.63,82.84) and (497.3,76.68) .. (505.53,76.68) .. controls (513.76,76.68) and (520.43,82.84) .. (520.43,90.45) .. controls (520.43,98.06) and (513.76,104.22) .. (505.53,104.22) .. controls (497.3,104.22) and (490.63,98.06) .. (490.63,90.45) -- cycle ;
%Straight Lines [id:da20690026255567973] 
\draw [line width=1.5]    (506,32) -- (505.57,72.68) ;
\draw [shift={(505.53,76.68)}, rotate = 270.6] [fill={rgb, 255:red, 0; green, 0; blue, 0 }  ][line width=0.08]  [draw opacity=0] (11.61,-5.58) -- (0,0) -- (11.61,5.58) -- cycle    ;
%Shape: Path Data [id:dp5923398308383396] 
\draw  [line width=1.5]  (283,59.09) -- (283,63.09) -- (323,63.09) -- (323,120) -- (282.17,120) -- (282.17,116) -- (242.17,116) -- (242.17,59.09) -- (283,59.09) -- cycle ;
%Straight Lines [id:da023018556967414172] 
\draw [line width=1.5]    (520.43,90.45) -- (545.44,90.88) ;
\draw [shift={(549.44,90.95)}, rotate = 180.99] [fill={rgb, 255:red, 0; green, 0; blue, 0 }  ][line width=0.08]  [draw opacity=0] (11.61,-5.58) -- (0,0) -- (11.61,5.58) -- cycle    ;
%Straight Lines [id:da43982489498145316] 
\draw  [dash pattern={on 4.5pt off 4.5pt}]  (166.5,91) -- (166.5,136) -- (241.5,138) -- (242.17,116) ;
%Shape: Rectangle [id:dp4251942074669439] 
\draw  [line width=1.5]  (182.17,74.09) -- (214,74.09) -- (214,104) -- (182.17,104) -- cycle ;
%Shape: Rectangle [id:dp8630536426570494] 
\draw  [line width=1.5]  (108.17,60.09) -- (153,60.09) -- (153,118) -- (108.17,118) -- cycle ;
%Straight Lines [id:da0661931984677957] 
\draw [line width=1.5]    (75.43,88.45) -- (106,88.94) ;
\draw [shift={(110,89)}, rotate = 180.91] [fill={rgb, 255:red, 0; green, 0; blue, 0 }  ][line width=0.08]  [draw opacity=0] (11.61,-5.58) -- (0,0) -- (11.61,5.58) -- cycle    ;
%Straight Lines [id:da4031019751407976] 
\draw [line width=1.5]    (151.5,90) -- (179.5,90) ;
\draw [shift={(183.5,90)}, rotate = 180] [fill={rgb, 255:red, 0; green, 0; blue, 0 }  ][line width=0.08]  [draw opacity=0] (11.61,-5.58) -- (0,0) -- (11.61,5.58) -- cycle    ;
%Straight Lines [id:da16415406983663483] 
\draw [line width=1.5]    (213.5,89) -- (240.5,89) ;
\draw [shift={(244.5,89)}, rotate = 180] [fill={rgb, 255:red, 0; green, 0; blue, 0 }  ][line width=0.08]  [draw opacity=0] (11.61,-5.58) -- (0,0) -- (11.61,5.58) -- cycle    ;

% Text Node
\draw (60.53,87.61) node  [font=\large,xscale=1.4,yscale=1.4]  {$-$};
% Text Node
\draw (15.84,72.76) node  [font=\large,xscale=1.4,yscale=1.4]  {$r$};
% Text Node
\draw (642.62,68.76) node  [font=\large,xscale=1.4,yscale=1.4]  {$y$};
% Text Node
\draw (205,124) node  [font=\large,xscale=1.4,yscale=1.4]  {$e_{r}$};
% Text Node
\draw (431,26) node  [font=\large,xscale=1.4,yscale=1.4] [align=left] {{\fontfamily{ptm}\selectfont {\large \textbf{ \ \ \ Liner }}}\\{\fontfamily{ptm}\selectfont {\large \textbf{Controller}}}};
% Text Node
\draw (574,40) node  [font=\large,xscale=1.4,yscale=1.4] [align=left] {{\fontfamily{ptm}\selectfont \textbf{{\large Plant}}}};
% Text Node
\draw (504.53,91.61) node  [font=\large,xscale=1.4,yscale=1.4]  {$+$};
% Text Node
\draw (506.84,15.76) node  [font=\large,xscale=1.4,yscale=1.4]  {$d$};
% Text Node
\draw (285,88) node  [font=\large,xscale=1.4,yscale=1.4]  {$\mathbf{C_{\mathbf{R}}}$};
% Text Node
\draw (286,33) node  [font=\large,xscale=1.4,yscale=1.4] [align=left] {{\fontfamily{ptm}\selectfont {\large \textbf{ \ \ \ Reset }}}\\{\fontfamily{ptm}\selectfont {\large \textbf{Controller}}}};
% Text Node
\draw (423.75,89) node  [font=\large,xscale=1.4,yscale=1.4]  {$\mathbf{C_{\mathbf{L}_{2}}}$};
% Text Node
\draw (344.35,74.76) node  [font=\large,xscale=1.4,yscale=1.4]  {$u_{r}$};
% Text Node
\draw (574.34,90.04) node  [font=\large,xscale=1.4,yscale=1.4]  {$\mathbf{G}$};
% Text Node
\draw (198.09,89.04) node  [font=\large,xscale=1.4,yscale=1.4]  {$\mathcal{C}_{s}$};
% Text Node
\draw (130.59,89.04) node  [font=\large,xscale=1.4,yscale=1.4]  {$\mathbf{C_{\mathbf{L}_{1}}}$};
% Text Node
\draw (88.84,72.76) node  [font=\large,xscale=1.4,yscale=1.4]  {$e$};
% Text Node
\draw (131,26) node  [font=\large,xscale=1.4,yscale=1.4] [align=left] {{\fontfamily{ptm}\selectfont {\large \textbf{ \ \ \ Liner }}}\\{\fontfamily{ptm}\selectfont {\large \textbf{Controller}}}};
\end{tikzpicture}}
 	\caption{The closed-loop architecture of a modified reset element}
 	\label{F_4-04}
 \end{figure}
\begin{corollary}\label{CO_42}
Let the NSV vector for the reset control system shown in Fig.~\ref{F_4-04} be
\setlength{\arraycolsep}{0.0em}
\begin{eqnarray}\label{E_4CO2}
\vv{\mathcal{N}}_{MF}(\omega)&{=}&[\mathcal{N}_{MF_\chi} \quad \mathcal{N}_{MF_\Upsilon}]^T\nonumber\\
&{=}&\left[\Re(\dfrac{L^\prime(j\omega)\kappa(j\omega)}{\mathcal{C}_s(j\omega)}) \quad \Re(\kappa(j\omega)C_R(j\omega))\right]^T,
\end{eqnarray}
\setlength{\arraycolsep}{5pt} in which $L^\prime(s)=C_{L_1}(s)C_R(s)C_{L_2}(s)\mathcal{C}_s(s)G(s)$. Then, the zero equilibrium of the reset control system (\ref{E_4-27}) in the configuration of Fig.~\ref{F_4-04} with GFORE~(\ref{E_4-22}), or CI~(\ref{E_4-2-23}), or PCI~(\ref{E_4-23}) is globally uniformly asymptotically stable when $w=0$, and the system has the UBIBS property for any input $w$ which is a Bohl function if all of the following conditions are satisfied.	
 	\begin{itemize}
 		\item The base linear system is stable and the open-loop transfer function does not have any pole-zero cancellation.
		\item In the case of CI~(\ref{E_4-2-23}), $C_{L_1}(s)C_{L_2}(s)G(s)$ does not have any pole at the origin and $n-m=2$. 
 		\item The reset control system (\ref{E_4-27}) is either of Type I and/or of Type II. 
		\item $A_\rho=\gamma,\ -1<\gamma<1.$
		\item $\mathcal{C}_s(s)=1$ and/or the reset instants have the well-posedness property.
 	\end{itemize}
\end{corollary}
\begin{pf}
See Appendix~\ref{aap_42}.
\end{pf}
%%%%%%%%%%%%%%%%%%%%%%%%%%%%%%%%%%%%%GSORE
\section{Stability analysis of reset control systems with second order reset elements}\label{sec_4:4}
\subsection{Reset control systems with GSORE}\label{sec_4:40}
In this section a frequency-domain method for assessing stability properties of the reset control system~(\ref{E_4-27}) with GSORE (\ref{E_4-24}), which has the canonical controllable form state-space realization~(\ref{E_4-25}), is proposed. In this method the $H_\beta$ condition is combined with optimization tools to provide sufficient conditions to guarantee stability properties of the reset control system~(\ref{E_4-27}). Before presenting the main result, one preliminary fact, which is useful for assessing stability properties of the reset control system~(\ref{E_4-27}) with GSORE (\ref{E_4-24}), is presented.
\begin{prop}\label{pro_41}
Let $\vv{\mathcal{Q}}\in\mathbb{R}^2$ and $\vv{\mathcal{F}}\in\mathbb{R}^2$ be defined as $\vv{\mathcal{Q}}=\begin{bmatrix}
Q_1 & Q_2
\end{bmatrix}^T$ and $\vv{\mathcal{F}}(\omega)=\begin{bmatrix}\mathcal{F}_1(\omega)& \mathcal{F}_2(\omega)\end{bmatrix}^T$. Let $\phase{\vv{\mathcal{Q}},\vv{\mathcal{F}}(\omega)}=\vartheta(\omega,\dfrac{Q_2}{Q_1})$,  $\omega_p=\{\omega\in\mathbb{R}^+|\ \mathcal{F}_3(\omega)\geq0 \}$, $\omega_N=\mathbb{R}^+-\omega_p$,
$$g_p=\left\{\dfrac{Q_2}{Q_1}\in\mathbb{R}|\ \forall\omega\in\omega_p:\ Q_1\mathcal{F}_1(\omega)+Q_2\mathcal{F}_2(\omega)>0 \right\},$$
and
$$g_N=\left\{\dfrac{Q_2}{Q_1}\in\mathbb{R}|\ \forall\omega\in\omega_N:\ Q_1\mathcal{F}_1(\omega)+Q_2\mathcal{F}_2(\omega)>0 \right\}.$$ Then the condition
\begin{equation}\label{p_4311}
Q_1\mathcal{F}_1(\omega)+Q_2\mathcal{F}_2(\omega)>\mathcal{F}_3(\omega), 
\end{equation}
holds for all $\omega\in\mathbb{R}$ if and only if
\begin{itemize}
	\item $\eta_1(\dfrac{Q_2}{Q_1})<\sqrt{Q_1^2+Q_2^2}<\eta_2(\dfrac{Q_2}{Q_1})$,
	\item $\dfrac{Q_2}{Q_1}\in\left\{\dfrac{Q_2}{Q_1}\in g_p|\ \eta_1(\dfrac{Q_2}{Q_1})<\eta_2(\dfrac{Q_2}{Q_1})\right\}$,
\end{itemize} 
where
{\footnotesize
\begin{align}\label{E_4CO2}
\eta_1(\dfrac{Q_2}{Q_1})&=\begin{cases}
-\infty &  \omega_p=\varnothing,\\
\underset{\omega\in\omega_p}{\max\ }\dfrac{\mathcal{F}_3(\omega)}{\cos(\vartheta)\sqrt{\mathcal{F}^2_1(\omega)+\mathcal{F}^2_2(\omega)}}, & \omega_p\neq\varnothing, 
\end{cases}\nonumber\\
\eta_2(\dfrac{Q_2}{Q_1})&=\begin{cases}
+\infty & \dfrac{Q_2}{Q_1}\in g_N,\lor\omega_N=\varnothing,\\
\underset{\omega\in\omega_N}{\min\ }\dfrac{\mathcal{F}_3(\omega)}{\cos(\vartheta)\sqrt{\mathcal{F}^2_1(\omega)+\mathcal{F}^2_2(\omega)}}, & \dfrac{Q_2}{Q_1}\notin g_N. 
\end{cases}
\end{align}\par}
\end{prop}
\begin{pf}
See Appendix~\ref{aa_43}.
\end{pf}
\begin{remark}
	{\rm The sets $g_p$ and $g_N$ can be easily obtained using the method described in~\cite{AliCDC}.} 
\end{remark}
Define now $\Gamma(\gamma_1,\gamma_2)=\dfrac{(\gamma_1\gamma_2-1)^2}{(\gamma_1^2-1)(\gamma_2^2-1)}$
\setlength{\arraycolsep}{0.0em}
\begin{eqnarray}\label{E_4CO2}
f_1(\mathcal{X}_1,\mathcal{X}_2,\mathcal{X}_3,\omega)&{=}&\mathcal{X}_1\Re(C_R(j\omega)\kappa(\omega)j\omega)\nonumber\\\
&&{+}\:\mathcal{X}_2\Re(C_R(j\omega)\kappa(\omega))\nonumber\\\
&&{+}\:\mathcal{X}_3\Re(\mathcal{C}_s(j\omega)(a^2+b^2+a)),\nonumber\\
f_2(\mathcal{X}_1,\mathcal{X}_2,\mathcal{X}_3,\omega)&{=}&\mathcal{X}_1\Re(C_R(j\omega)\kappa(\omega)(j\omega+2\xi\omega_r))\nonumber\\
&&{+}\:\mathcal{X}_3\Re(L(j\omega)\kappa(\omega)\mathcal{C}_s(j\omega)(j\omega+2\xi\omega_r))\nonumber\\
&&{+}\:\mathcal{X}_2\Re(C_R(j\omega)\kappa(\omega)(2j\xi\omega_r\omega-\omega^2)\nonumber\\
&&{-}\:(a+1)^2-b^2),\nonumber
\end{eqnarray}
{\scriptsize
\setlength{\arraycolsep}{0.0em}
\begin{eqnarray}\label{E_4CO22}
G_1(Q_1,Q_2,Q_3,Q_4)&{=}&\displaystyle\sup_{\omega\in(0,\infty)}\Bigg[\dfrac{f_1(Q_1,Q_2,1,\omega)}{f_1(Q_2,\dfrac{Q_2Q_3}{Q_4},\dfrac{Q_2}{Q_4},\omega)+f_2(Q_2,Q_1,1,\omega)}\nonumber\\
&&{\times}\dfrac{f_2(Q_3,Q_4,1,\omega)}{f_1(Q_4,Q_3,1,\omega)+f_2(Q_4,\dfrac{Q_1Q_4}{Q_2},\dfrac{Q_4}{Q_2},\omega)}\Bigg],\nonumber\\
G_2(Q_1,Q_2,Q_3,Q_4)&{=}&\displaystyle\sup_{\omega\in[0,\infty)}\Bigg[\dfrac{f_1(1,Q_2,Q_1,\omega)}{f_1(Q_2,\dfrac{Q_2}{Q_4},\dfrac{Q_2Q_3}{Q_4},\omega)+f_2(Q_2,1,Q_1,\omega)}\nonumber\\
&&{\times}\dfrac{f_2(1,Q_4,Q_3,\omega)}{f_1(Q_4,1,Q_3,\omega)+f_2(Q_4,\dfrac{Q_4}{Q_2},\dfrac{Q_1Q_4}{Q_2},\omega)}\Bigg].\nonumber\\
\end{eqnarray}
\par}\setlength{\arraycolsep}{5pt}We define systems of Type III, of Type IV, and of Type V to assess stability properties of the reset control system (\ref{E_4-27}) with GSORE (\ref{E_4-25}).\begin{definition}\label{D_44}
	The reset control system (\ref{E_4-27}) with GSORE (\ref{E_4-25}) is of Type III if the following conditions hold. 
	\begin{enumerate}[(1)]
		\item $M<4$, where $M=\underset{Q_1,Q_2,Q_3,Q_4}{\min} \ G_1(Q_1,Q_2,Q_3,Q_4)$, in which $Q_1$, $Q_2$, $Q_3$, and $Q_4$ are such that the following constraints hold:
		\begin{equation}\label{E_433-033}
		\begin{aligned}
		\mathcal{S}_1: \ & \forall\omega\in(0,\infty):\ K_{s_0}f_1(Q_1,Q_2,1,\omega)>0,\\
		\mathcal{S}_2: \ & \forall\omega\in(0,\infty):\ K_{s_0}f_2(Q_3,Q_4,1,\omega)>0,\\
		\mathcal{S}_3: \ & K_{s_0}\left(\dfrac{2\xi\omega_r}{Q_1}+\dfrac{Q_2}{Q_1Q_4}+\dfrac{2}{Q_1}\sqrt{\dfrac{2Q_2\xi\omega_r}{Q_4}-\dfrac{Q_2}{K_{s_0}}}\right)>1,\nonumber\\
		\mathcal{S}_4: \ & K_{s_0}\left(\dfrac{2\xi\omega_r}{Q_1}+\dfrac{Q_2}{Q_1Q_4}-\dfrac{2}{Q_1}\sqrt{\dfrac{2Q_2\xi\omega_r}{Q_4}-\dfrac{Q_2}{K_{s_0}}}\right)<1,\nonumber\\
		\mathcal{S}_5: \ &  \dfrac{\omega_r^2Q_1}{Q_2}+2\omega_r\left(\xi+2\sqrt{\dfrac{2Q_1\xi\omega_r}{Q_2}-1}\right)>\dfrac{Q_3}{Q_4},\nonumber
		\end{aligned}
		\end{equation}
		\begin{equation}\label{E_433-033}
		\begin{aligned}
		\mathcal{S}_6: \ &  \dfrac{\omega_r^2Q_1}{Q_2}+2\omega_r\left(\xi-2\sqrt{\dfrac{2Q_1\xi\omega_r}{Q_2}-1}\right)<\dfrac{Q_3}{Q_4},\\
                 \mathcal{S}_7: \ & K_{s_0}Q_i>0,\ 2\xi\omega_r>\dfrac{Q_4}{K_{s_0}},\ 2\xi\omega_r>\dfrac{Q_2}{Q_1},\\
                 \mathcal{S}_8: \ &\dfrac{Q_1Q_3}{Q_2Q_4}>\Gamma(\gamma_1,\gamma_2).		\end{aligned}
		\end{equation}
		\item The pairs $(\bar{A},C_0)$ and $(\bar{A},B_0)$ where $B_0=\begin{bmatrix}0_{n_p\times2} \\ I_2\end{bmatrix}$ and $C_0=\left[\begin{bmatrix}1 \\ \dfrac{Q_2}{Q_4} \end{bmatrix}\bar{C}_e\quad \begin{bmatrix}Q_1 & Q_2\\Q_2 & \dfrac{Q_2Q_3}{Q_4} \end{bmatrix}\right]$ are controllable and observable, respectively. 
		\item The open-loop system has at least one pole at the origin and $K_{s_0}\neq0$.
	\end{enumerate}	
\end{definition}
\begin{definition}\label{D_45}
	The reset control system (\ref{E_4-27}) with GSORE (\ref{E_4-25}) is of Type IV if the following conditions hold. 
	\begin{enumerate}[(1)]
		\item $M<4$, where $M=\underset{Q_1,Q_2,Q_3,Q_4}{\min} \ G_2(Q_1,Q_2,Q_3,Q_4)$, in which $Q_1$, $Q_2$, $Q_3$, and $Q_4$ are such that the following constraints hold:
		\begin{equation}\label{E_433-044}
		\begin{aligned}
		\mathcal{S}_1: \ & \forall\omega\in[0,\infty):\ f_1(1,Q_2,Q_1,\omega)>0,\\
		\mathcal{S}_2: \ & \forall\omega\in[0,\infty):\ f_2(1,Q_4,Q_3,\omega)>0,\\
		\mathcal{S}_3: \ &  \omega_r^2+2\omega_r\left(\xi Q_2+2\sqrt{2Q_2\xi\omega_r-Q_2^2}\right)>\dfrac{Q_2}{Q_4},\\
		\mathcal{S}_4: \ &  \omega_r^2+2\omega_r\left(\xi Q_2-2\sqrt{2Q_2\xi\omega_r-Q_2^2}\right)<\dfrac{Q_2}{Q_4},\\
                 \mathcal{S}_5: \ & Q_4>0,\ 0<Q_2<2\xi\omega_r,\ Q_2Q_4<\dfrac{1}{\Gamma(\gamma_1,\gamma_2)}.
		\end{aligned}
		\end{equation}
		\item The pairs $(\bar{A},C_0)$ and $(\bar{A},B_0)$ where $B_0=\begin{bmatrix}0_{n_p\times2} \\ I_2\end{bmatrix}$ and $C_0=\left[\begin{bmatrix}Q_1 \\ \dfrac{Q_2Q_3}{Q_4} \end{bmatrix}\bar{C}_e\quad \begin{bmatrix}1 & Q_2\\Q_2 & \dfrac{Q_2}{Q_4} \end{bmatrix}\right]$ are controllable and observable, respectively. 
		\item The open-loop system does not have any pole at the origin.
		\item $n-m>3$. 
	\end{enumerate}	
\end{definition}
\begin{definition}\label{D_46}
	The reset control system (\ref{E_4-27}) with GSORE (\ref{E_4-25}) is of Type V if the following conditions hold. 
	\begin{enumerate}[(1)]
		\item $M<4$, where $M=\underset{Q_1,Q_2,Q_3,Q_4}{\min} \ G_2(Q_1,Q_2,Q_3,Q_4)$, in which $Q_1$, $Q_2$, $Q_3$, and $Q_4$ are such that the following constraints hold:
		\begin{equation}\label{E_433-055}
		\begin{aligned}
                 \mathcal{S}_1: \ & \forall\omega\in[0,\infty):\ f_1(1,Q_2,Q_1,\omega)>0,\\
		\mathcal{S}_2: \ & \forall\omega\in[0,\infty):\ f_2(1,Q_4,Q_3,\omega)>0,\\
		\mathcal{S}_3: \ &  \resizebox{0.85\columnwidth}{!}{$\omega_r^2-K_nQ_1+2\xi\omega_rQ_2+2\sqrt{2\xi\omega_r^3Q_2+\dfrac{Q_2^2Q_3K_n}{Q_4}-\omega_r^2Q_2^2}>\dfrac{Q_2}{Q_4},$}\\
		\mathcal{S}_4: \ &  \resizebox{0.85\columnwidth}{!}{$\omega_r^2-K_nQ_1+2\xi\omega_rQ_2-2\sqrt{2\xi\omega_r^3Q_2+\dfrac{Q_2^2Q_3K_n}{Q_4}-\omega_r^2Q_2^2}<\dfrac{Q_2}{Q_4},$}\\
        \mathcal{S}_5: \ & 2\xi\omega_r^3Q_2+\dfrac{Q_2^2Q_3K_n}{Q_4}>\omega_r^2Q_2^2,\\
        \mathcal{S}_6: \ & Q_2<2\xi\omega_r,\ K_nQ_3<\omega_r^2Q_4,\ 0<Q_2Q_4<\dfrac{1}{\Gamma(\gamma_1,\gamma_2)}.        
		\end{aligned}
		\end{equation}
		\item The pairs $(\bar{A},C_0)$ and $(\bar{A},B_0)$ where $B_0=\begin{bmatrix}0_{n_p\times2} \\ I_2\end{bmatrix}$ and $C_0=\left[\begin{bmatrix}Q_1 \\ \dfrac{Q_2Q_3}{Q_4} \end{bmatrix}\bar{C}_e\quad \begin{bmatrix}1 & Q_2\\Q_2 & \dfrac{Q_2}{Q_4} \end{bmatrix}\right]$ are observable and controllable, respectively. 
		\item The open-loop system does not have any pole at the origin.
		\item $n-m=3$.
	\end{enumerate}	
\end{definition}
 \begin{thm}\label{T_43}
	The zero equilibrium of the reset control system (\ref{E_4-27}) with GSORE (\ref{E_4-25}) is globally uniformly asymptotically stable when $w=0$, and the system has the UBIBS property for any input $w$ which is a Bohl function if all of the following conditions are satisfied.	
	\begin{itemize}
		\item The base linear system is stable.
		\item $A_\rho=\begin{bmatrix}\gamma_1 & 0\\ 0 & \gamma_2\end{bmatrix}$ and $-1<\gamma_i<1$, for $i=1,\ 2$.
		\item The reset control system is either of Type III, or of Type IV, or of Type V. 
		\item $\mathcal{C}_s(s)=1$ and/or the reset instants have the well-posedness property.
	\end{itemize}
\end{thm}
\begin{pf}
	Theorem \ref{T_43} is proved in the following steps.
	\begin{itemize}
		\item Step 1: The transfer function $H(s)$ in (\ref{E_4-29}) for the reset control system~(\ref{E_4-27}) with GSORE (\ref{E_4-25}) is calculated. Then, it is shown that $A_\rho^T \Varrho A_\rho-\Varrho<0$.
		\item Step 2: It is shown that $\displaystyle\lim_{\omega\to \infty} \omega^2(H(j\omega)+H(-j\omega)^T)>0.$ 
		\item Step 3: For systems with poles at the origin it is shown that $\displaystyle\lim_{\omega\to 0} H(j\omega)+H(-j\omega)^T>0$. 
		\item Step 4: It is shown that $H(j\omega)+H(-j\omega)^T>0$, for all $\omega\in\mathbb{R}^+$.  
	\end{itemize}  
	Step 1: In the case of GSORE let $\beta=-\begin{bmatrix}
    \beta_1& \beta_2
	\end{bmatrix}$ and $\Varrho=\begin{bmatrix}
\Varrho_1& \Varrho_2\\ \Varrho_2& \Varrho_3
	\end{bmatrix}>0$ be such that
	\begin{equation}\label{E_433-0444}
	\beta_i\in\mathbb{R},\ \Varrho_3>0,\ \Varrho_1>0,\ \Varrho_1\Varrho_3>\Varrho_2^2. 
	\end{equation}
	In addition, since $A_\rho=\begin{bmatrix}\gamma_1& 0\\ 0& \gamma_2\end{bmatrix}$, we have the condition
	\begin{equation}\label{E_433-1}
	A_\rho^T \Varrho A_\rho-\Varrho=\begin{bmatrix} (\gamma_1^2-1)\Varrho_1 & (\gamma_1\gamma_2-1)\Varrho_2\\ (\gamma_1\gamma_2-1)\Varrho_2 & (\gamma_2^2-1)\Varrho_3\end{bmatrix}<0.
	\end{equation} 
	Since $-1<\gamma_i<1$, using (\ref{E_433-0444}) and (\ref{E_433-1}), yields
	\begin{equation}\label{E_433-16}
	\dfrac{\Varrho_1\Varrho_3}{\Varrho_2^2}>\Gamma(\gamma_1,\gamma_2)=\dfrac{(\gamma_1\gamma_2-1)^2}{(\gamma_1^2-1)(\gamma_2^2-1)}\geq1.
	\end{equation}
	With the considered matrix $\Varrho$ and vector $\beta$, $H(s)$ in~(\ref{E_4-29}) with $C_0$ as in~(\ref{E_4-F1}) is equal to (see also Fig.~\ref{F_4-06})
	\begin{equation}\label{E_433-04}
	H(s)=\begin{bmatrix}
	H_{11}(s)&   H_{12}(s) \\ H_{21}(s)& H_{22}(s) 
	\end{bmatrix}.
	%H(s)=\begin{bmatrix}\dfrac{\beta_1L(s)\mathcal{C}_s(s)+C_R(s)(\Varrho_1s+\Varrho_2)}{1+L(s)} &  \dfrac{(s+2\xi\omega_r)(\beta_1L(s)\mathcal{C}_(s)+C_R(s)(\Varrho_1s+\Varrho_2))}{1+L(s)}-\Varrho_1 \\
	%\dfrac{\beta_2L(s)\mathcal{C}_s(s)+C_R(s)(\Varrho_2s+\Varrho_3)}{1+L(s)} &  \dfrac{(s+2\xi\omega_r)(\beta_2L(s)\mathcal{C}_(s)+C_R(s)(\Varrho_2s+\Varrho_3))}{1+L(s)}-\Varrho_2
	%\end{bmatrix}
	\end{equation} 
%%%%%%%%%%%%%%%%%%%%%%%%%%%%%%%%%%%%%%%%%Figure
\begin{figure}
  \centering
  \begin{subfigure}{0.485\textwidth}
    \centering
\resizebox{\textwidth}{!}{   
\tikzset{every picture/.style={line width=0.75pt}} %set default line width to 0.75pt        
\begin{tikzpicture}[x=0.75pt,y=0.75pt,yscale=-1,xscale=1]
%uncomment if require: \path (0,266); %set diagram left start at 0, and has height of 266
%Shape: Rectangle [id:dp3558696855554735] 
\draw  [line width=1.5]  (363,102) -- (401.5,102) -- (401.5,146) -- (363,146) -- cycle ;
%Shape: Ellipse [id:dp605355155007869] 
\draw  [line width=1.5]  (34.63,120.45) .. controls (34.63,112.84) and (41.3,106.68) .. (49.53,106.68) .. controls (57.76,106.68) and (64.43,112.84) .. (64.43,120.45) .. controls (64.43,128.06) and (57.76,134.22) .. (49.53,134.22) .. controls (41.3,134.22) and (34.63,128.06) .. (34.63,120.45) -- cycle ;
%Straight Lines [id:da8035539331450979] 
\draw [line width=1.5]    (538.5,125) -- (538.5,255) -- (49.5,253) ;
%Straight Lines [id:da10064598351922416] 
\draw [line width=1.5]    (2.5,120) -- (30.63,120.39) ;
\draw [shift={(34.63,120.45)}, rotate = 180.8] [fill={rgb, 255:red, 0; green, 0; blue, 0 }  ][line width=0.08]  [draw opacity=0] (11.61,-5.58) -- (0,0) -- (11.61,5.58) -- cycle    ;
%Straight Lines [id:da053830610313541416] 
\draw [line width=1.5]    (403.77,124.37) -- (425,124.06) ;
\draw [shift={(429,124)}, rotate = 539.1600000000001] [fill={rgb, 255:red, 0; green, 0; blue, 0 }  ][line width=0.08]  [draw opacity=0] (11.61,-5.58) -- (0,0) -- (11.61,5.58) -- cycle    ;
%Straight Lines [id:da3228273022349576] 
\draw [line width=1.5]    (261.3,124.37) -- (284.5,124.05) ;
\draw [shift={(288.5,124)}, rotate = 539.22] [fill={rgb, 255:red, 0; green, 0; blue, 0 }  ][line width=0.08]  [draw opacity=0] (11.61,-5.58) -- (0,0) -- (11.61,5.58) -- cycle    ;
%Straight Lines [id:da344361302290097] 
\draw [line width=1.5]    (215.5,126) -- (214.5,187) -- (199.5,187) ;
%Straight Lines [id:da33022400931887896] 
\draw [line width=1.5]    (64.43,120.45) -- (87.5,120.38) ;
\draw [shift={(91.5,120.37)}, rotate = 539.8299999999999] [fill={rgb, 255:red, 0; green, 0; blue, 0 }  ][line width=0.08]  [draw opacity=0] (11.61,-5.58) -- (0,0) -- (11.61,5.58) -- cycle    ;
%Shape: Ellipse [id:dp9293274033414575] 
\draw  [line width=1.5]  (91.5,122.37) .. controls (91.5,114.76) and (98.17,108.6) .. (106.4,108.6) .. controls (114.63,108.6) and (121.3,114.76) .. (121.3,122.37) .. controls (121.3,129.97) and (114.63,136.14) .. (106.4,136.14) .. controls (98.17,136.14) and (91.5,129.97) .. (91.5,122.37) -- cycle ;
%Shape: Rectangle [id:dp8281312117405866] 
\draw  [line width=1.5]  (148.17,94.09) -- (186.5,94.09) -- (186.5,152) -- (148.17,152) -- cycle ;
%Straight Lines [id:da6736196050401356] 
\draw [line width=1.5]    (121.5,123) -- (144.5,123) ;
\draw [shift={(148.5,123)}, rotate = 180] [fill={rgb, 255:red, 0; green, 0; blue, 0 }  ][line width=0.08]  [draw opacity=0] (11.61,-5.58) -- (0,0) -- (11.61,5.58) -- cycle    ;
%Straight Lines [id:da3981723043709309] 
\draw [line width=1.5]    (188.5,125) -- (214.49,125.28) -- (229.5,125.06) ;
\draw [shift={(233.5,125)}, rotate = 539.1600000000001] [fill={rgb, 255:red, 0; green, 0; blue, 0 }  ][line width=0.08]  [draw opacity=0] (11.61,-5.58) -- (0,0) -- (11.61,5.58) -- cycle    ;
%Shape: Rectangle [id:dp5499525015603304] 
\draw  [line width=1.5]  (288.17,94.09) -- (326.5,94.09) -- (326.5,152) -- (288.17,152) -- cycle ;
%Straight Lines [id:da6559801315350425] 
\draw    (142.5,159) -- (194.78,84.46) ;
\draw [shift={(196.5,82)}, rotate = 485.04] [fill={rgb, 255:red, 0; green, 0; blue, 0 }  ][line width=0.08]  [draw opacity=0] (8.93,-4.29) -- (0,0) -- (8.93,4.29) -- cycle    ;
%Shape: Triangle [id:dp09144611596273] 
\draw  [line width=1.5]  (219.09,223.15) -- (278.5,198) -- (278.61,246.43) -- cycle ;
%Straight Lines [id:da40302773266796976] 
\draw [line width=1.5]    (132.99,188.13) -- (121.5,188.03) ;
\draw [shift={(117.5,188)}, rotate = 360.5] [fill={rgb, 255:red, 0; green, 0; blue, 0 }  ][line width=0.08]  [draw opacity=0] (11.61,-5.58) -- (0,0) -- (11.61,5.58) -- cycle    ;
%Shape: Ellipse [id:dp6744490028368978] 
\draw  [line width=1.5]  (94.5,186.3) .. controls (94.5,179.28) and (99.87,173.6) .. (106.5,173.6) .. controls (113.13,173.6) and (118.5,179.28) .. (118.5,186.3) .. controls (118.5,193.31) and (113.13,199) .. (106.5,199) .. controls (99.87,199) and (94.5,193.31) .. (94.5,186.3) -- cycle ;
%Straight Lines [id:da045329198187440634] 
\draw [line width=1.5]    (219.5,223) -- (106.5,223) -- (106.5,203) ;
\draw [shift={(106.5,199)}, rotate = 450] [fill={rgb, 255:red, 0; green, 0; blue, 0 }  ][line width=0.08]  [draw opacity=0] (11.61,-5.58) -- (0,0) -- (11.61,5.58) -- cycle    ;
%Straight Lines [id:da7672950068580145] 
\draw [line width=1.5]    (106.5,173.6) -- (106.41,140.14) ;
\draw [shift={(106.4,136.14)}, rotate = 449.85] [fill={rgb, 255:red, 0; green, 0; blue, 0 }  ][line width=0.08]  [draw opacity=0] (11.61,-5.58) -- (0,0) -- (11.61,5.58) -- cycle    ;
%Straight Lines [id:da09105117201000468] 
\draw [line width=1.5]    (360.5,124) -- (328.5,124) ;
\draw [shift={(364.5,124)}, rotate = 180] [fill={rgb, 255:red, 0; green, 0; blue, 0 }  ][line width=0.08]  [draw opacity=0] (11.61,-5.58) -- (0,0) -- (11.61,5.58) -- cycle    ;
%Straight Lines [id:da289353530454163] 
\draw [line width=1.5]    (342.5,123) -- (341.5,222) -- (282.5,221.06) ;
\draw [shift={(278.5,221)}, rotate = 360.90999999999997] [fill={rgb, 255:red, 0; green, 0; blue, 0 }  ][line width=0.08]  [draw opacity=0] (11.61,-5.58) -- (0,0) -- (11.61,5.58) -- cycle    ;
%Straight Lines [id:da03924130623164013] 
\draw    (278.5,156) -- (330.78,81.46) ;
\draw [shift={(332.5,79)}, rotate = 485.04] [fill={rgb, 255:red, 0; green, 0; blue, 0 }  ][line width=0.08]  [draw opacity=0] (8.93,-4.29) -- (0,0) -- (8.93,4.29) -- cycle    ;
%Straight Lines [id:da7137180224329931] 
\draw [line width=1.5]    (49.5,253) -- (49.53,138.22) ;
\draw [shift={(49.53,134.22)}, rotate = 450.01] [fill={rgb, 255:red, 0; green, 0; blue, 0 }  ][line width=0.08]  [draw opacity=0] (11.61,-5.58) -- (0,0) -- (11.61,5.58) -- cycle    ;
%Shape: Ellipse [id:dp40584224718490813] 
\draw  [line width=1.5]  (231.5,124.37) .. controls (231.5,116.76) and (238.17,110.6) .. (246.4,110.6) .. controls (254.63,110.6) and (261.3,116.76) .. (261.3,124.37) .. controls (261.3,131.97) and (254.63,138.14) .. (246.4,138.14) .. controls (238.17,138.14) and (231.5,131.97) .. (231.5,124.37) -- cycle ;
%Straight Lines [id:da2795771951052588] 
\draw [line width=1.5]    (468.77,125.37) -- (490,125.06) ;
\draw [shift={(494,125)}, rotate = 539.1600000000001] [fill={rgb, 255:red, 0; green, 0; blue, 0 }  ][line width=0.08]  [draw opacity=0] (11.61,-5.58) -- (0,0) -- (11.61,5.58) -- cycle    ;
%Straight Lines [id:da9370658968302499] 
\draw [line width=1.5]    (531.77,125.37) -- (552.5,125.06) ;
\draw [shift={(556.5,125)}, rotate = 539.15] [fill={rgb, 255:red, 0; green, 0; blue, 0 }  ][line width=0.08]  [draw opacity=0] (11.61,-5.58) -- (0,0) -- (11.61,5.58) -- cycle    ;
%Shape: Rectangle [id:dp419555887879832] 
\draw  [line width=1.5]  (554,101) -- (590.5,101) -- (590.5,146) -- (554,146) -- cycle ;
%Shape: Triangle [id:dp45645024657239697] 
\draw  [line width=1.5]  (322.58,29.91) -- (285.39,51.95) -- (285.57,6.53) -- cycle ;
%Straight Lines [id:da9892730504898035] 
\draw [line width=1.5]    (590.77,124.37) -- (605.5,124.08) ;
\draw [shift={(609.5,124)}, rotate = 538.87] [fill={rgb, 255:red, 0; green, 0; blue, 0 }  ][line width=0.08]  [draw opacity=0] (11.61,-5.58) -- (0,0) -- (11.61,5.58) -- cycle    ;
%Shape: Ellipse [id:dp334412232926359] 
\draw  [line width=1.5]  (659.5,124.37) .. controls (659.5,116.76) and (666.17,110.6) .. (674.4,110.6) .. controls (682.63,110.6) and (689.3,116.76) .. (689.3,124.37) .. controls (689.3,131.97) and (682.63,138.14) .. (674.4,138.14) .. controls (666.17,138.14) and (659.5,131.97) .. (659.5,124.37) -- cycle ;
%Straight Lines [id:da4706171268677113] 
\draw [line width=1.5]    (643.59,124.35) -- (656.5,124.08) ;
\draw [shift={(660.5,124)}, rotate = 538.8199999999999] [fill={rgb, 255:red, 0; green, 0; blue, 0 }  ][line width=0.08]  [draw opacity=0] (11.61,-5.58) -- (0,0) -- (11.61,5.58) -- cycle    ;
%Straight Lines [id:da24673206885357013] 
\draw [line width=1.5]    (246.5,67) -- (246.41,106.6) ;
\draw [shift={(246.4,110.6)}, rotate = 270.13] [fill={rgb, 255:red, 0; green, 0; blue, 0 }  ][line width=0.08]  [draw opacity=0] (11.61,-5.58) -- (0,0) -- (11.61,5.58) -- cycle    ;
%Straight Lines [id:da2155928673705345] 
\draw [line width=1.5]    (215.5,126) -- (215.5,28) -- (281.5,28.94) ;
\draw [shift={(285.5,29)}, rotate = 180.82] [fill={rgb, 255:red, 0; green, 0; blue, 0 }  ][line width=0.08]  [draw opacity=0] (11.61,-5.58) -- (0,0) -- (11.61,5.58) -- cycle    ;
%Straight Lines [id:da5753610592145828] 
\draw [line width=1.5]    (342.5,123) -- (342.5,72) -- (363.5,72) ;
\draw [shift={(367.5,72)}, rotate = 180] [fill={rgb, 255:red, 0; green, 0; blue, 0 }  ][line width=0.08]  [draw opacity=0] (11.61,-5.58) -- (0,0) -- (11.61,5.58) -- cycle    ;
%Straight Lines [id:da4921680215355613] 
\draw [line width=1.5]    (320.5,29) -- (438.5,29) -- (438.41,53.6) ;
\draw [shift={(438.4,57.6)}, rotate = 270.2] [fill={rgb, 255:red, 0; green, 0; blue, 0 }  ][line width=0.08]  [draw opacity=0] (11.61,-5.58) -- (0,0) -- (11.61,5.58) -- cycle    ;
%Shape: Ellipse [id:dp5734223791604907] 
\draw  [line width=1.5]  (423.5,71.37) .. controls (423.5,63.76) and (430.17,57.6) .. (438.4,57.6) .. controls (446.63,57.6) and (453.3,63.76) .. (453.3,71.37) .. controls (453.3,78.97) and (446.63,85.14) .. (438.4,85.14) .. controls (430.17,85.14) and (423.5,78.97) .. (423.5,71.37) -- cycle ;
%Straight Lines [id:da5899254947560695] 
\draw [line width=1.5]    (400.59,71.35) -- (420.5,71.06) ;
\draw [shift={(424.5,71)}, rotate = 539.1600000000001] [fill={rgb, 255:red, 0; green, 0; blue, 0 }  ][line width=0.08]  [draw opacity=0] (11.61,-5.58) -- (0,0) -- (11.61,5.58) -- cycle    ;
%Straight Lines [id:da7963312286707658] 
\draw [line width=1.5]    (453.3,71.37) -- (672.5,71) -- (672.41,106.6) ;
\draw [shift={(672.4,110.6)}, rotate = 270.15] [fill={rgb, 255:red, 0; green, 0; blue, 0 }  ][line width=0.08]  [draw opacity=0] (11.61,-5.58) -- (0,0) -- (11.61,5.58) -- cycle    ;
%Straight Lines [id:da6534119475527871] 
\draw [line width=1.5]    (689.3,124.37) -- (717.5,124.05) ;
\draw [shift={(721.5,124)}, rotate = 539.3399999999999] [fill={rgb, 255:red, 0; green, 0; blue, 0 }  ][line width=0.08]  [draw opacity=0] (11.61,-5.58) -- (0,0) -- (11.61,5.58) -- cycle    ;
%Shape: Rectangle [id:dp49027646238639555] 
\draw  [line width=1.5]  (429,103) -- (467.5,103) -- (467.5,147) -- (429,147) -- cycle ;
%Shape: Rectangle [id:dp1750775476177684] 
\draw  [line width=1.5]  (492,103) -- (530.5,103) -- (530.5,147) -- (492,147) -- cycle ;
%Shape: Triangle [id:dp5539449317965373] 
\draw  [line width=1.5]  (132.99,188.13) -- (198.45,162.97) -- (198.56,211.4) -- cycle ;
%Shape: Triangle [id:dp5525954247840159] 
\draw  [line width=1.5]  (402.58,70.91) -- (365.39,92.95) -- (365.57,47.53) -- cycle ;
%Shape: Triangle [id:dp9441099905583554] 
\draw  [line width=1.5]  (643.59,124.35) -- (606.39,146.39) -- (606.58,100.97) -- cycle ;

% Text Node
\draw (49.53,118.61) node  [font=\large,xscale=1.5,yscale=1.5]  {$-$};
% Text Node
\draw (18.84,100.76) node  [font=\large,xscale=1.5,yscale=1.5]  {$r_{1}$};
% Text Node
\draw (705.62,102.76) node  [font=\large,xscale=1.5,yscale=1.5]  {$y_{1}$};
% Text Node
\draw (75.35,104.76) node  [font=\large,xscale=1.5,yscale=1.5]  {$e$};
% Text Node
\draw (107.4,120.37) node  [font=\large,xscale=1.5,yscale=1.5]  {$-$};
% Text Node
\draw (167.34,123.04) node  [font=\large,xscale=1.5,yscale=1.5]  {$\dfrac{1}{s}$};
% Text Node
\draw (177,187) node  [font=\large,xscale=1.2,yscale=1.2]  {$2\omega _{r\alpha }$};
% Text Node
\draw (307.34,123.04) node  [font=\large,xscale=1.5,yscale=1.5]  {$\dfrac{1}{s}$};
% Text Node
\draw (184,70) node  [font=\large,xscale=1.5,yscale=1.5]  {$\gamma _{1}$};
% Text Node
\draw (260,221) node  [font=\large,xscale=1.5,yscale=1.5]  {$\omega ^{2}_{r\alpha }$};
% Text Node
\draw (106.5,185.3) node  [font=\large,xscale=1.5,yscale=1.5]  {$+$};
% Text Node
\draw (383.25,123) node  [font=\large,xscale=1.5,yscale=1.5]  {$C_{L_{2}}$};
% Text Node
\draw (448.25,125) node  [font=\large,xscale=1.5,yscale=1.5]  {$G$};
% Text Node
\draw (247.4,122.37) node  [font=\large,xscale=1.5,yscale=1.5]  {$+$};
% Text Node
\draw (511.25,125) node  [font=\large,xscale=1.5,yscale=1.5]  {$C_{L_{1}}$};
% Text Node
\draw (576.25,123.5) node  [font=\large,xscale=1.5,yscale=1.5]  {$\mathcal{C}_{s}$};
% Text Node
\draw (299.25,25) node  [font=\large,xscale=1.5,yscale=1.5]  {$\Varrho _{1}$};
% Text Node
\draw (621.25,121) node  [font=\large,xscale=1.5,yscale=1.5]  {$\beta _{1}$};
% Text Node
\draw (675.4,122.37) node  [font=\large,xscale=1.5,yscale=1.5]  {$+$};
% Text Node
\draw (259.84,69.76) node  [font=\large,xscale=1.5,yscale=1.5]  {$r_{2}$};
% Text Node
\draw (439.4,69.37) node  [font=\large,xscale=1.5,yscale=1.5]  {$+$};
% Text Node
\draw (322,67) node  [font=\large,xscale=1.5,yscale=1.5]  {$\gamma _{2}$};
% Text Node
\draw (379.25,66) node  [font=\large,xscale=1.5,yscale=1.5]  {$\Varrho _{2}$};
\end{tikzpicture}}
  \end{subfigure}
\begin{subfigure}{0.485\textwidth}
    \centering
\resizebox{\textwidth}{!}{   
\tikzset{every picture/.style={line width=0.75pt}} %set default line width to 0.75pt        
\begin{tikzpicture}[x=0.75pt,y=0.75pt,yscale=-1,xscale=1]
%uncomment if require: \path (0,266); %set diagram left start at 0, and has height of 266
%Shape: Rectangle [id:dp3558696855554735] 
\draw  [line width=1.5]  (363,102) -- (401.5,102) -- (401.5,146) -- (363,146) -- cycle ;
%Shape: Ellipse [id:dp605355155007869] 
\draw  [line width=1.5]  (34.63,120.45) .. controls (34.63,112.84) and (41.3,106.68) .. (49.53,106.68) .. controls (57.76,106.68) and (64.43,112.84) .. (64.43,120.45) .. controls (64.43,128.06) and (57.76,134.22) .. (49.53,134.22) .. controls (41.3,134.22) and (34.63,128.06) .. (34.63,120.45) -- cycle ;
%Straight Lines [id:da8035539331450979] 
\draw [line width=1.5]    (538.5,125) -- (538.5,255) -- (49.5,253) ;
%Straight Lines [id:da10064598351922416] 
\draw [line width=1.5]    (2.5,120) -- (30.63,120.39) ;
\draw [shift={(34.63,120.45)}, rotate = 180.8] [fill={rgb, 255:red, 0; green, 0; blue, 0 }  ][line width=0.08]  [draw opacity=0] (11.61,-5.58) -- (0,0) -- (11.61,5.58) -- cycle    ;
%Straight Lines [id:da053830610313541416] 
\draw [line width=1.5]    (403.77,124.37) -- (425,124.06) ;
\draw [shift={(429,124)}, rotate = 539.1600000000001] [fill={rgb, 255:red, 0; green, 0; blue, 0 }  ][line width=0.08]  [draw opacity=0] (11.61,-5.58) -- (0,0) -- (11.61,5.58) -- cycle    ;
%Straight Lines [id:da3228273022349576] 
\draw [line width=1.5]    (261.3,124.37) -- (284.5,124.05) ;
\draw [shift={(288.5,124)}, rotate = 539.22] [fill={rgb, 255:red, 0; green, 0; blue, 0 }  ][line width=0.08]  [draw opacity=0] (11.61,-5.58) -- (0,0) -- (11.61,5.58) -- cycle    ;
%Straight Lines [id:da344361302290097] 
\draw [line width=1.5]    (215.5,126) -- (214.5,187) -- (199.5,187) ;
%Straight Lines [id:da33022400931887896] 
\draw [line width=1.5]    (64.43,120.45) -- (87.5,120.38) ;
\draw [shift={(91.5,120.37)}, rotate = 539.8299999999999] [fill={rgb, 255:red, 0; green, 0; blue, 0 }  ][line width=0.08]  [draw opacity=0] (11.61,-5.58) -- (0,0) -- (11.61,5.58) -- cycle    ;
%Shape: Ellipse [id:dp9293274033414575] 
\draw  [line width=1.5]  (91.5,122.37) .. controls (91.5,114.76) and (98.17,108.6) .. (106.4,108.6) .. controls (114.63,108.6) and (121.3,114.76) .. (121.3,122.37) .. controls (121.3,129.97) and (114.63,136.14) .. (106.4,136.14) .. controls (98.17,136.14) and (91.5,129.97) .. (91.5,122.37) -- cycle ;
%Shape: Rectangle [id:dp8281312117405866] 
\draw  [line width=1.5]  (148.17,94.09) -- (186.5,94.09) -- (186.5,152) -- (148.17,152) -- cycle ;
%Straight Lines [id:da6736196050401356] 
\draw [line width=1.5]    (121.5,123) -- (144.5,123) ;
\draw [shift={(148.5,123)}, rotate = 180] [fill={rgb, 255:red, 0; green, 0; blue, 0 }  ][line width=0.08]  [draw opacity=0] (11.61,-5.58) -- (0,0) -- (11.61,5.58) -- cycle    ;
%Straight Lines [id:da3981723043709309] 
\draw [line width=1.5]    (188.5,125) -- (214.49,125.28) -- (229.5,125.06) ;
\draw [shift={(233.5,125)}, rotate = 539.1600000000001] [fill={rgb, 255:red, 0; green, 0; blue, 0 }  ][line width=0.08]  [draw opacity=0] (11.61,-5.58) -- (0,0) -- (11.61,5.58) -- cycle    ;
%Shape: Rectangle [id:dp5499525015603304] 
\draw  [line width=1.5]  (288.17,94.09) -- (326.5,94.09) -- (326.5,152) -- (288.17,152) -- cycle ;
%Straight Lines [id:da6559801315350425] 
\draw    (142.5,159) -- (194.78,84.46) ;
\draw [shift={(196.5,82)}, rotate = 485.04] [fill={rgb, 255:red, 0; green, 0; blue, 0 }  ][line width=0.08]  [draw opacity=0] (8.93,-4.29) -- (0,0) -- (8.93,4.29) -- cycle    ;
%Shape: Triangle [id:dp09144611596273] 
\draw  [line width=1.5]  (219.09,223.15) -- (278.5,198) -- (278.61,246.43) -- cycle ;
%Straight Lines [id:da40302773266796976] 
\draw [line width=1.5]    (132.99,188.13) -- (121.5,188.03) ;
\draw [shift={(117.5,188)}, rotate = 360.5] [fill={rgb, 255:red, 0; green, 0; blue, 0 }  ][line width=0.08]  [draw opacity=0] (11.61,-5.58) -- (0,0) -- (11.61,5.58) -- cycle    ;
%Shape: Ellipse [id:dp6744490028368978] 
\draw  [line width=1.5]  (94.5,186.3) .. controls (94.5,179.28) and (99.87,173.6) .. (106.5,173.6) .. controls (113.13,173.6) and (118.5,179.28) .. (118.5,186.3) .. controls (118.5,193.31) and (113.13,199) .. (106.5,199) .. controls (99.87,199) and (94.5,193.31) .. (94.5,186.3) -- cycle ;
%Straight Lines [id:da045329198187440634] 
\draw [line width=1.5]    (219.5,223) -- (106.5,223) -- (106.5,203) ;
\draw [shift={(106.5,199)}, rotate = 450] [fill={rgb, 255:red, 0; green, 0; blue, 0 }  ][line width=0.08]  [draw opacity=0] (11.61,-5.58) -- (0,0) -- (11.61,5.58) -- cycle    ;
%Straight Lines [id:da7672950068580145] 
\draw [line width=1.5]    (106.5,173.6) -- (106.41,140.14) ;
\draw [shift={(106.4,136.14)}, rotate = 449.85] [fill={rgb, 255:red, 0; green, 0; blue, 0 }  ][line width=0.08]  [draw opacity=0] (11.61,-5.58) -- (0,0) -- (11.61,5.58) -- cycle    ;
%Straight Lines [id:da09105117201000468] 
\draw [line width=1.5]    (360.5,124) -- (328.5,124) ;
\draw [shift={(364.5,124)}, rotate = 180] [fill={rgb, 255:red, 0; green, 0; blue, 0 }  ][line width=0.08]  [draw opacity=0] (11.61,-5.58) -- (0,0) -- (11.61,5.58) -- cycle    ;
%Straight Lines [id:da289353530454163] 
\draw [line width=1.5]    (342.5,123) -- (341.5,222) -- (282.5,221.06) ;
\draw [shift={(278.5,221)}, rotate = 360.90999999999997] [fill={rgb, 255:red, 0; green, 0; blue, 0 }  ][line width=0.08]  [draw opacity=0] (11.61,-5.58) -- (0,0) -- (11.61,5.58) -- cycle    ;
%Straight Lines [id:da03924130623164013] 
\draw    (278.5,156) -- (330.78,81.46) ;
\draw [shift={(332.5,79)}, rotate = 485.04] [fill={rgb, 255:red, 0; green, 0; blue, 0 }  ][line width=0.08]  [draw opacity=0] (8.93,-4.29) -- (0,0) -- (8.93,4.29) -- cycle    ;
%Straight Lines [id:da7137180224329931] 
\draw [line width=1.5]    (49.5,253) -- (49.53,138.22) ;
\draw [shift={(49.53,134.22)}, rotate = 450.01] [fill={rgb, 255:red, 0; green, 0; blue, 0 }  ][line width=0.08]  [draw opacity=0] (11.61,-5.58) -- (0,0) -- (11.61,5.58) -- cycle    ;
%Shape: Ellipse [id:dp40584224718490813] 
\draw  [line width=1.5]  (231.5,124.37) .. controls (231.5,116.76) and (238.17,110.6) .. (246.4,110.6) .. controls (254.63,110.6) and (261.3,116.76) .. (261.3,124.37) .. controls (261.3,131.97) and (254.63,138.14) .. (246.4,138.14) .. controls (238.17,138.14) and (231.5,131.97) .. (231.5,124.37) -- cycle ;
%Straight Lines [id:da2795771951052588] 
\draw [line width=1.5]    (468.77,125.37) -- (490,125.06) ;
\draw [shift={(494,125)}, rotate = 539.1600000000001] [fill={rgb, 255:red, 0; green, 0; blue, 0 }  ][line width=0.08]  [draw opacity=0] (11.61,-5.58) -- (0,0) -- (11.61,5.58) -- cycle    ;
%Straight Lines [id:da9370658968302499] 
\draw [line width=1.5]    (531.77,125.37) -- (552.5,125.06) ;
\draw [shift={(556.5,125)}, rotate = 539.15] [fill={rgb, 255:red, 0; green, 0; blue, 0 }  ][line width=0.08]  [draw opacity=0] (11.61,-5.58) -- (0,0) -- (11.61,5.58) -- cycle    ;
%Shape: Rectangle [id:dp419555887879832] 
\draw  [line width=1.5]  (554,101) -- (590.5,101) -- (590.5,146) -- (554,146) -- cycle ;
%Shape: Triangle [id:dp45645024657239697] 
\draw  [line width=1.5]  (322.58,29.91) -- (285.39,51.95) -- (285.57,6.53) -- cycle ;
%Straight Lines [id:da9892730504898035] 
\draw [line width=1.5]    (590.77,124.37) -- (605.5,124.08) ;
\draw [shift={(609.5,124)}, rotate = 538.87] [fill={rgb, 255:red, 0; green, 0; blue, 0 }  ][line width=0.08]  [draw opacity=0] (11.61,-5.58) -- (0,0) -- (11.61,5.58) -- cycle    ;
%Shape: Ellipse [id:dp334412232926359] 
\draw  [line width=1.5]  (659.5,124.37) .. controls (659.5,116.76) and (666.17,110.6) .. (674.4,110.6) .. controls (682.63,110.6) and (689.3,116.76) .. (689.3,124.37) .. controls (689.3,131.97) and (682.63,138.14) .. (674.4,138.14) .. controls (666.17,138.14) and (659.5,131.97) .. (659.5,124.37) -- cycle ;
%Straight Lines [id:da4706171268677113] 
\draw [line width=1.5]    (643.59,124.35) -- (656.5,124.08) ;
\draw [shift={(660.5,124)}, rotate = 538.8199999999999] [fill={rgb, 255:red, 0; green, 0; blue, 0 }  ][line width=0.08]  [draw opacity=0] (11.61,-5.58) -- (0,0) -- (11.61,5.58) -- cycle    ;
%Straight Lines [id:da24673206885357013] 
\draw [line width=1.5]    (246.5,67) -- (246.41,106.6) ;
\draw [shift={(246.4,110.6)}, rotate = 270.13] [fill={rgb, 255:red, 0; green, 0; blue, 0 }  ][line width=0.08]  [draw opacity=0] (11.61,-5.58) -- (0,0) -- (11.61,5.58) -- cycle    ;
%Straight Lines [id:da2155928673705345] 
\draw [line width=1.5]    (215.5,126) -- (215.5,28) -- (281.5,28.94) ;
\draw [shift={(285.5,29)}, rotate = 180.82] [fill={rgb, 255:red, 0; green, 0; blue, 0 }  ][line width=0.08]  [draw opacity=0] (11.61,-5.58) -- (0,0) -- (11.61,5.58) -- cycle    ;
%Straight Lines [id:da5753610592145828] 
\draw [line width=1.5]    (342.5,123) -- (342.5,72) -- (363.5,72) ;
\draw [shift={(367.5,72)}, rotate = 180] [fill={rgb, 255:red, 0; green, 0; blue, 0 }  ][line width=0.08]  [draw opacity=0] (11.61,-5.58) -- (0,0) -- (11.61,5.58) -- cycle    ;
%Straight Lines [id:da4921680215355613] 
\draw [line width=1.5]    (320.5,29) -- (438.5,29) -- (438.41,53.6) ;
\draw [shift={(438.4,57.6)}, rotate = 270.2] [fill={rgb, 255:red, 0; green, 0; blue, 0 }  ][line width=0.08]  [draw opacity=0] (11.61,-5.58) -- (0,0) -- (11.61,5.58) -- cycle    ;
%Shape: Ellipse [id:dp5734223791604907] 
\draw  [line width=1.5]  (423.5,71.37) .. controls (423.5,63.76) and (430.17,57.6) .. (438.4,57.6) .. controls (446.63,57.6) and (453.3,63.76) .. (453.3,71.37) .. controls (453.3,78.97) and (446.63,85.14) .. (438.4,85.14) .. controls (430.17,85.14) and (423.5,78.97) .. (423.5,71.37) -- cycle ;
%Straight Lines [id:da5899254947560695] 
\draw [line width=1.5]    (400.59,71.35) -- (420.5,71.06) ;
\draw [shift={(424.5,71)}, rotate = 539.1600000000001] [fill={rgb, 255:red, 0; green, 0; blue, 0 }  ][line width=0.08]  [draw opacity=0] (11.61,-5.58) -- (0,0) -- (11.61,5.58) -- cycle    ;
%Straight Lines [id:da7963312286707658] 
\draw [line width=1.5]    (453.3,71.37) -- (672.5,71) -- (672.41,106.6) ;
\draw [shift={(672.4,110.6)}, rotate = 270.15] [fill={rgb, 255:red, 0; green, 0; blue, 0 }  ][line width=0.08]  [draw opacity=0] (11.61,-5.58) -- (0,0) -- (11.61,5.58) -- cycle    ;
%Straight Lines [id:da6534119475527871] 
\draw [line width=1.5]    (689.3,124.37) -- (717.5,124.05) ;
\draw [shift={(721.5,124)}, rotate = 539.3399999999999] [fill={rgb, 255:red, 0; green, 0; blue, 0 }  ][line width=0.08]  [draw opacity=0] (11.61,-5.58) -- (0,0) -- (11.61,5.58) -- cycle    ;
%Shape: Rectangle [id:dp49027646238639555] 
\draw  [line width=1.5]  (429,103) -- (467.5,103) -- (467.5,147) -- (429,147) -- cycle ;
%Shape: Rectangle [id:dp1750775476177684] 
\draw  [line width=1.5]  (492,103) -- (530.5,103) -- (530.5,147) -- (492,147) -- cycle ;
%Shape: Triangle [id:dp5539449317965373] 
\draw  [line width=1.5]  (132.99,188.13) -- (198.45,162.97) -- (198.56,211.4) -- cycle ;
%Shape: Triangle [id:dp5525954247840159] 
\draw  [line width=1.5]  (402.58,70.91) -- (365.39,92.95) -- (365.57,47.53) -- cycle ;
%Shape: Triangle [id:dp9441099905583554] 
\draw  [line width=1.5]  (643.59,124.35) -- (606.39,146.39) -- (606.58,100.97) -- cycle ;

% Text Node
\draw (49.53,118.61) node  [font=\large,xscale=1.5,yscale=1.5]  {$-$};
% Text Node
\draw (18.84,100.76) node  [font=\large,xscale=1.5,yscale=1.5]  {$r_{1}$};
% Text Node
\draw (705.62,102.76) node  [font=\large,xscale=1.5,yscale=1.5]  {$y_{2}$};
% Text Node
\draw (75.35,104.76) node  [font=\large,xscale=1.5,yscale=1.5]  {$e$};
% Text Node
\draw (107.4,120.37) node  [font=\large,xscale=1.5,yscale=1.5]  {$-$};
% Text Node
\draw (167.34,123.04) node  [font=\large,xscale=1.5,yscale=1.5]  {$\dfrac{1}{s}$};
% Text Node
\draw (177,187) node  [font=\large,xscale=1.2,yscale=1.2]  {$2\omega _{r\alpha }$};
% Text Node
\draw (307.34,123.04) node  [font=\large,xscale=1.5,yscale=1.5]  {$\dfrac{1}{s}$};
% Text Node
\draw (184,70) node  [font=\large,xscale=1.5,yscale=1.5]  {$\gamma _{1}$};
% Text Node
\draw (260,221) node  [font=\large,xscale=1.5,yscale=1.5]  {$\omega ^{2}_{r\alpha }$};
% Text Node
\draw (106.5,185.3) node  [font=\large,xscale=1.5,yscale=1.5]  {$+$};
% Text Node
\draw (383.25,123) node  [font=\large,xscale=1.5,yscale=1.5]  {$C_{L_{2}}$};
% Text Node
\draw (448.25,125) node  [font=\large,xscale=1.5,yscale=1.5]  {$G$};
% Text Node
\draw (247.4,122.37) node  [font=\large,xscale=1.5,yscale=1.5]  {$+$};
% Text Node
\draw (511.25,125) node  [font=\large,xscale=1.5,yscale=1.5]  {$C_{L_{1}}$};
% Text Node
\draw (576.25,123.5) node  [font=\large,xscale=1.5,yscale=1.5]  {$\mathcal{C}_{s}$};
% Text Node
\draw (299.25,25) node  [font=\large,xscale=1.5,yscale=1.5]  {$\Varrho _{2}$};
% Text Node
\draw (621.25,121) node  [font=\large,xscale=1.5,yscale=1.5]  {$\beta _{2}$};
% Text Node
\draw (675.4,122.37) node  [font=\large,xscale=1.5,yscale=1.5]  {$+$};
% Text Node
\draw (259.84,69.76) node  [font=\large,xscale=1.5,yscale=1.5]  {$r_{2}$};
% Text Node
\draw (439.4,69.37) node  [font=\large,xscale=1.5,yscale=1.5]  {$+$};
% Text Node
\draw (322,67) node  [font=\large,xscale=1.5,yscale=1.5]  {$\gamma _{2}$};
% Text Node
\draw (379.25,66) node  [font=\large,xscale=1.5,yscale=1.5]  {$\Varrho _{3}$};
\end{tikzpicture}}
  \end{subfigure}
  \caption{The block diagram of the $H_\beta$ condition for the closed-loop architecture Fig.~\ref{F_4-01} with GSORE}  
  \label{F_4-06}
\end{figure}
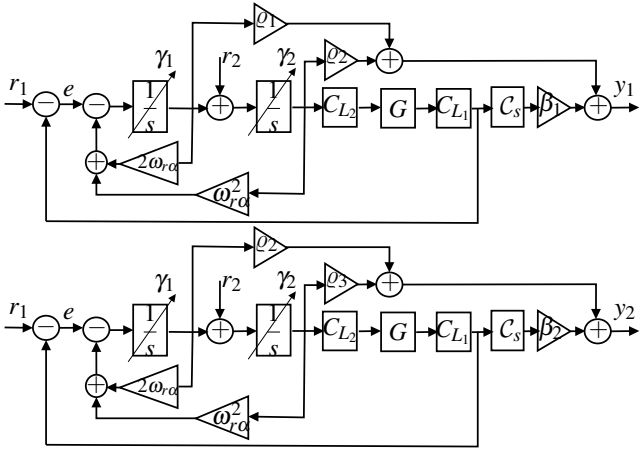 
in which $H_{ij}(s)$ with $i,j=1,2$ is transfer function from $r_j$ to $y_i$. Thus, $H(j\omega)+H(-j\omega)^T$ is equal to
\begin{equation}
\begin{array}{*{35}{c}}
\begin{bmatrix}
	2\Re(H_{11}(j\omega)) & \Re(H_{12}(j\omega)+H_{21}(j\omega)) \\ \Re(H_{12}(j\omega)+H_{21}(j\omega)) & 2\Re(H_{22}(j\omega))
	\end{bmatrix}>0\Rightarrow
\end{array}
\end{equation}
\begin{multline}\label{E_433-05}
	\dfrac{1}{|\kappa(\omega)|^2}\left[
	\begin{matrix}
	2f_1(\Varrho_1,\Varrho_2,\beta_1,\omega)  \\ f_1(\Varrho_2,\Varrho_1,\beta_2,\omega)+f_2(\Varrho_2,\Varrho_1,\beta_2,\omega)  
	\end{matrix}\right.
	\\
\left.
\begin{matrix}
 f_1(\Varrho_2,\Varrho_1,\beta_2,\omega)+f_2(\Varrho_2,\Varrho_1,\beta_2,\omega)\\
2f_2(\Varrho_3,\Varrho_2,\beta_2,\omega)
\end{matrix}\right]>0.
\end{multline}
Step 2: Since the transfer functions $\dfrac{y_i}{r_j}$, with $i,j=1,2$, are strictly proper, $\displaystyle\lim_{s\to \infty}H(s)=0$. Therefore, it is necessary to have $\displaystyle\lim_{\omega\to \infty} \omega^2(H(j\omega)+H(-j\omega)^T)>0.$ Note that in the case of SORE, $n-m\geq3$. By~(\ref{E_433-05}), if $n-m>3$, $\displaystyle\lim_{\omega\to \infty} \omega^2(H(j\omega)+H(-j\omega)^T)$ is equal to
\begin{equation}\label{E_433-06}
\begin{bmatrix}
4\Varrho_1\xi\omega_r-2\Varrho_2 & \omega_r^2\Varrho_1+2\Varrho_2\xi\omega_r-\Varrho_3 \\  \omega_r^2\Varrho_1+2\Varrho_2\xi\omega_r-\Varrho_3  &  2\omega_r^2\Varrho_2
\end{bmatrix}.
\end{equation}
Therefore, the condition $\displaystyle\lim_{\omega\to \infty} \omega^2(H(j\omega)+H(-j\omega)^T)>0$ is equivalent to
\begin{equation}\label{E_433-067}
2\Varrho_1\xi\omega_r>\Varrho_2,\quad \Varrho_2>0,
\end{equation}
and
\begin{equation}\label{E_433-07}\resizebox{\columnwidth}{!}{$
\begin{array}{*{35}{c}}
4(2\Varrho_1\xi\omega_r-\Varrho_2)(\omega_r^2\Varrho_2)>(\omega_r^2\Varrho_1+2\Varrho_2\xi\omega_r-\Varrho_3)^2\\ \Downarrow\\
\Varrho_3^2-2\Varrho_3(\omega_r^2\Varrho_1+2\xi\omega_r\Varrho_2)+(\omega_r^2\Varrho_1-2\xi\omega_r\Varrho_2)^2+4\omega_r^2\Varrho_2^2<0\\ \Downarrow\\
\left(\Varrho_3>(\omega_r^2\Varrho_1+2\xi\omega_r\Varrho_2)-2\omega_r\sqrt{2\xi\omega_r\Varrho_1\Varrho_2-\Varrho_2^2}\right)\ \land\\
\left(\Varrho_3<(\omega_r^2\Varrho_1+2\xi\omega_r\Varrho_2)+2\omega_r\sqrt{2\xi\omega_r\Varrho_1\Varrho_2-\Varrho_2^2}\right).
\end{array}$}
\end{equation}
When $n-m=3$, condition (\ref{E_433-06}) is re-written as
\begin{equation}\label{E_433-08}\resizebox{\columnwidth}{!}{$
\begin{bmatrix}
4\Varrho_1\xi\omega_r-2\Varrho_2 & \omega_r^2\Varrho_1+2\Varrho_2\xi\omega_r-\Varrho_3-K_n\beta_1 \\  \omega_r^2\Varrho_1+2\Varrho_2\xi\omega_r-\Varrho_3-K_n\beta_1  &  2\omega_r^2\Varrho_2-2K_n\beta_2
\end{bmatrix}>0,$}
\end{equation}
which is equivalent to
\begin{equation}\label{E_433-098}
2\Varrho_1\xi\omega_r>\Varrho_2,\quad \omega_r^2\Varrho_2>K_n\beta_2,
\end{equation}
and
\begin{equation}\label{E_433-09}\resizebox{\columnwidth}{!}{$
\begin{array}{*{35}{c}}
4(2\Varrho_1\xi\omega_r-\Varrho_2)(\omega_r^2\Varrho_2-K_n\beta_2)>(\omega_r^2\Varrho_1+2\Varrho_2\xi\omega_r-\Varrho_3-K_n\beta_1)^2\\ \Downarrow\\
%\Varrho_3^2-2\Varrho_3(\omega_r^2\Varrho_1-K_n\beta_1+2\xi\omega_r\Varrho_2)+(\omega_r^2\Varrho_1-K_n\beta_1+2\xi\omega_r\Varrho_2)^2-8\xi\omega_r^2\Varrho_1\Varrho_2+4\omega_r^2\Varrho_2^2-4K_n\Varrho_2\beta_2<0\\
\left(\Varrho_3>(\omega_r^2\Varrho_1-K_n\beta_1+2\xi\omega_r\Varrho_2)-2\sqrt{2\xi\omega_r^3\Varrho_1\Varrho_2+K_n\Varrho_2\beta_2-\omega_r^2\Varrho_2^2}\right)\ \land\\
\left(\Varrho_3<(\omega_r^2\Varrho_1-K_n\beta_1+2\xi\omega_r\Varrho_2)+2\sqrt{2\xi\omega_r^3\Varrho_1\Varrho_2+K_n\Varrho_2\beta_2-\omega_r^2\Varrho_2^2}\right)\ \land\\
\left(2\xi\omega_r^3\Varrho_1\Varrho_2+K_n\Varrho_2\beta_2>\omega_r^2\Varrho_2^2\right).
\end{array}$}
\end{equation}
Step 3: When $L(s)$ has at least one pole at the origin, by~(\ref{E_433-05}) $displaystyle\lim_{\omega\to 0} H(j\omega)+H(-j\omega)^T$ is equal to 
\begin{equation}\label{E_433-10}%\resizebox{\columnwidth}{!}{$
\begin{bmatrix}
2K_{s_0}\beta_1 & K_{s_0}\beta_2+2K_{s_0}\beta_1\xi\omega_r-\Varrho_1 \\  K_{s_0}\beta_2+2K_{s_0}\beta_1\xi\omega_r-\Varrho_1  &  4K_{s_0}\beta_2\xi\omega_r-2\Varrho_2
\end{bmatrix}>0,%$}
\end{equation}
which is equivalent to 
\begin{equation}\label{E_433-1145}
K_{s_0}\beta_1>0,\quad 2K_{s_0}\beta_2\xi\omega_r>\Varrho_2,
\end{equation}
and
\begin{equation}\label{E_433-11}\resizebox{\columnwidth}{!}{$
\begin{array}{*{35}{c}}
4(K_{s_0}\beta_1)(2K_{s_0}\beta_2\xi\omega_r-\Varrho_2)>(K_{s_0}\beta_2+2K_{s_0}\beta_1\xi\omega_r-\Varrho_1)^2\\ \Downarrow\\
%\Varrho_1^2-2\Varrho_1(2K_{s_0}\beta_1\xi\omega_r+K_{s_0}\beta_2)+(2K_{s_0}\beta_1\xi\omega_r-K_{s_0}\beta_2)^2+4K_{s_0}\beta_1\Varrho_2<0\\ \Downarrow\\
\left(\Varrho_1>K_{s_0}(2\beta_1\xi\omega_r+\beta_2)-2\sqrt{2K_{s_0}^2\xi\omega_r\beta_1\beta_2-K_{s_0}\beta_1\Varrho_2}\right)\ \land\\
\left(\Varrho_1<K_{s_0}(2\beta_1\xi\omega_r+\beta_2)+2\sqrt{2K_{s_0}^2\xi\omega_r\beta_1\beta_2-K_{s_0}\beta_1\Varrho_2}\right).
\end{array}$}
\end{equation}
Step 4: In the case in which $L(s)$ has poles at the origin, denote $Q_1=\dfrac{\Varrho_1}{\beta_1}$, $Q_2=\dfrac{\Varrho_2}{\beta_1}$, $Q_3=\dfrac{\Varrho_3}{\beta_2}$ and $Q_4=\dfrac{\Varrho_2}{\beta_2}$. Furthermore, since $K_{s_0}\beta_1$, $K_{s_0}\beta_2$, and $|\kappa(\omega)|^2$ are positive, condition (\ref{E_433-05}) is equal to
\begin{multline}\label{E_433-12}
%\resizebox{\columnwidth}{!}{$
\left[
\begin{matrix}
2K_{s_0}f_1(Q_1,Q_2,1,\omega)   \\ f_1(Q_4,Q_3,1,\omega)+f_2(Q_4,\dfrac{Q_1Q_4}{Q_2},\dfrac{Q_4}{Q_2},\omega) 
\end{matrix}\right.\\
%$}
\left.
\begin{matrix}
f_1(Q_2,\dfrac{Q_2Q_3}{Q_4},\dfrac{Q_2}{Q_4},\omega)+f_2(Q_2,Q_1,1,\omega)\\
 \dfrac{2}{K_{s_0}}f_2(Q_3,Q_4,1,\omega)
\end{matrix}\right]
>0.
\end{multline}
Therefore, for all $\omega\in(0,\infty)$, there exist $Q_1,\ Q_2,\ Q_3$, and $Q_4$ such that
\begin{equation}\label{E_433-139}
K_{s_0}f_1(Q_1,Q_2,1,\omega)>0,\ K_{s_0}f_2(Q_3,Q_4,1,\omega)>0,
\end{equation}
and since $f_1(Q_1,Q_2,1,\omega)f_2(Q_3,Q_4,1,\omega)>0$,
\setlength{\arraycolsep}{0.0em}
\begin{eqnarray}\label{E_433-13}
4&{>}&\dfrac{f_1(Q_2,\dfrac{Q_2Q_3}{Q_4},\dfrac{Q_2}{Q_4},\omega)+f_2(Q_2,Q_1,1,\omega)}{f_1(Q_1,Q_2,1,\omega)}\nonumber\\
&&{\times}\dfrac{f_1(Q_4,Q_3,1,\omega)+f_2(Q_4,\dfrac{Q_1Q_4}{Q_2},\dfrac{Q_4}{Q_2},\omega)}{f_2(Q_3,Q_4,1,\omega)}.\\
\end{eqnarray}
\setlength{\arraycolsep}{5pt}Thus, since the condition~(\ref{E_433-13}) must hold for all $\omega\in(0,\infty)$, $\underset{Q_i}{\min}\ G_1(Q_1,Q_2,Q_3,Q_4)<4$, with $i=1,2,3,4$. Moreover, re-writing equations~(\ref{E_433-0444}), (\ref{E_433-16}) (\ref{E_433-07}), and (\ref{E_433-11}) using the variables $Q_1,\ Q_2,\ Q_3$, and $Q_4$, the constraints $\mathcal{S}_3-\mathcal{S}_8$ of Definition \ref{D_44} are obtained.
\newline When $L(s)$ does not have any pole at the origin, let $Q^\prime_1=\dfrac{\beta_1}{\Varrho_1}$, $Q^\prime_2=\dfrac{\Varrho_2}{\Varrho_1}$, $Q^\prime_3=\dfrac{\beta_2}{\Varrho_3}$ and $Q^\prime_4=\dfrac{\Varrho_2}{\Varrho_3}$. With this change of variables, since $\Varrho_3$, $\Varrho_1$ and $|\kappa(\omega)|^2$ are positive, condition (\ref{E_433-05}) is equivalent to
\begin{multline}\label{E_433-14}
%\resizebox{\columnwidth}{!}{$
\left[
\begin{matrix}
2f_1(1,Q^\prime_2,Q^\prime_1,\omega)  \\  f_1(Q^\prime_4,1,Q^\prime_3,\omega)+f_2(Q^\prime_4,\dfrac{Q^\prime_4}{Q^\prime_2},\dfrac{Q^\prime_1Q^\prime_4}{Q^\prime_2},\omega)
\end{matrix}\right.\\
%$}
\left.
\begin{matrix}
f_1(Q^\prime_2,\dfrac{Q^\prime_2}{Q^\prime_4},\dfrac{Q^\prime_2Q^\prime_3}{Q^\prime_4},\omega)+f_2(Q^\prime_2,1,Q^\prime_1,\omega)\\
2f_2(1,Q^\prime_4,Q^\prime_3,\omega)
\end{matrix}\right]
>0.
\end{multline}
This implies that $
f_1(1,Q^\prime_2,Q^\prime_1,\omega)>0,\ f_2(1,Q^\prime_4,Q^\prime_3,\omega)>0,$ and since $f_1(1,Q^\prime_2,Q^\prime_1,\omega)f_2(1,Q^\prime_4,Q^\prime_3,\omega)>0$,
\setlength{\arraycolsep}{0.0em}
\begin{eqnarray}\label{E_433-15}
4&{>}&\dfrac{f_1(Q^\prime_2,\dfrac{Q^\prime_2}{Q^\prime_4},\dfrac{Q^\prime_2Q^\prime_3}{Q^\prime_4},\omega)+f_2(Q^\prime_2,1,Q^\prime_1,\omega)}{f_1(1,Q^\prime_2,Q^\prime_1,\omega)}\nonumber\\
&&{\times}\dfrac{f_1(Q^\prime_4,1,Q^\prime_3,\omega)+f_2(Q^\prime_4,\dfrac{Q^\prime_4}{Q^\prime_2},\dfrac{Q^\prime_1Q^\prime_4}{Q^\prime_2},\omega)}{f_2(1,Q^\prime_4,Q^\prime_3,\omega)}.\\
\end{eqnarray}
\setlength{\arraycolsep}{5pt}Therefore, since condition~(\ref{E_433-15}) must hold for all $\omega\in[0,\infty)$, $\underset{Q^\prime_i}{\min}\ G_2(Q^\prime_1,Q^\prime_2,Q^\prime_3,Q^\prime_4)<4$, with $i=1,2,3,4$. Re-writing equations~(\ref{E_433-16}) and (\ref{E_433-07}) with the variables $Q^\prime_1$, $Q^\prime_2$, $Q^\prime_3$, and $Q^\prime_4$, the constraints $\mathcal{S}_3-\mathcal{S}_5$ of Definition \ref{D_45} are achieved. Similarly, using these variables in equations~(\ref{E_433-16}) and (\ref{E_433-09}), the constraints $\mathcal{S}_3-\mathcal{S}_6$ of Definition \ref{D_46} are obtained. 
\newline By Steps 1-4, $A_\rho^T\Varrho A_\rho-\Varrho<0$, $H(s)$ is SPR~\cite{khalil2002nonlinear}, $(\bar{A},C_0)$ is observable and $(\bar{A},B_0)$ is controllable, and the base linear system is stable. Thus, the $H_\beta$ condition is satisfied for the reset control system~(\ref{E_4-27}) with GSORE (\ref{E_4-25}). Hence, the zero equilibrium of the system is globally uniformly asymptotically stable when $w=0$ and according to Lemma \ref{L_41}, it has the UBIBS property for any initial condition $x_0$ and any input $w$ which is a Bohl function. 
\end{pf}
\begin{remark}\label{R_44}
{\rm The minimum value of the function $\Gamma(\gamma_1,\gamma_2)$ occurs when $\gamma_1=\gamma_2$. In other words, if $\gamma_1=\gamma_2\Rightarrow\Gamma(\gamma_1,\gamma_2)=\underset{\gamma_1,\ \gamma_2}{\min\ }\Gamma(\gamma_1,\gamma_2)=1.$
Thus, if Theorem \ref{T_43} holds for a pair of ($\gamma_1$, $\gamma_2$), it also holds for $A_\rho=\gamma I,\ -1<\gamma<1$.} 
\end{remark}
Note that unlike linear controllers, the GSORE~(\ref{E_4-24}) with a different state-space realization yields different performance, and Theorem \ref{T_43} can not be used for such realizations. For example, the GSORE~(\ref{E_4-24}) can also be realized in observable canonical form, that is with
\begin{equation}\label{E_433-19}
A_r=\begin{bmatrix} 0 & -\omega_r^2 \\ 1 & -2\xi\omega_r 
\end{bmatrix},\ B_r=\begin{bmatrix} 1 \\ 0
\end{bmatrix},\ C_r=\begin{bmatrix} 0 & 1
\end{bmatrix},\ D_r=0,
\end{equation} 
or it can be realized with two GFORE yielding the realization (provided $\xi\geq1$)
\begin{equation}\label{E_433-20}
\begin{array}{*{35}{c}}
A_r=\begin{bmatrix} -\omega_{r_1} & 0  \\ 1 & -\omega_{r_2} 
\end{bmatrix},\ B_r=\begin{bmatrix} 1 \\ 0
\end{bmatrix},\ C_r=\begin{bmatrix} 0 & 1
\end{bmatrix},\ D_r=0,\\
\omega_{r_1}+\omega_{r_2}=2\xi\omega_r,\ \omega_{r_1}\omega_{r_2}=\omega_r^2,
\end{array}
\end{equation} 
which results in different closed-loop performance.
\begin{corollary}\label{CO_44}
Suppose hypotheses of Theorem \ref{T_43} hold for the reset control system~(\ref{E_4-27}) with the GSORE~(\ref{E_4-24}) in the controllable canonical form~(\ref{E_4-25}) for a pair ($\gamma_1$, $\gamma_2$). Then the reset control system (\ref{E_4-27}) with GSORE (\ref{E_4-24}) with realization (\ref{E_433-19}) or (\ref{E_433-20}) and $A_\rho=\gamma I,\ -1<\gamma<1$ has the following property. For each initial condition $x_0$ such that $x_0=\begin{bmatrix}0 &0 & \zeta_0^T\end{bmatrix}^T$ and each bounded input $w$ which is a Bohl function, there exists $\epsilon>0$ such that $||x(t,x_0,w(t))||<\epsilon$ for $t\geq0$.
\end{corollary}
\begin{pf}
See Appendix~\ref{AA_44}.
\end{pf}
%%%%%%%%%%%%%%%%%%%%%%%%%%%%%%%%%%%%%Modified GSORE
\subsection{Reset control systems with (SOSRE)}\label{sec_4:41}
In this section stability analysis for the reset control system (\ref{E_4-27}) with the SOSRE~\cite{Nima} is presented. In~\cite{Nima} GSORE (\ref{E_4-25}) with $A_\rho=\begin{bmatrix}\gamma & 0\\0 &1\end{bmatrix}$, which is termed SOSRE, is used to improve the performance of the reset control system (\ref{E_4-27}). In the case of SOSRE one state of GSORE is reset and the other state is utilized to reduce the high order harmonics of the reset element. 
\begin{corollary}\label{CO_45}
Consider the reset control system (\ref{E_4-27}) with SOSRE. Define the NSV vector as
$$\vv{\mathcal{N}}_{SOS}(\omega)=[\mathcal{N}_{SOS_\chi} \quad \mathcal{N}_{SOS_\Upsilon}]^T=$$  
$$\begin{bmatrix}\Re(L(j\omega)\kappa(j\omega)\mathcal{C}_s(j\omega)) & -\Im(\omega\kappa(j\omega)C_R(j\omega))\end{bmatrix}^T.$$    
Suppose that the reset instants have the well-posedness property and $-1<\gamma<1$. Then, with this definition of NSV the zero equilibrium of the reset control system (\ref{E_4-27}) with SOSRE is globally uniformly asymptotically stable when $w=0$, and the system has the UBIBS property for any input $w$ which is a Bohl function if all of the following conditions are satisfied.	
 	\begin{itemize}
 		\item The base linear system is stable and the open-loop transfer function does not have any pole-zero cancellation.
 		\item The reset control system (\ref{E_4-27}) is either of Type I and/or of Type II. 
 	\end{itemize}
\end{corollary}
\begin{pf}
Let $\beta^\prime=-\beta$. The transfer function~(\ref{E_4-29}) with $C_0$ as in~(\ref{E_4-F1}) can be rewritten as (see also Fig.~\ref{F_4-06}, transfer function from $r_1$ to $y_1$ with $\Varrho_2=0$)
\begin{equation}\label{CO_45-01}
H(s)=\dfrac{\beta^\prime L(s)\mathcal{C}_s(s)+\Varrho s C_R(s)}{1+L(s)}.
\end{equation}
Step 1 and Step 4 of the proof of Theorem~\ref{T_41} are repeated with small modifications. When the open-loop system has poles at the origin
\begin{equation}\label{CO_45-02}
%\resizebox{\columnwidth}{!}{$
	\displaystyle\lim_{\omega\to 0}\Re(H(j\omega))=K_{s_0}\beta^\prime>0.
	%$}
\end{equation}
In the case of SOSRE one has $n-m\geq3$. Consequently,
\begin{equation}\label{CO_45-03}
\displaystyle\lim_{\omega\to \infty} \omega^2\Re(H(j\omega))=2\Varrho\xi\omega_r>0,
\end{equation}  
and the proof is complete. 
\end{pf}
Note that it is impossible to satisfy Assumption~\ref{AS_42} for this configuration. Thus, the reset instants must have the well-posedness property. 
%%%%%%%%%%%%%%%%%%%%%%%%%%%%%%%%%%%%%%%%%%%%%%%%%%%%%%%%%%%%%%%%%%%%%%%%%%%Illustrative Example
\section{Illustrative Examples}\label{sec_4:5}
\begin{figure}[!t]
\centering
\includegraphics[width=0.9\columnwidth]{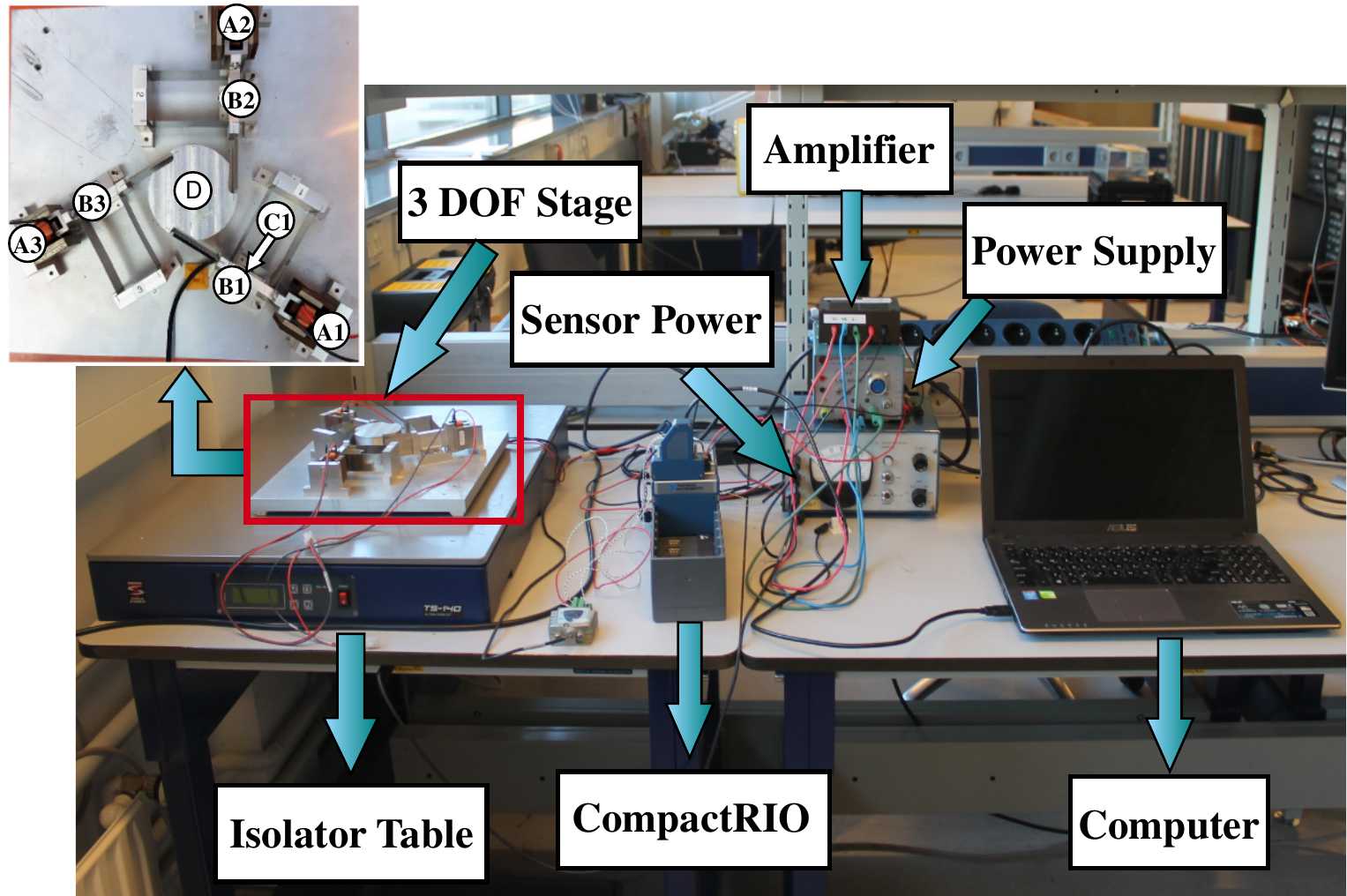}
\caption{The whole setup including computer, CompactRio, power supply, sensor power, amplifier, isolator, sensor and, stage}
\label{F_4-0755}
\end{figure}

\begin{figure}[!t]
	\centering
	\includegraphics[width=0.9\columnwidth]{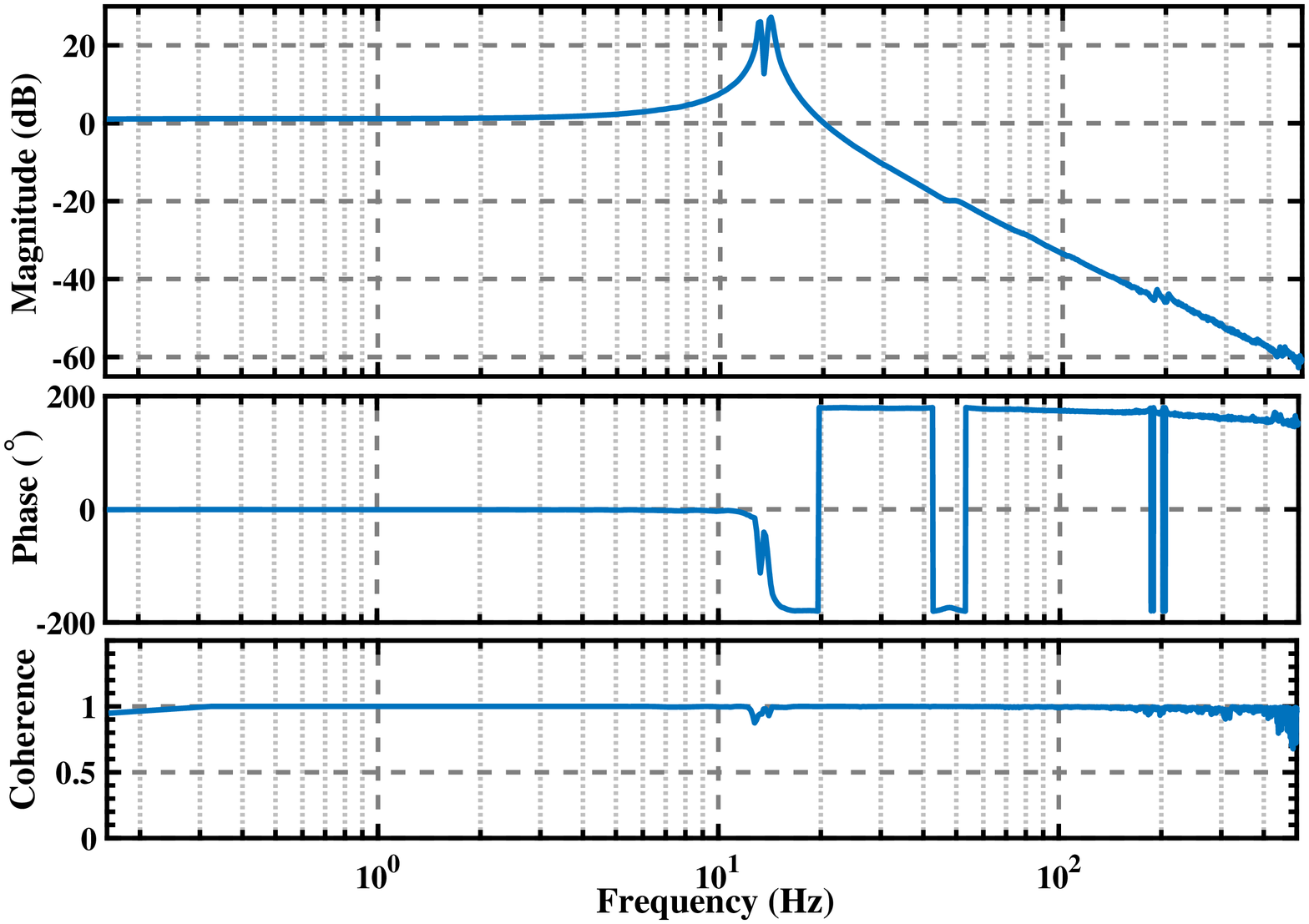}
	\caption{FRF measurement of considered SISO position of the Spyder stage}
\label{F_4-08}
\end{figure}

In this section two examples showing how the proposed methods can be used to study stability properties of reset control systems are presented. In particular, stability properties of a precision positioning system~\cite{saikumar2019constant} (knows as a spider stage) controlled by a reset controller are considered. In this system (see Fig.~\ref{F_4-0755}), three actuators are angularly spaced to actuate three masses (labeled as B1, B2, and B3) which are constrained by parallel flexures and connected to the central mass D through leaf flexures. Only one of the actuators (A1) is considered and used for controlling the position of the mass B1 attached to the same actuator, which results in a SISO system. For using these stability methods the FRF measurement of the plant (Fig.~\ref{F_4-08}) is needed. 
In~\cite{saikumar2019constant} a non-linear phase compensator, which is termed ``Constant in gain Lead in phase" (CgLp) (for more detail see~\cite{palanikumar2018no,chen2019development,saikumar2019constant}), has been used to improve the performance of this precision positioning stage. CgLp compensators, consisting of a first/second order lead filter and a GFORE/GSORE, have been utilized along with a PID controller to enhance the precision of the system. In the following, stability properties of two CgLp+PID controllers, one of which has GSORE and the other has SOSRE, are assessed with the proposed methods. The general structure of the controller is      
\setlength{\arraycolsep}{0pt}
\begin{eqnarray}\label{E_45-01}
C(s) &{=}& K_{p}\underbrace{\overbrace{\left(\cancelto{A_\rho}{\dfrac{1}{s^2+2\xi\omega_rs+\omega_r^2}}\right)}^{\text{GSORE}}\overbrace{\left(\dfrac{s^2+2\xi_d\omega_ds+\omega_d^2}{s^2+20\omega_c+100\omega_c^2}\right)}^{\text{Lead}}}_{\text{CgLp}}\nonumber\\
&&{\times}\underbrace{\overbrace{\left(1+\frac{\omega_c}{10s}\right)}^{\text{PI}}\overbrace{\left(\dfrac{\frac{3s}{\omega_c}+1}{\frac{s}{3\omega_c}+1}\right)}^{\text{Lead}}}_{\text{PID}},
\end{eqnarray}
\setlength{\arraycolsep}{5pt}in which $\omega_c$ is the cross-over frequency and $K_{p}$, $\gamma$, $\omega_d$, $\omega_r$, $\xi$, and $\xi_d$ are tuning parameters. The PID part is tuned on the basis of~\cite{schmidt2014design,dastjerdi2018tuning} and the CgLp part is tuned on the basis of~\cite{saikumar2019constant,Nima,Nimi}, and $K_{p}$ is set so that $\omega_c=200\pi$, considering the Describing Function (DF) method~\cite{saikumar2019constant}. In addition, no shaping filter is used for modifying the performance of the reset controller (i.e. $\mathcal{C}_s(s)=1$). Note that the tuning of the CgLp compensator is not within the scope of this paper, and we only discuss how to assess stability properties of reset control systems with these compensators.  
\begin{remark}\label{R_45}
{\rm Suppose that the $H_\beta$ condition is/is not satisfied for the reset control system~(\ref{E_4-27}) with $\mathcal{C}_s(s)$, $C_{L_1}(s)$, $C_{L_2}(s)$, $C_{R}(s)$, $G(s)$, and $A_\rho$. Then the $H_\beta$ condition is/is not satisfied for the reset control system~(\ref{E_4-27}) with $\mathcal{C}_s(s)$, $C^\prime_{L_1}(s)$, $C^\prime_{L_2}(s)$, $C_{R}(s)$, $G^\prime(s)$, and $A_\rho$ if $C^\prime_{L_1}(s)C^\prime_{L_2}(s)G^\prime(s)=C_{L_1}(s)C_{L_2}(s)G(s)$ and $G^\prime(s)$ is strictly proper. In other words, the ``position" of the reset element does not change in the $H_\beta$ condition. However, the ``position" of the reset element has effects on the performance of the reset control systems~\cite{Caipaper}. In the two following examples, the sequence of control filters is such that the tracking error is the input of the reset element and other linear parts following in series.}
\end{remark}
%%%%%%%%%%%%%%%%%%%%%%%%GSORE
\subsection{A Reset control system with GSORE}\label{sec_4:51}
\begin{figure}[!t]
	\centering
	\includegraphics[width=0.95\columnwidth]{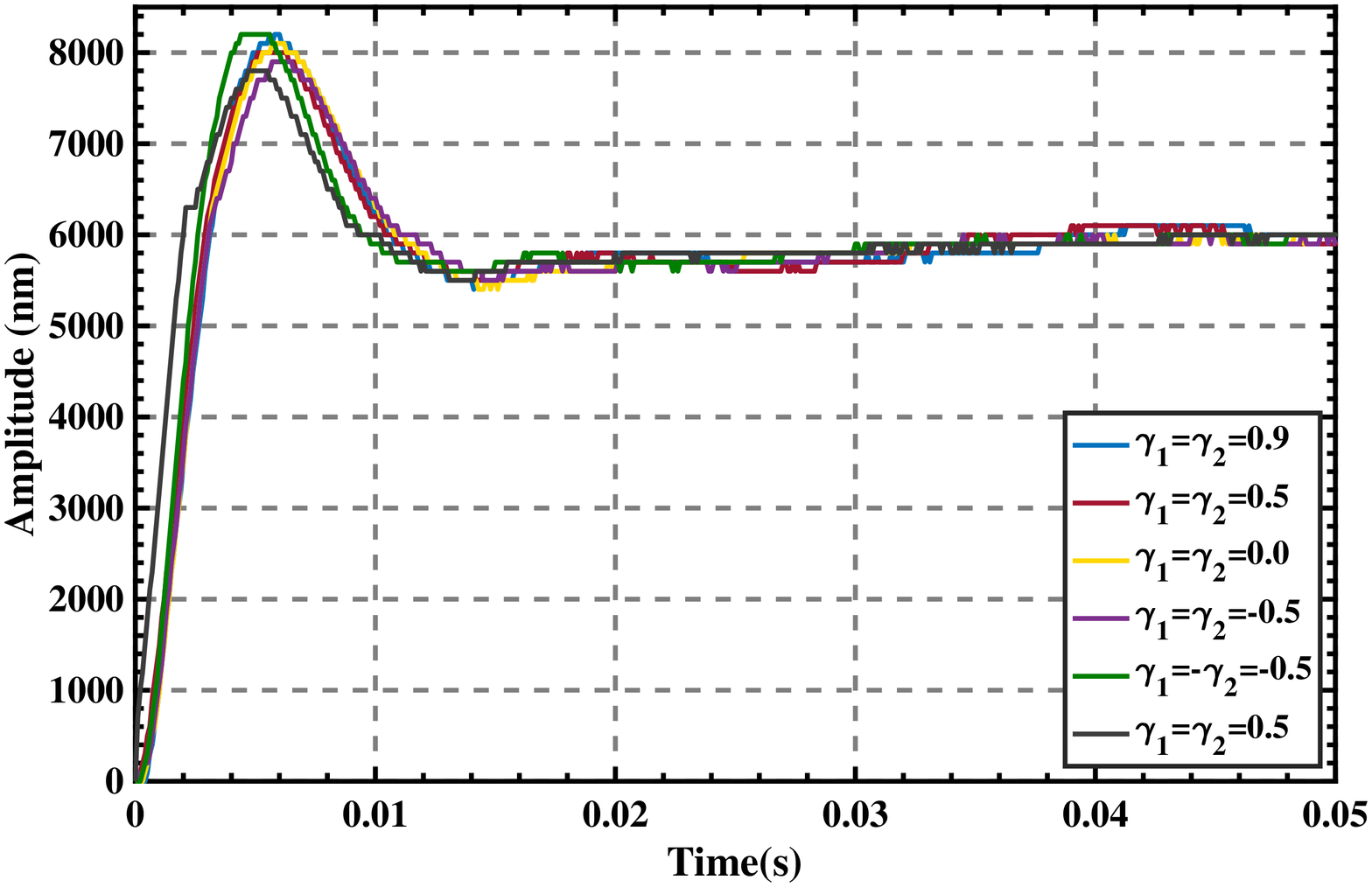}
	\caption{Step response of the closed-loop system with the designed GSORE for different values of $\gamma_i$}
\label{F_4-09}
\end{figure}

In the case of GSORE, the control parameters are $\gamma_1=\gamma_2=0.5$, $\omega_r=800\pi$, $\omega_d=720\pi$, $K_p=8.5273e^7$, and $\xi=\xi_d=1$. Since the controller has a pole at the origin, we use Definition \ref{D_44} to assess stability properties of this reset control system. Using Proposition \ref{pro_41} yields $340<\dfrac{Q_2}{Q_1}<5057 $ and $1132<\dfrac{Q_3}{Q_4}$ for $\mathcal{S}_1$ and $\mathcal{S}_2$, respectively. Thus, we have to solve the optimization problem $M=\underset{Q_1,Q_2,Q_3,Q_4}{\min} \ G_1(Q_1,Q_2,Q_3,Q_4)$ such that the following constraints hold
 \begin{equation}\label{E_451-01}\resizebox{\columnwidth}{!}{$
		\begin{aligned}
		\mathcal{S}_1: \ & \forall\omega\in(0,\infty):\ f_1(Q_1,Q_2,1,\omega)>0\\
		\mathcal{S}_2: \ & \forall\omega\in(0,\infty):\ f_2(Q_3,Q_4,1,\omega)>0\\
		\mathcal{S}_3: \ & \dfrac{1600\pi}{Q_1}+\dfrac{Q_2}{Q_1Q_4}+\dfrac{2}{Q_1}\sqrt{\dfrac{1600\pi Q_2}{Q_4}-Q_2}>1\\
		\mathcal{S}_4: \ & \dfrac{1600\pi}{Q_1}+\dfrac{Q_2}{Q_1Q_4}-\dfrac{2}{Q_1}\sqrt{\dfrac{1600\pi Q_2}{Q_4}-Q_2}<1\\
		\mathcal{S}_5: \ &  \dfrac{640000\pi^2Q_1}{Q_2}+1600\pi\left(1+2\sqrt{\dfrac{1600\pi Q_1}{Q_2}-1}\right)>\dfrac{Q_3}{Q_4}\\
		\mathcal{S}_6: \ &  \dfrac{640000\pi^2Q_1}{Q_2}+1600\pi\left(1-2\sqrt{\dfrac{1600\pi Q_1}{Q_2}-1}\right)<\dfrac{Q_3}{Q_4}\\
                 \mathcal{S}_7: \ & Q_i>0,\ 1600\pi>Q_4,\ 1600\pi<\dfrac{Q_2}{Q_1}<5057,\ 1132<\dfrac{Q_3}{Q_4},\\ 	
                 \mathcal{S}_8: \ &  \dfrac{Q_1Q_3}{Q_2Q_4}>1,     \end{aligned}$}
		\end{equation}
This optimization problem is solved using Genetic Algorithm and Proposition \ref{pro_41}. The optimal solution is $Q_1=13172$, $Q_2=12001144$, $Q_3=8113151$, and $Q_4=1055$, yielding $M=3.5$ (note that it is not necessary to find the global minimum in these methods). Furthermore, $(\bar{A},C_0)$ is observable and $(\bar{A},B_0)$ is controllable. Hence, the reset control system is of Type III and using Theorem \ref{T_43} this GSORE has the UBIBS property for $A_\rho=\gamma I,\ -1<\gamma<1$. Furthermore, since $\dfrac{Q_1Q_3}{Q_2Q_4}>\Gamma(-0.5,0.5)$ and $\dfrac{Q_1Q_3}{Q_2Q_4}>\Gamma(0.5,-0.5)$, Theorem \ref{T_43} holds for the considered closed-loop system with $A_\rho=\begin{bmatrix}0.5 & 0\\0 &-0.5\end{bmatrix}$ or $A_\rho=\begin{bmatrix}-0.5 & 0\\0 &0.5\end{bmatrix}$. In Fig.~\ref{F_4-09} the step responses of the closed-loop Spider stage (Fig.~\ref{F_4-0755}) with the designed controller for different values of $\gamma_i$ are displayed. As it can be observed, the values of $\gamma_i$ have effect on the performance of the system. In the sense of transient response, the reset controller with $\gamma_1=\gamma_2=0.5$ has better performance among other configurations (for more detail see~\cite{saikumar2019constant,Nima}). 
%%%%%%%%%%%%%%%%%%%%%%%%ModifiedGSORE
\subsection{Reset control system with SOSRE}\label{sec_4:51}
\begin{figure}[!t]
 	\centering
 	\resizebox{0.95\columnwidth}{!}{
 		\begin{tikzpicture}
 		\node[anchor=south west,inner sep=0] at (0,0) {\includegraphics[width=0.7\columnwidth]{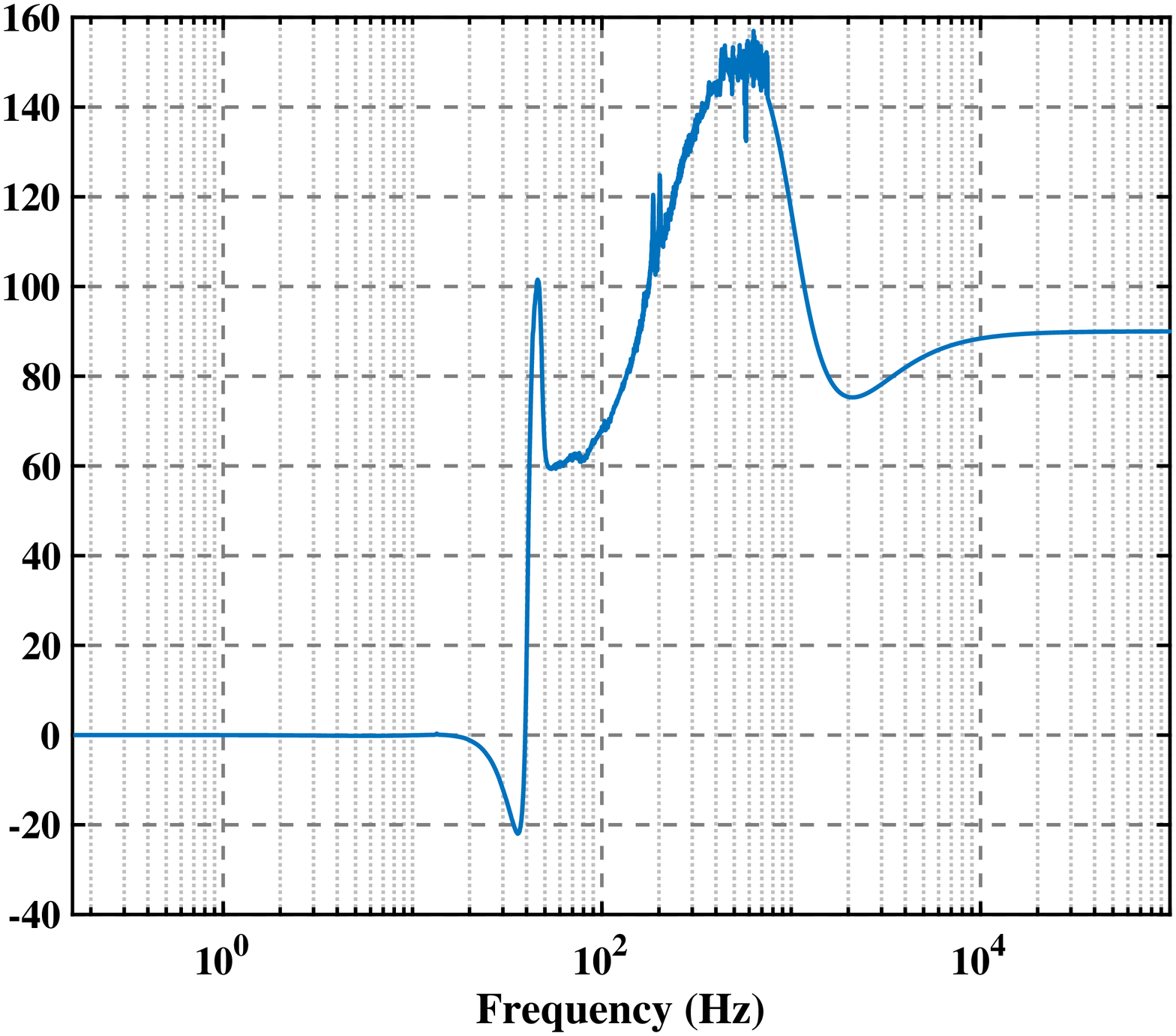}};
 		% Text Node
 		\draw (0,2.6) node [scale=0.6]  {$\boldmath{\phase{\vv{\mathcal{N}}(\omega)}(^\circ)}$};
 		\end{tikzpicture}}
 		\caption{$\phase{\vec{\mathcal{N}}(\omega)}$ for the reset control systems with SOSORE}
 		\label{F_4-910}
 	\end{figure}

\begin{figure}[!t]
	\centering
	\includegraphics[width=0.95\columnwidth]{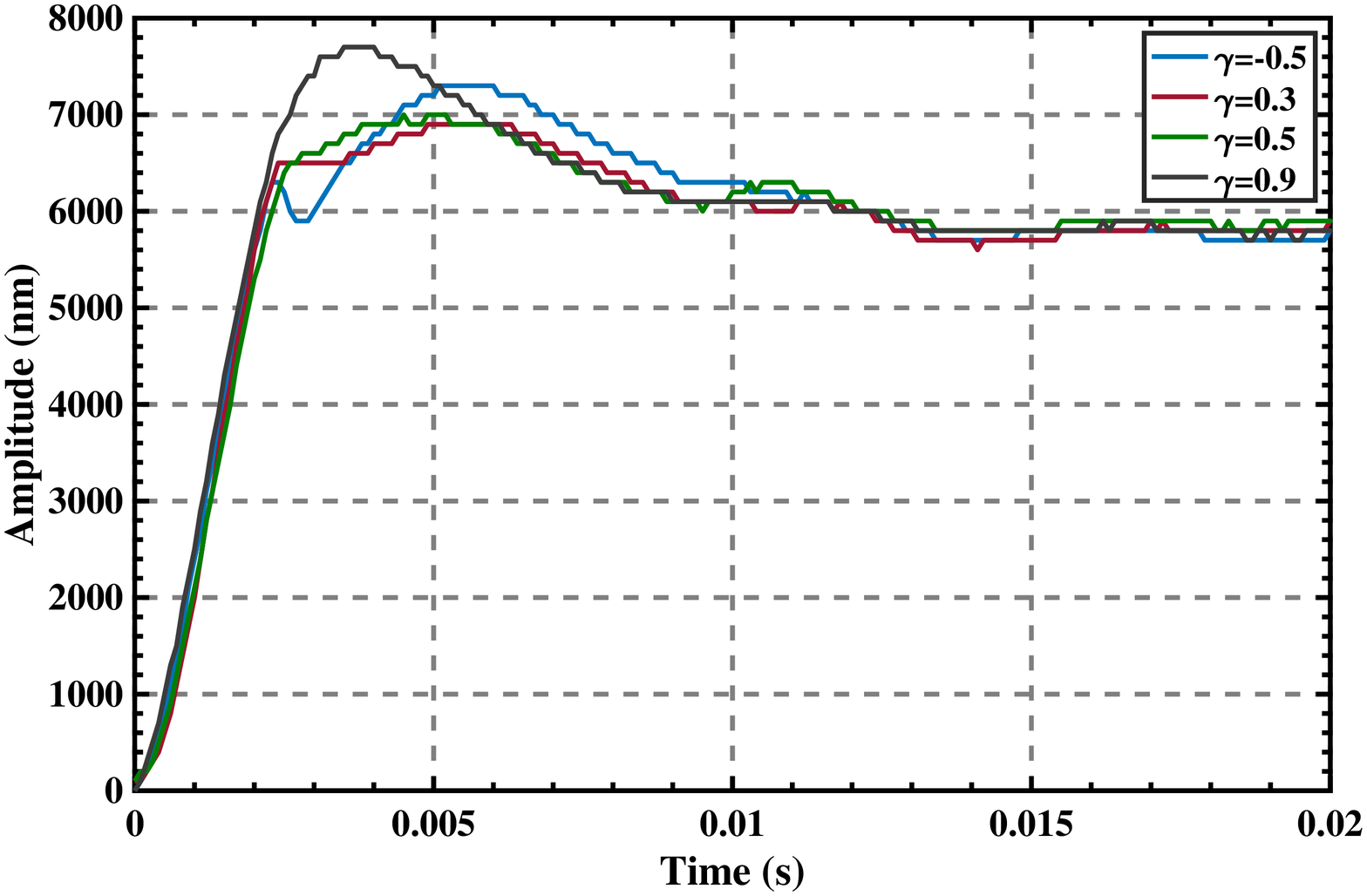}
	\caption{Step response of the closed-loop system with the designed SOSRE for different values of $\gamma$}
\label{F_4-10}
\end{figure}

In the case in which the controller is a SOSRE the control parameters are $-1<\gamma<1$, $\omega_r=150\pi$, $\omega_d=96\pi$, $K_p=1.135e^6$, and $\xi=\xi_d=1$. Since the controller has a pole at the origin, we use Definition \ref{D_42} with the NSV defined in Corollary \ref{CO_45} to assess stability properties. The phase of the NSV for this example is shown in Fig.~\ref{F_4-910}. Since the phase of the NSV for this example is between $(-\dfrac{\pi}{2},\pi)$ and the difference between its maximum and its minimum is less than $\pi$, by Remark~\ref{R_4s1} the reset control system is of Type I. Moreover, the time regularization technique (to prevent successive reset instants, i.e. if the reset happened at one sample time before, the system does not reset) is used to guarantee the well-posedness property. Consequently, by Corollary \ref{CO_45} the designed SOSRE yields a closed-loop system which has the UBIBS property. The step responses of the closed-loop Spider stage (Fig.~\ref{F_4-0755}) with the designed controller for different values of $\gamma$ are shown in Fig.~\ref{F_4-10}. In the sense of transient response, reset control system with $\gamma=0.5$ has better performance among other controllers. For deeper insights on the performance of closed-loop reset control systems with SOSRE see~\cite{Nimi,Nima}. 
%%%%%%%%%%%%%%%%%%%%%%%%%%%%%%%%%%%%%%%%%%%%%%%%%%%%%%%%%%%%%%%%%%%%%%%%%%%%%Conclusion
\section{Conclusion}\label{sec_4:6}
In this paper a novel frequency-domain approach based on the $H_\beta$ condition for assessing stability properties of reset control systems has been proposed. This method can be used to determine stability properties of control systems with first and second order reset elements using FRF measurements of their base linear open-loop system. Consequently, the methods do not need an accurate parametric model of the system and the solution of LMIs. In addition, these methods are applicable to the case in which partial reset techniques are used. The effectiveness of the proposed methods have been illustrated with a practical example.
%%%%%%%%%%%%%%%%%%%%%%%%%Acknowledgment 
\begin{ack}                               % Place acknowledgements
This work has been partially supported by NWO through OTP TTW project $\#$16335, by the EACEA, by the European Union's Horizon 2020 Research and Innovation Programme under grant agreement No 739551 (KIOS CoE), and by the Italian Ministry for Research in the framework of the 2017 Program for Research Projects of National Interest (PRIN), Grant no. 2017YKXYXJ.  % here.
\end{ack}
\bibliographystyle{plain}        % Include this if you use bibtex 
\bibliography{phd_4}           

\begin{thebibliography}{10}

\bibitem{AliCDC}
A.~{A. Dastjerdi}, A.~{Astolfi}, and S.~H. {HosseinNia}.
\newblock A frequency-domain stability method for reset systems.
\newblock In {\em IEEE 59th Conference on Decision and Control}, 2020.

\bibitem{aangenent2010performance}
W.~H. T.~M. Aangenent, G.~Witvoet, W.~P. M.~H. Heemels, M.~J.~G. van~de
  Molengraft, and M.~Steinbuch.
\newblock Performance analysis of reset control systems.
\newblock {\em International Journal of Robust and Nonlinear Control},
  20(11):1213--1233, 2010.

\bibitem{banos2011reset}
Alfonso Ba{\~n}os and Antonio Barreiro.
\newblock {\em Reset control systems}.
\newblock Springer Science $\&$ Business Media, 2011.

\bibitem{banos2007reset}
Alfonso Banos, Joaquin Carrasco, and Antonio Barreiro.
\newblock Reset times-dependent stability of reset control with unstable base
  systems.
\newblock In {\em 2007 IEEE International Symposium on Industrial Electronics},
  pages 163--168. IEEE, 2007.

\bibitem{banos2010reset}
Alfonso Ba{\~n}os, Joaqu{\'\i}n Carrasco, and Antonio Barreiro.
\newblock Reset times-dependent stability of reset control systems.
\newblock {\em IEEE Transactions on Automatic Control}, 56(1):217--223, 2010.

\bibitem{banos2014tuning}
Alfonso Ba{\~n}os and Miguel~A Dav{\'o}.
\newblock Tuning of reset proportional integral compensators with a variable
  reset ratio and reset band.
\newblock {\em IET Control Theory \& Applications}, 8(17):1949--1962, 2014.

\bibitem{banos2016impulsive}
Alfonso Banos, Juan~I Mulero, Antonio Barreiro, and Miguel~A Davo.
\newblock An impulsive dynamical systems framework for reset control systems.
\newblock {\em International Journal of Control}, 89(10):1985--2007, 2016.

\bibitem{barreiro2014reset}
Antonio Barreiro, Alfonso Ba{\~n}os, Sebasti{\'a}n Dormido, and Jos{\'e}~A
  Gonz{\'a}lez-Prieto.
\newblock Reset control systems with reset band: Well-posedness, limit cycles
  and stability analysis.
\newblock {\em Systems \& Control Letters}, 63:1--11, 2014.

\bibitem{muri3}
R.~Beerens, A.~Bisoffi, L.~Zaccarian, W.P.M.H. Heemels, H.~Nijmeijer, and
  N.~{van de Wouw}.
\newblock Reset integral control for improved settling of pid-based motion
  systems with friction.
\newblock {\em Automatica}, 107:483 -- 492, 2019.

\bibitem{muri4}
R.~Beerens, A.~Bisoffi, L.~Zaccarian, W.P.M.H. Heemels, H.~Nijmeijer, and
  N.~{van de Wouw}.
\newblock Reset integral control for improved settling of pid-based motion
  systems with friction.
\newblock {\em Automatica}, 107:483 -- 492, 2019.

\bibitem{beker1999stability}
O~Beker, CV~Hollot, Q~Chen, and Y~Chait.
\newblock Stability of a reset control system under constant inputs.
\newblock In {\em Proceedings of the 1999 American Control Conference (Cat. No.
  99CH36251)}, volume~5, pages 3044--3045. IEEE, 1999.

\bibitem{beker2004fundamental}
Orhan Beker, C.V. Hollot, Y.~Chait, and H.~Han.
\newblock Fundamental properties of reset control systems.
\newblock {\em Automatica}, 40(6):905 -- 915, 2004.

\bibitem{muri2}
A.~Bisoffi, R.~Beerens, W.P.M.H. Heemels, H.~Nijmeijer, N.~{van de Wouw}, and
  L.~Zaccarian.
\newblock To stick or to slip: A reset pid control perspective on positioning
  systems with friction.
\newblock {\em Annual Reviews in Control}, 49:37 -- 63, 2020.

\bibitem{Caipaper}
Chengwei Cai, Ali~Ahmadi Dastjerdi, Niranjan Saikumar, and S.H. HosseinNia.
\newblock The optimal sequence for reset controllers.
\newblock In {\em $18^{th}$ European Control Conference (ECC 2020)}, 2020.

\bibitem{carrasco2010passivity}
Joaqu{\'\i}n Carrasco, Alfonso Ba{\~n}os, and Arjan van~der Schaft.
\newblock A passivity-based approach to reset control systems stability.
\newblock {\em Systems \& Control Letters}, 59(1):18--24, 2010.

\bibitem{chen2019development}
L.~{Chen}, N.~{Saikumar}, and S.~H. {HosseinNia}.
\newblock Development of robust fractional-order reset control.
\newblock {\em IEEE Transactions on Control Systems Technology},
  28(4):1404--1417, 2020.

\bibitem{clegg1958nonlinear}
J.~C. {Clegg}.
\newblock A nonlinear integrator for servomechanisms.
\newblock {\em Transactions of the American Institute of Electrical Engineers,
  Part II: Applications and Industry}, 77(1):41--42, 1958.

\bibitem{dastjerdi2018tuning}
Ali~Ahmadi Dastjerdi, Niranjan Saikumar, and S.~Hassan HosseinNia.
\newblock Tuning guidelines for fractional order {PID} controllers: Rules of
  thumb.
\newblock {\em Mechatronics}, 56:26 -- 36, 2018.

\bibitem{forni2011reset}
Fulvio Forni, Dragan Ne{\v{s}}i{\'c}, and Luca Zaccarian.
\newblock Reset passivation of nonlinear controllers via a suitable
  time-regular reset map.
\newblock {\em Automatica}, 47(9):2099 -- 2106, 2011.

\bibitem{griggs2007stability}
Wynita~M Griggs, Brian~DO Anderson, Alexander Lanzon, and Michael~C Rotkowitz.
\newblock A stability result for interconnections of nonlinear systems with
  ``mixed'' small gain and passivity properties.
\newblock In {\em 2007 46th IEEE Conference on Decision and Control}, pages
  4489--4494. IEEE, 2007.

\bibitem{guo2009frequency}
Yuqian Guo, Youyi Wang, and Lihua Xie.
\newblock Frequency-domain properties of reset systems with application in
  hard-disk-drive systems.
\newblock {\em IEEE Transactions on Control Systems Technology},
  17(6):1446--1453, 2009.

\bibitem{guo2015analysis}
Yuqian Guo, Lihua Xie, and Youyi Wang.
\newblock {\em Analysis and Design of Reset Control Systems}.
\newblock Institution of Engineering and Technology, 2015.

\bibitem{hazeleger2016second}
L.~{Hazeleger}, M.~{Heertjes}, and H.~{Nijmeijer}.
\newblock Second-order reset elements for stage control design.
\newblock In {\em American Control Conference (ACC)}, pages 2643--2648, 2016.

\bibitem{muris}
M.F. Heertjes, K.G.J. Gruntjens, S.J.L.M. van Loon, N.~Kontaras, and W.P.M.H.
  Heemels.
\newblock Design of a variable gain integrator with reset.
\newblock In {\em American Control Conference (ACC), Chicago, USA}, pages
  2155--2160, 2015.

\bibitem{hollot2001establishing}
CV~Hollot, Orhan Beker, Yossi Chait, and Qian Chen.
\newblock On establishing classic performance measures for reset control
  systems.
\newblock In {\em Perspectives in robust control}, pages 123--147. Springer,
  2001.

\bibitem{hollot1997stability}
CV~Hollot, Y~Zheng, and Y~Chait.
\newblock Stability analysis for control systems with reset integrators.
\newblock In {\em Proceedings of the 36th IEEE Conference on Decision and
  Control}, volume~2, pages 1717--1719. IEEE, 1997.

\bibitem{horowitz1975non}
Isaac Horowitz and Patrick Rosenbaum.
\newblock Non-linear design for cost of feedback reduction in systems with
  large parameter uncertainty.
\newblock {\em International Journal of Control}, 21(6):977--1001, 1975.

\bibitem{hosseinnia2013fractional}
S~Hassan HosseinNia, In{\'e}s Tejado, and Blas~M Vinagre.
\newblock Fractional-order reset control: Application to a servomotor.
\newblock {\em Mechatronics}, 23(7):781--788, 2013.

\bibitem{Nima}
Nima Karbasizadeh, Ali Ahmadi~Dastjerdi, Niranjan Saikumar, Duarte Valerio, and
  S.H. HosseinNia.
\newblock Benefiting from linear behaviour of a nonlinear reset-based element
  at certain frequencies.
\newblock In {\em Australian and New Zealand Control Conference (ANZCC)}, 2020.

\bibitem{mechNima}
Nima Karbasizadeh, Ali~Ahmadi Dastjerdi, Niranjan Saikumar, and S~Hassan
  HosseinNia.
\newblock and-passing nonlinearity in reset elements.
\newblock {\em arXiv preprint arXiv:2005.02887}, 2020.

\bibitem{Nimi}
Nima Karbasizadeh, Niranjan Saikumar, and S~Hassan Hoseinnia.
\newblock Fractional-order single state reset element.
\newblock {\em Nonliner Dynamics}, 2021.

\bibitem{khalil2002nonlinear}
Hassan~K Khalil and Jessy~W Grizzle.
\newblock {\em Nonlinear systems}, volume~3.
\newblock Prentice hall Upper Saddle River, NJ, 2002.

\bibitem{onevsic2008stability}
Dragan Ne{\v{s}}i{\'c}, Luca Zaccarian, and Andrew~R Teel.
\newblock Stability properties of reset systems.
\newblock {\em Automatica}, 44(8):2019--2026, 2008.

\bibitem{paesa2011design}
D~Paesa, J~Carrasco, O~Lucia, and C~Sagues.
\newblock On the design of reset systems with unstable base: A fixed reset-time
  approach.
\newblock In {\em IECON 2011-37th Annual Conference of the IEEE Industrial
  Electronics Society}, pages 646--651. IEEE, 2011.

\bibitem{palanikumar2018no}
A.~{Palanikumar}, N.~{Saikumar}, and S.~H. {HosseinNia}.
\newblock No more differentiator in {PID}: Development of nonlinear lead for
  precision mechatronics.
\newblock In {\em European Control Conference (ECC)}, pages 991--996, 2018.

\bibitem{polenkova2012stability}
Svetlana Polenkova, Jan~W Polderman, and Romanus Langerak.
\newblock Stability of reset systems.
\newblock In {\em Proceedings of the 20th International Symposium on
  Mathematical Theory of Networks and Systems}, pages 9--13, 2012.

\bibitem{rifai2006compositional}
KE~Rifai and J-JE Slotine.
\newblock Compositional contraction analysis of resetting hybrid systems.
\newblock {\em IEEE Transactions on Automatic Control}, 51(9):1536--1541, 2006.

\bibitem{saikumar2019constant}
N.~{Saikumar}, R.~K. {Sinha}, and S.~H. {HosseinNia}.
\newblock `{C}onstant in gain {L}ead in phase' element-application in precision
  motion control.
\newblock {\em IEEE/ASME Transactions on Mechatronics}, 24(3):1176--1185, 2019.

\bibitem{schmidt2014design}
R~Munnig Schmidt, Georg Schitter, and Adrian Rankers.
\newblock {\em The Design of High Performance Mechatronics High-Tech
  Functionality by Multidisciplinary System Integration}.
\newblock IOS Press, 2014.

\bibitem{valerio2019reset}
Duarte Val{\'e}rio, Niranjan Saikumar, Ali~Ahmadi Dastjerdi, Nima Karbasizadeh,
  and S~Hassan HosseinNia.
\newblock Reset control approximates complex order transfer functions.
\newblock {\em Nonlinear Dynamics}, pages 1--15, 2019.

\bibitem{van2018hybrid}
S.~J. A.~M. {Van den Eijnden}, Y.~{Knops}, and M.~F. {Heertjes}.
\newblock A hybrid integrator-gain based low-pass filter for nonlinear motion
  control.
\newblock In {\em IEEE Conference on Control Technology and Applications
  (CCTA)}, pages 1108--1113, 2018.

\bibitem{van2017frequency}
S.J.L.M. {van Loon}, K.G.J. Gruntjens, M.F. Heertjes, N.~{van de Wouw}, and
  W.P.M.H. Heemels.
\newblock Frequency-domain tools for stability analysis of reset control
  systems.
\newblock {\em Automatica}, 82:101 -- 108, 2017.

\bibitem{vettori2014geometric}
Paolo Vettori, Jan~Willem Polderman, and Rom Langerak.
\newblock A geometric approach to stability of linear reset systems.
\newblock {\em Proceedings of the 21st Mathematical Theory of Networks and
  Systems}, 2014.

\bibitem{villaverde2011reset}
A.~F. {Villaverde}, A.~B. {Blas}, J.~{Carrasco}, and A.~B. {Torrico}.
\newblock Reset control for passive bilateral teleoperation.
\newblock {\em IEEE Transactions on Industrial Electronics}, 58(7):3037--3045,
  2011.

\bibitem{zaccarian2005first}
L.~{Zaccarian}, D.~{Nesic}, and A.~R. {Teel}.
\newblock First order reset elements and the {C}legg integrator revisited.
\newblock In {\em American Control Conference}, pages 563--568 vol. 1, 2005.

\bibitem{zheng2007improved}
Jinchuan Zheng, Yuqian Guo, Minyue Fu, Youyi Wang, and Lihua Xie.
\newblock Improved reset control design for a pzt positioning stage.
\newblock In {\em 2007 IEEE International Conference on Control Applications},
  pages 1272--1277. IEEE, 2007.

\end{thebibliography}
\appendix
%\begin{subappendices}
\section{Proof of Lemma \ref{L_41}}\label{aap_41}
It has been shown in~\cite{beker2004fundamental} that when $A_\rho=0$, $\mathcal{C}_s(s)=1$, Assumption~\ref{AS_42} holds, and the $H_\beta$ condition is satisfied, the reset control system has  the UBIBS property. In what follows, we provide a slight modification of the proof in~\cite{beker2004fundamental} to deal with the case $A_\rho\neq0$. The base linear dynamic of the reset control system is given by
\begin{equation}\label{Ap_4-1}
\begin{cases} 
\dot{x}_l(t)=\bar{A}x_l(t)+\bar{B}w(t),\\
y_l(t)=\bar{C}x_l(t),
\end{cases}
\end{equation} 
where $x_l(t)=[x_{r_l}(t)^T\ \zeta_l(t)^T]^T\in\mathbb{R}^{n_p+n_r}$. Denoting $z(t):x(t)-x_l(t)=[z_p(t)^T\ \ z_r(t)^T]^T$, yields
\begin{equation}\label{Ap_4_2}
\begin{cases} 
\dot{z}(t)=\bar{A}z(t), & e(t)\neq 0,\\
z(t^+)=\bar{A}_\rho z(t)+(\bar{A}_\rho-I)x_l(t), & e(t)=0.  \\
\end{cases}
\end{equation}
According to~\cite{beker2004fundamental}, it is sufficient to show that $z(t)$ is bounded. Since the $H_\beta$ condition is satisfied, there exists a matrix $P=P^T>0$ such that
\begin{equation}\label{Ap_4_2}
P=\begin{bmatrix}
P_1 & (\beta \bar{C}_e)^T\\
\beta \bar{C}_e & \Varrho
\end{bmatrix},\ P_1=P_1^T>0.
\end{equation}
Consider now the quadratic Lyapunov function $V(t)=z(t)^TPz(t)$. Using the same procedure as in~\cite{beker2004fundamental} yields
\begin{equation}\label{Ap_4_3}
V(t)\leq e^{-\epsilon (t-t_i)}V(t_i),\quad t\in(t_i,t_{i+1}],\ \epsilon>0,
\end{equation}
and 
\setlength{\arraycolsep}{0pt}
\begin{eqnarray}\label{Ap_4_4}
V(t_i^+) &{=}& V(t_i)+x_r^T(t_i)(A_\rho^T\Varrho A_\rho-\Varrho)x_r(t_i)\nonumber\\
&&{+}\:2(A_\rho^T-I)x_r^T(t_i)\beta \bar{C}_ez_p(t_i)-2x^T_r(t_i)A_\rho^T\Varrho x_{r_l}(t_i),\nonumber\\
\end{eqnarray}
\setlength{\arraycolsep}{5pt}in which $t_i$ are the reset instants. Now, let the maximum eigenvalue of $A_\rho^T\Varrho A_\rho-\Varrho$ be $\lambda_{\max}$ and note that $\lambda_{\max}<0$ since $A_\rho^T\Varrho A_\rho-\Varrho<0$. As a result
\setlength{\arraycolsep}{0pt}
\begin{eqnarray}\label{Ap_4_5}
V(t_i^+)&{\leq}& V(t_i)-|\lambda_{\max}|x_r^T(t_i)x_r(t_i)+2(A_\rho^T-I)x_r^T(t_i)\beta \bar{C}_ez_p(t_i)\nonumber\\
&&{-}\:2x^T_r(t_i)A_\rho^T\Varrho x_{r_l}(t_i)\Rightarrow\nonumber\\
V(t_i^+)&{\leq}& V(t_i)+2\norm{x_r(t_i)}(||A_\rho^T-I||\norm{\beta \bar{C}_ez_p(t_i)}\nonumber\\
&&{+}\:||A_\rho\Varrho x_{r_l}(t_i)||).
\end{eqnarray} 
\setlength{\arraycolsep}{5pt}At the reset instants $|\bar{C}_ez_p(t_i)|\leq|D_er(t)|$ which implies that $|\bar{C}_ez_p(t_i)|$ is bounded. Moreover, since the base linear system is stable, $x_{r_l}(t_i)$ is bounded. Assume now that $x_r(t_i)$ is unbounded. By~(\ref{Ap_4_3}) and (\ref{Ap_4_5}), we conclude that $\displaystyle\lim_{i\to\infty}V(t_i)=0$. This is a contradiction because $z(t)=0\Rightarrow x(t)=x_l(t)$ which implies that the system is a stable linear system with bounded state. Therefore, $x_r(t_i)$ is bounded. Now, we prove that $\dot{x}_r(t_i)$ is bounded. If reset happens when the input of the reset element is zero (i.e. $\mathcal{C}_s(s)=1$) and Assumption~\ref{AS_42} holds, then  
\setlength{\arraycolsep}{0pt}
\begin{eqnarray}\label{E_4-51}
\dfrac{dx_r(t)}{dt}\Big|_{t=t_i^-}&{=}&\resizebox{0.78\columnwidth}{!}{$A_r\left(e^{A_r(t_i-t_{i-1})}x_r(t_{i-1})+\displaystyle\int_{t_{i-1}}^{t_i}e^{A_r(t_i-\tau)}B_re(\tau)d\tau\right)$}\nonumber\\
&{=}&A_rx_r(t_i)\Rightarrow\abs{\dot{x}_r(t_i^-)}=\abs{A_rx_r(t_i)}.
\end{eqnarray} 
\setlength{\arraycolsep}{5pt}Thus, since $|x_r(t_i)|$ is bounded, $\abs{\dot{x}_r(t_i^-)}$ is bounded. As a result, since $|x_r(t_{i-1}^+)|\leq|A_\rho||x_r(t_{i-1})|$, $|x_r(t_{i})|$ and $\abs{\dot{x}_r(t_i^-)}$ are bounded,
\begin{equation}\label{E_4-52}
\exists\ K_1>0,\ \alpha>0\ \text{such that } |x_r(t_{i})|\leq K_1\left(1-e^{\alpha(t_i-t_{i-1})}\right),\ \forall\ t_i.
\end{equation}  
Now assume that there exist $t_i$ and $t_{i-1}$ such that for any $\epsilon>0$, $t_i-t_{i-1}<\epsilon$. Thus, by~(\ref{E_4-52}) and for sufficient small $\epsilon$, $x_r(t_{i})\to0$. This is a contradiction because $(I-\bar{A}_\rho)x_r(t_{i})\to0$ which means that $t_{i}$ is not a reset instant. Thus, there exists $\lambda>0$ such that, for all $k\in\mathbb{N}$, $\lambda\leq t_{k+1}-t_k$. Therefore, the reset instants have the the well-posedness property (see Definition~\ref{D_4D0}). 
\newline In the case in which $\mathcal{C}_s=1$ or Assumption~\ref{AS_42} does not hold,~(\ref{E_4-51}) can not be concluded. However, if the well-posedness property of the reset instants holds, then there exists $\lambda>0$ such that, for all $k\in\mathbb{N}$, $\lambda\leq t_{k+1}-t_k$. In addition, since $|x_r(t_{i-1}^+)|\leq|A_\rho||x_r(t_{i-1})|$ and $|x_r(t_{i})|$ are bounded, we conclude~(\ref{E_4-52}). Since the system has the well-posedness property, the reset control system (\ref{E_4-27}) has an unique well-defined solution for any initial condition $x_0$ and any input $w$ which is a Bohl function~\cite{banos2016impulsive}. The rest of the proof is the same as the proof in~\cite{beker2004fundamental}.
%%%%%%%%%%%%%%%%%%%%%%%%%%%%%%%%%%%%%%% Proof of Corollary 2
\section{Proof of Corollary \ref{CO_42}}\label{aap_42}
Let $\beta^\prime=-\beta$ and $\Varrho^\prime=\dfrac{\Varrho}{C_r}$. By the proof of the $H_\beta$ condition in~\cite{beker2004fundamental} the transfer function (\ref{E_4-29}) for the configuration shown in Fig.~\ref{F_4-04}) can be rewritten as (see also Fig.~\ref{F_4-05}) 
\begin{equation}\label{E_4E-332}
H(s)=\dfrac{\beta^\prime \dfrac{L^\prime(s)}{\mathcal{C}_s(s)}+\Varrho^\prime C_R(s)}{1+L^\prime(s)}.
\end{equation}
Let $C_{L_1}(s)C_{L_2}(s)C_R(s)G(s)=\dfrac{k_ms^m+k_{m-1}s^{m-1}+...+k_0}{s^n+k^\prime_{n-1}s^{n-1}+...+k^\prime_0}$. Using the NSV defined in~(\ref{E_4CO2}), one could repeat Steps 1 to 4 of the proof of Theorem~\ref{T_41}. Note that $K_{s_0}\beta^\prime$ in (\ref{E_4E-44})-(\ref{E_4E-446}) and~(\ref{Ap_43-2}) has to be replaced by $\dfrac{\beta^\prime}{K_{s_0}}$ and $K_n$ has also to be replaced by $k_n$ in~(\ref{E_4E-52}).
  \begin{figure}[!t]
 	\centering
	\resizebox{0.9\columnwidth}{!}{%
\tikzset{every picture/.style={line width=0.75pt}} %set default line width to 0.75pt        
\begin{tikzpicture}[x=0.75pt,y=0.75pt,yscale=-1,xscale=1]
%uncomment if require: \path (0,226); %set diagram left start at 0, and has height of 226
%Shape: Rectangle [id:dp8872257254473093] 
\draw  [line width=1.5]  (229.5,64) -- (289,64) -- (289,117) -- (229.5,117) -- cycle ;
%Shape: Ellipse [id:dp19980167715656805] 
\draw  [line width=1.5]  (38.63,89.45) .. controls (38.63,81.84) and (45.3,75.68) .. (53.53,75.68) .. controls (61.76,75.68) and (68.43,81.84) .. (68.43,89.45) .. controls (68.43,97.06) and (61.76,103.22) .. (53.53,103.22) .. controls (45.3,103.22) and (38.63,97.06) .. (38.63,89.45) -- cycle ;
%Straight Lines [id:da0385434449901394] 
\draw [line width=1.5]    (479,98) -- (480.5,198) -- (295.5,199.96) ;
\draw [shift={(291.5,200)}, rotate = 359.39] [fill={rgb, 255:red, 0; green, 0; blue, 0 }  ][line width=0.08]  [draw opacity=0] (11.61,-5.58) -- (0,0) -- (11.61,5.58) -- cycle    ;
%Straight Lines [id:da6065506865121402] 
\draw [line width=1.5]    (289.49,92.5) -- (314.5,92.93) ;
\draw [shift={(318.5,93)}, rotate = 180.99] [fill={rgb, 255:red, 0; green, 0; blue, 0 }  ][line width=0.08]  [draw opacity=0] (11.61,-5.58) -- (0,0) -- (11.61,5.58) -- cycle    ;
%Straight Lines [id:da00759412524357761] 
\draw [line width=1.5]    (186.49,92.5) -- (224,92.95) ;
\draw [shift={(228,93)}, rotate = 180.69] [fill={rgb, 255:red, 0; green, 0; blue, 0 }  ][line width=0.08]  [draw opacity=0] (11.61,-5.58) -- (0,0) -- (11.61,5.58) -- cycle    ;
%Shape: Path Data [id:dp5923398308383396] 
\draw  [line width=1.5]  (145,62.09) -- (145,66.09) -- (185,66.09) -- (185,123) -- (144.17,123) -- (144.17,119) -- (104.17,119) -- (104.17,62.09) -- (145,62.09) -- cycle ;
%Straight Lines [id:da0661931984677957] 
\draw [line width=1.5]    (68.43,91.45) -- (99,91.94) ;
\draw [shift={(103,92)}, rotate = 180.91] [fill={rgb, 255:red, 0; green, 0; blue, 0 }  ][line width=0.08]  [draw opacity=0] (11.61,-5.58) -- (0,0) -- (11.61,5.58) -- cycle    ;
%Straight Lines [id:da34467405320406397] 
\draw [line width=1.5]    (376.49,94.5) -- (403,94.93) ;
\draw [shift={(407,95)}, rotate = 180.94] [fill={rgb, 255:red, 0; green, 0; blue, 0 }  ][line width=0.08]  [draw opacity=0] (11.61,-5.58) -- (0,0) -- (11.61,5.58) -- cycle    ;
%Straight Lines [id:da022191737056530547] 
\draw [line width=1.5]    (464.49,97.5) -- (497,97.05) ;
\draw [shift={(501,97)}, rotate = 539.22] [fill={rgb, 255:red, 0; green, 0; blue, 0 }  ][line width=0.08]  [draw opacity=0] (11.61,-5.58) -- (0,0) -- (11.61,5.58) -- cycle    ;
%Straight Lines [id:da2013061122362101] 
\draw [line width=1.5]    (1,90) -- (36,90) ;
\draw [shift={(40,90)}, rotate = 180] [fill={rgb, 255:red, 0; green, 0; blue, 0 }  ][line width=0.08]  [draw opacity=0] (11.61,-5.58) -- (0,0) -- (11.61,5.58) -- cycle    ;
%Shape: Rectangle [id:dp5064137799607067] 
\draw  [line width=1.5]  (316.5,66) -- (376,66) -- (376,119) -- (316.5,119) -- cycle ;
%Shape: Rectangle [id:dp5099807740867237] 
\draw  [line width=1.5]  (404.5,69) -- (464,69) -- (464,122) -- (404.5,122) -- cycle ;
%Shape: Triangle [id:dp6872168669491976] 
\draw  [line width=1.5]  (532.6,98.75) -- (500.4,123.87) -- (500.6,72.93) -- cycle ;
%Shape: Triangle [id:dp06219002161491982] 
\draw  [line width=1.5]  (415.6,30.75) -- (383.4,55.87) -- (383.6,4.93) -- cycle ;
%Straight Lines [id:da3832375292532042] 
\draw [line width=1.5]    (199,91) -- (200,29) -- (378.5,29.98) ;
\draw [shift={(382.5,30)}, rotate = 180.31] [fill={rgb, 255:red, 0; green, 0; blue, 0 }  ][line width=0.08]  [draw opacity=0] (11.61,-5.58) -- (0,0) -- (11.61,5.58) -- cycle    ;
%Straight Lines [id:da7623898981871877] 
\draw [line width=1.5]    (415.6,30.75) -- (569.5,32) -- (568.61,79.68) ;
\draw [shift={(568.53,83.68)}, rotate = 271.07] [fill={rgb, 255:red, 0; green, 0; blue, 0 }  ][line width=0.08]  [draw opacity=0] (11.61,-5.58) -- (0,0) -- (11.61,5.58) -- cycle    ;
%Shape: Ellipse [id:dp9645205509132346] 
\draw  [line width=1.5]  (553.63,97.45) .. controls (553.63,89.84) and (560.3,83.68) .. (568.53,83.68) .. controls (576.76,83.68) and (583.43,89.84) .. (583.43,97.45) .. controls (583.43,105.06) and (576.76,111.22) .. (568.53,111.22) .. controls (560.3,111.22) and (553.63,105.06) .. (553.63,97.45) -- cycle ;
%Straight Lines [id:da4432835278242917] 
\draw [line width=1.5]    (532.6,98.75) -- (549.64,97.7) ;
\draw [shift={(553.63,97.45)}, rotate = 536.46] [fill={rgb, 255:red, 0; green, 0; blue, 0 }  ][line width=0.08]  [draw opacity=0] (11.61,-5.58) -- (0,0) -- (11.61,5.58) -- cycle    ;
%Straight Lines [id:da43486905417643107] 
\draw [line width=1.5]    (583.43,97.45) -- (610,97.06) ;
\draw [shift={(614,97)}, rotate = 539.1600000000001] [fill={rgb, 255:red, 0; green, 0; blue, 0 }  ][line width=0.08]  [draw opacity=0] (11.61,-5.58) -- (0,0) -- (11.61,5.58) -- cycle    ;
%Shape: Rectangle [id:dp24836675603721092] 
\draw  [line width=1.5]  (258.17,185.09) -- (290,185.09) -- (290,215) -- (258.17,215) -- cycle ;
%Straight Lines [id:da3055159021674416] 
\draw [line width=1.5]    (258.5,200) -- (53.5,201) -- (53.53,107.22) ;
\draw [shift={(53.53,103.22)}, rotate = 450.02] [fill={rgb, 255:red, 0; green, 0; blue, 0 }  ][line width=0.08]  [draw opacity=0] (11.61,-5.58) -- (0,0) -- (11.61,5.58) -- cycle    ;

% Text Node
\draw (53.53,90.61) node  [font=\large,xscale=1.4,yscale=1.4]  {$-$};
% Text Node
\draw (597.62,77.76) node  [font=\large,xscale=1.4,yscale=1.4]  {$y$};
% Text Node
\draw (147,91) node  [font=\large,xscale=1.4,yscale=1.4]  {$\mathbf{C_{\mathbf{R}}}$};
% Text Node
\draw (259.25,90.5) node  [font=\large,xscale=1.4,yscale=1.4]  {$\mathbf{C_{\mathbf{L}_{2}}}$};
% Text Node
\draw (346.25,92.5) node  [font=\large,xscale=1.4,yscale=1.4]  {$\mathbf{G}$};
% Text Node
\draw (434.25,95.5) node  [font=\large,xscale=1.4,yscale=1.4]  {$\mathbf{C_{\mathbf{L}_{1}}}$};
% Text Node
\draw (12.62,71.76) node  [font=\large,xscale=1.4,yscale=1.4]  {$r$};
% Text Node
\draw (511.55,99.47) node  [font=\large,xscale=1.3,yscale=1.3]  {$\beta ^{\prime }$};
% Text Node
\draw (394.55,31.47) node  [font=\large,xscale=1.4,yscale=1.4]  {$\Varrho ^{\prime }$};
% Text Node
\draw (568.53,98.61) node  [font=\large,xscale=1.4,yscale=1.4]  {$+$};
% Text Node
\draw (274.09,200.04) node  [font=\large,xscale=1.4,yscale=1.4]  {$\mathcal{C}_{s}$};
\end{tikzpicture}}
 	\caption{The block diagram of $H_\beta$ condition for the modified architecture Fig.~\ref{F_4-04} with GFORE or PCI}
 	\label{F_4-05}
 \end{figure}   
%%%%%%%%%%%%%%%%%%%%%%%%%%%%%%%%%%%%%%%Proof of Preposition 1
\section{Proof of Proposition \ref{pro_41}}\label{aa_43}
Consider $Q_1\mathcal{F}_1(\omega)+Q_2\mathcal{F}_2(\omega)$ as the scalar product of the two vectors $\vv{\mathcal{F}}(\omega)$ and $\vv{\mathcal{Q}}$. Thus, for all $\omega\in\mathbb{R}^+$, the condition~(\ref{p_4311}) can be re-written as
\begin{equation}\label{Ap_4-02}
\sqrt{Q_1^2+Q_2^2}\sqrt{\mathcal{F}_1^2(\omega)+\mathcal{F}_2^2(\omega)}\cos(\vartheta)>\mathcal{F}_3(\omega).
\end{equation}
As a result, when $\mathcal{F}_3(\omega)\geq0$, $\cos(\vartheta)$ must be positive and
\begin{equation}\label{Ap_4-03}
\sqrt{Q_1^2+Q_2^2}>\underset{\omega\in\omega_p}{\max\ }\dfrac{\mathcal{F}_3(\omega)}{\cos(\vartheta)\sqrt{\mathcal{F}^2_1(\omega)+\mathcal{F}^2_2(\omega)}}=\eta_1(\dfrac{Q_2}{Q_1}).
\end{equation}
Positivity of $\cos(\vartheta)$ implies $\dfrac{Q_2}{Q_1}\in g_p$. When $\mathcal{F}_3(\omega)<0$, there are two solutions for condition~(\ref{Ap_4-02}). $\cos(\vartheta)\geq0$ which requiers $\dfrac{Q_2}{Q_1}\in g_N$, or
\begin{equation}\label{Ap_4-04}
\sqrt{Q_1^2+Q_2^2}<\underset{\omega\in\omega_N}{\min\ }\dfrac{\mathcal{F}_3(\omega)}{\cos(\vartheta)\sqrt{\mathcal{F}^2_1(\omega)+\mathcal{F}^2_2(\omega)}}=\eta_2(\dfrac{Q_2}{Q_1}).
\end{equation}
Therefore, by (\ref{Ap_4-03}) and (\ref{Ap_4-04}) $\eta_2(\dfrac{Q_2}{Q_1})>\eta_1(\dfrac{Q_2}{Q_1})$ and the proof is complete.
 %%%%%%%%%%%%%%%%%%%%%%%%%%%%%%%%%%%%%%% Proof of Corollary 4
\section{Proof of Corollary \ref{CO_44}}\label{AA_44}
First, a preliminary result is stated and proved.
\begin{lemma}\label{L_42}
Consider the reset controllers $C_{R_1}$ and $C_{R_2}$ shown in Fig.~\ref{F_4-07}. Suppose $C_{R_1}$ and $C_{R_2}$ have the same base linear system, are strictly proper, and have different state-space realizations. Then if $A_\rho=\gamma I$ and their initial conditions are zero, $y_1(t)=y_2(t)$, for $t\geq0$. 
\end{lemma}
 \begin{figure}[!t]
 	\centering
	\resizebox{0.3\columnwidth}{!}{%
\tikzset{every picture/.style={line width=0.75pt}} %set default line width to 0.75pt        
\begin{tikzpicture}[x=0.75pt,y=0.75pt,yscale=-1,xscale=1]
%uncomment if require: \path (0,177); %set diagram left start at 0, and has height of 177
%Shape: Path Data [id:dp5923398308383396] 
\draw  [line width=1.5]  (114,6.09) -- (114,10.09) -- (154,10.09) -- (154,67) -- (113.17,67) -- (113.17,63) -- (73.17,63) -- (73.17,6.09) -- (114,6.09) -- cycle ;
%Straight Lines [id:da21224497609988013] 
\draw [line width=1.5]    (2.5,87) -- (45.63,87.45) -- (46,33) -- (68,33.85) ;
\draw [shift={(72,34)}, rotate = 182.2] [fill={rgb, 255:red, 0; green, 0; blue, 0 }  ][line width=0.08]  [draw opacity=0] (11.61,-5.58) -- (0,0) -- (11.61,5.58) -- cycle    ;
%Shape: Path Data [id:dp4537871375673477] 
\draw  [line width=1.5]  (114,106.09) -- (114,110.09) -- (154,110.09) -- (154,167) -- (113.17,167) -- (113.17,163) -- (73.17,163) -- (73.17,106.09) -- (114,106.09) -- cycle ;
%Straight Lines [id:da1342683265261655] 
\draw [line width=1.5]    (45.63,87.45) -- (46,136) -- (67,136) ;
\draw [shift={(71,136)}, rotate = 180] [fill={rgb, 255:red, 0; green, 0; blue, 0 }  ][line width=0.08]  [draw opacity=0] (11.61,-5.58) -- (0,0) -- (11.61,5.58) -- cycle    ;
%Straight Lines [id:da7115202272243253] 
\draw [line width=1.5]    (155.5,37) -- (209,37) ;
\draw [shift={(213,37)}, rotate = 180] [fill={rgb, 255:red, 0; green, 0; blue, 0 }  ][line width=0.08]  [draw opacity=0] (11.61,-5.58) -- (0,0) -- (11.61,5.58) -- cycle    ;
%Straight Lines [id:da4039663983630011] 
\draw [line width=1.5]    (153.5,138) -- (207,138) ;
\draw [shift={(211,138)}, rotate = 180] [fill={rgb, 255:red, 0; green, 0; blue, 0 }  ][line width=0.08]  [draw opacity=0] (11.61,-5.58) -- (0,0) -- (11.61,5.58) -- cycle    ;

% Text Node
\draw (116,35) node  [font=\large,xscale=1.4,yscale=1.4]  {$\mathbf{C_{\mathbf{R}_{1}}}$};
% Text Node
\draw (22.84,71.76) node  [font=\large,xscale=1.4,yscale=1.4]  {$u( t)$};
% Text Node
\draw (116,135) node  [font=\large,xscale=1.4,yscale=1.4]  {$\mathbf{C_{\mathbf{R}_{2}}}$};
% Text Node
\draw (184.84,16.76) node  [font=\large,xscale=1.4,yscale=1.4]  {$y_{1}( t)$};
% Text Node
\draw (185.84,115.76) node  [font=\large,xscale=1.4,yscale=1.4]  {$y_{2}( t)$};
\end{tikzpicture}}
 	\caption{Two reset controllers $C_{R_1}$ and $C_{R_2}$ with different realizations}
 	\label{F_4-07}
 \end{figure}   
\begin{pf}
Let $A_{r_1}$, $B_{r_1}$, and $C_{r_1}$ be the state-space realization of $C_{R_1}(s)$ and $A_{r_2}$, $B_{r_2}$, and $C_{r_2}$ be the state-space realization of $C_{R_2}(s)$. Since the base transfer function of $C_{R_1}(s)$ and $C_{R_2}(s)$ are the same,
\begin{equation}\label{Ap_4-05}
 C_{r_1}\displaystyle\int_{t_0}^{t}e^{A_{r_1}(t-\tau)}B_{r_1}u(\tau)d\tau=C_{r_2}\displaystyle\int_{t_0}^{t}e^{A_{r_2}(t-\tau)}B_{r_2}u(\tau)d\tau.
\end{equation}
Let $t_k,\ k\in\mathbb{N}$, be the reset instants of $C_{R_1}$ and $C_{R_2}$. Since the initial condition is zero, 
\setlength{\arraycolsep}{0pt}
\begin{eqnarray}\label{E_4-07}
 y_1(t)&{=}&\resizebox{0.88\columnwidth}{!}{$C_{r_1}\Big(\displaystyle\int_{t_k}^{t}e^{A_{r_1}(t-\tau)}B_{r_1}u(\tau)d\tau+\gamma\displaystyle\int_{t_{k-1}}^{t_k}e^{A_{r_1}(t-\tau)}B_{r_1}u(\tau)d\tau$}\nonumber\\
 &&{+}\:\cdots+\gamma^k\displaystyle\int_{t_0}^{t_{1}}e^{A_{r_1}(t-\tau)}B_{r_1}u(\tau)d\tau\Big),
\end{eqnarray}
\begin{eqnarray}\label{Ap_4-08}
 y_2(t)&{=}&\resizebox{0.88\columnwidth}{!}{$C_{r_2}\Big(\displaystyle\int_{t_k}^{t}e^{A_{r_2}(t-\tau)}B_{r_2}u(\tau)d\tau+\gamma\displaystyle\int_{t_{k-1}}^{t_k}e^{A_{r_2}(t-\tau)}B_{r_2}u(\tau)d\tau$}\nonumber\\
 &&{+}\: \cdots+\gamma^k\displaystyle\int_{t_0}^{t_{1}}e^{A_{r_2}(t-\tau)}B_{r_2}u(\tau)d\tau\Big).
\end{eqnarray}
\setlength{\arraycolsep}{5pt}As a result, $y_1(t)=y_2(t)$, for all $t\geq0$ by (\ref{Ap_4-05}).
\end{pf} 
The corollary is now proved. Using Lemma \ref{L_42}, since the initial condition of the reset part are zero and $A_\rho=\gamma I$, the states $\zeta(t)$ are the same for different state-space realizations of $C_R(s)$ of the reset control system~(\ref{E_4-27}). Thus, since Theorem~\ref{T_43} holds for the reset control system~(\ref{E_4-27}) with GSORE (\ref{E_4-24}) with the controllable realization (\ref{E_4-25}), $\zeta(t)$, which contains $u_1(t)$ and $u_r(t)$, is also bounded in the reset control system (\ref{E_4-27}) with GSORE (\ref{E_4-24}) with realization configurations (\ref{E_433-19}) and (\ref{E_433-20}). Therefore, it is just needed to show that the reset state $x_r(t)$ is bounded. Consider the reset controller~(\ref{E_4-21}). Now assume that $x_r(t)$ is unbounded, since $u_1(t)$ is bounded and by state-space realization (\ref{E_433-19}) and (\ref{E_433-20}), $u_r(t)$ becomes unbounded which is a contradiction. Therefore, $x_r(t)$ must be bounded and the proof is complete. 
\end{document}